\documentclass[11pt,english]{article}
\usepackage[latin9]{inputenc}
\usepackage[letterpaper]{geometry}
\usepackage{babel}
\usepackage{array}
\usepackage{mathtools}
\usepackage{bm}
\usepackage{multirow}
\usepackage{amsmath}
\usepackage{amsthm}
\usepackage{amssymb}
\usepackage{graphicx}
\usepackage{setspace}
\usepackage[authoryear]{natbib}
\usepackage[unicode=true]
 {hyperref}

\makeatletter

\providecommand{\tabularnewline}{\\}

\theoremstyle{plain}
\newtheorem{assumption}{\protect\assumptionname}
\theoremstyle{plain}
\newtheorem{thm}{\protect\theoremname}
\theoremstyle{plain}
\newtheorem{cor}{\protect\corollaryname}

\usepackage{amsfonts}
\usepackage{amsthm}
\usepackage{fancyhdr}
\usepackage{lastpage}
\usepackage{extramarks}
\usepackage{chngpage}
\usepackage{soul}
\usepackage{color}
\usepackage{graphics}
\usepackage{lscape}
\usepackage{url}

\usepackage{bm}
\usepackage{enumerate}
\usepackage{verbatim}
\usepackage{babel}
\usepackage{titlesec}

\renewcommand{\thefigure}{\arabic{figure}}

\def\ci{\perp\!\!\!\perp}

\usepackage{pifont}

\titleformat{\subsection}
  {\normalfont\large\bfseries}{\thesubsection}{1em}{}

\usepackage{pifont}
\usepackage{pgf}
\usepackage{tikz}\usetikzlibrary{positioning}
\usetikzlibrary{arrows}

\usepackage{footmisc}

\allowdisplaybreaks

\AtBeginDocument{%
  \addtolength\abovedisplayskip{-0.2\baselineskip}%
  \addtolength\belowdisplayskip{-0.2\baselineskip}%
}

\providecommand{\keywords}[1]
{
  \textbf{Keywords:} #1
}

\makeatother

\providecommand{\assumptionname}{Assumption}
\providecommand{\corollaryname}{Corollary}
\providecommand{\theoremname}{Theorem}

\begin{document}
\title{{\Large{}{}\vspace{-60pt}
}Semiparametric Estimation for Causal Mediation Analysis with Multiple
Causally Ordered Mediators\thanks{\protect\normalsize Direct all correspondence to Xiang Zhou, Department
of Sociology, Harvard University, 33 Kirkland Street, Cambridge MA
02138; email: xiang\_zhou@fas.harvard.edu. The author thanks the Editor,
the Associate Editor, two anonymous reviewers, two reviewers from
the Alexander and Diviya Magaro Peer Pre-Review Program, and Aleksei
Opacic for helpful comments.}}
\author{Xiang Zhou\\
 Harvard University}
\date{}
\maketitle
\begin{abstract}
\noindent \sloppy Causal mediation analysis concerns the pathways
through which a treatment affects an outcome. While most of the mediation
literature focuses on settings with a single mediator, a flourishing
line of research has examined settings involving multiple mediators,
under which path-specific effects (PSEs) are often of interest. We
consider estimation of PSEs when the treatment effect operates through
$K(\geq1)$ causally ordered, possibly multivariate mediators. In
this setting, the PSEs for many causal paths are not nonparametrically
identified, and we focus on a set of PSEs that are identified under
Pearl's nonparametric structural equation model. These PSEs are defined
as contrasts between the expectations of $2^{K+1}$ potential outcomes
and identified via what we call the generalized mediation functional
(GMF). We introduce an array of regression-imputation, weighting,
and \textquotedblleft hybrid\textquotedblright{} estimators, and,
in particular, two $K+2$-robust and locally semiparametric efficient
estimators for the GMF. The latter estimators are well suited to the
use of data-adaptive methods for estimating their nuisance functions.
We establish the rate conditions required of the nuisance functions
for semiparametric efficiency. We also discuss how our framework applies
to several estimands that may be of particular interest in empirical
applications. The proposed estimators are illustrated with a simulation
study and an empirical example.
\end{abstract}

\clearpage\sloppy

\section{Introduction}

Causal mediation analysis aims to disentangle the pathways through
which a treatment affects an outcome. While traditional approaches
to mediation analysis have relied on linear structural equation models,
along with their stringent parametric assumptions, to define and estimate
direct and indirect effects (e.g., \citealt{baron1986moderator}),
a large body of research has emerged within the causal inference literature
that disentangles the tasks of definition, identification, and estimation
in the study of causal mechanisms. Using the potential outcomes framework
(\citealt{neyman1923application,Rubin1974688}), this body of research
has provided model-free definitions of direct and indirect effects
(\citealt{robins1992direct,pearl2001direct}), established the assumptions
needed for nonparametric identification (\citealt{robins1992direct,pearl2001direct,robins2003semantics,petersen2006estimation,imai2010identification,hafeman2011alternative,vanderweele2015explanation}),
and developed an array of imputation, weighting, and multiply robust
methods for estimation (e.g., \citealt{goetgeluk2008estimation,albert2012mediation,tchetgen2012semiparametric,vansteelandt2012imputation,zheng2012targeted,tchetgen2013inverse,vanderweele2015explanation,wodtke2020effect}).

While the bulk of the causal mediation literature focuses on settings
with a single mediator (or a set of mediators considered as a whole),
a flourishing line of research has studied settings that involve multiple
causally dependent mediators, under which a set of path-specific effects
(PSEs) are often of interest (\citealt{avin2005identifiability,albert2011generalized,shpitser2013counterfactual,vanderweele2014mediation,vandwerweele2014effectdecomp,daniel2015causal,lin2017interventional,miles2017quantifying,steen2017flexible,vansteelandt2017interventional,miles2020semiparametric}).
In particular, \citet{daniel2015causal} demonstrated a large number
of ways in which the total effect of a treatment can be decomposed
into PSEs, established the assumptions under which a subset of these
PSEs are identified, and provided a parametric method for estimating
these effects (see also \citealt{albert2011generalized}). More recently,
for a particular PSE in the case of two causally ordered mediators,
\citet{miles2020semiparametric} offered an in-depth discussion of
alternative estimation methods, and, utilizing the efficient influence
function of its identification formula, developed a triply robust
and locally semiparametric efficient estimator. This estimator, by
virtue of its multiple robustness, is well suited to the use of data-adaptive
methods for estimating its nuisance functions.

To date, most of the literature on PSEs has focused on the case of
two mediators, and it remains underexplored how the estimation methods
developed in previous studies, such as those in \citet{vandwerweele2014effectdecomp}
and \citet{miles2020semiparametric}, generalize to the case of $K(\geq1)$
causally ordered mediators. This article aims to bridge this gap.
First, we observe that despite a multitude of ways in which a PSE
can be defined for each causal path from the treatment to the outcome,
most of these PSEs are not identified under Pearl's nonparametric
structural equation model. This observation leads us to focus on the
much smaller set of PSEs that \textit{can be }nonparametrically identified.
These PSEs are defined as contrasts between the expectations of $2^{K+1}$
potential outcomes, which, in turn, are identified through a formula
that can be viewed as an extension of \citeauthor{pearl2001direct}'s
(2001) and \citeauthor{daniel2015causal}'s (2015) mediation formulae
to the case of $K$ causally ordered mediators. Following \citet{tchetgen2012semiparametric},
we refer to the identification formula for these expected potential
outcomes as the \textit{generalized mediation functional} (GMF).

We then show that the GMF can be estimated via an array of regression,
weighting, and ``hybrid'' estimators. More important, building on
its efficient influence function (EIF), we develop two multiply robust
and locally semiparametric efficient estimators for the GMF. Both
of these estimators are $K+2$-robust, in the sense that they are
consistent provided that one of $K+2$ sets of nuisance functions
is correctly specified and consistently estimated. These multiply
robust estimators are well suited to the use of data-adaptive methods
for estimating the nuisance functions. We establish rate conditions
for consistency and semiparametric efficiency when data-adaptive methods
and cross-fitting (\citealt{zheng2011cross,chernozhukov2018double})
are used to estimate the nuisance functions.

Compared with existing estimators that have been proposed for causal
mediation analysis, the methodology proposed in this article is distinct
in its generality. In fact, the doubly robust estimator for the mean
of an incomplete outcome (\citealt{scharfstein1999adjusting}), the
triply robust estimator developed by \citet{tchetgen2012semiparametric}
for the mediation functional in the one-mediator setting (see also
\citealt{zheng2012targeted}), and the estimator proposed by \citet{miles2020semiparametric}
for their particular PSE, can all be viewed as special cases of the
$K+2$-robust estimators --- when $K=0,1,2$, respectively. Yet,
our framework also encompasses important estimands for which semiparametric
estimators have not been proposed. To demonstrate the generality of
our framework, we show how our multiply robust semiparametric estimators
apply to several estimands that may be of particular interest in empirical
applications, including the natural direct effect (NDE), the natural/total
indirect effect (NIE/TIE), the natural path-specific effect (nPSE),
and the cumulative path-specific effect (cPSE). In Supplementary Material
\ref{sec:Decomposition-of-Between-group}, we discuss how our framework
can also be employed to estimate noncausal decompositions of between-group
disparities that are widely used in social science research (\citealt{fortin2011decomposition}).

Before proceeding, we note that in a separate strand of literature,
the term ``multiple robustness'' has been used to characterize a
class of estimators for the mean of incomplete data that are consistent
if one of several working models for the propensity score or one of
several working models for the outcome is correctly specified (e.g.,
\citealt{han2013estimation,han2014multiply}). In this paper, we use
``$V$-robustness'' to characterize estimators that require modeling
\textit{multiple parts of the observed data likelihood} and are consistent
provided that one of $V$ sets of the corresponding models is correctly
specified, in keeping with the terminology in the causal mediation
literature. This definition of ``multiple robustness'' does not
imply that a ``$K+2$-robust'' estimator is necessarily more robust
than, for example, a ``$K+1$-robust'' estimator. First, they may
correspond to different estimands that require modeling different
parts of the likelihood. For example, the doubly robust estimator
of the average treatment effect only involves a propensity score model
and an outcome model; it is thus less demanding than \citeauthor{tchetgen2012semiparametric}'s
(2012) triply robust estimator of the mediation functional, which
involves an additional model for the mediator. Second, for our semiparametric
estimators of the GMF, the ``$K+2$-robustness'' property is not
``sharp'' because it can be tightened in various special cases.
As we demonstrate in Section \ref{sec:Special-Cases} and Supplementary
Material \ref{sec:Decomposition-of-Between-group}, such a tightening
may result in a lower $V$ (as in the case of NDE, NIE/TIE, nPSE,
and cPSE), or a higher $V$ (as in the case of noncausal decompositions
of between-group disparities).

The rest of the paper is organized as follows. In Section 2, we define
the PSEs of interest, lay out their identification assumptions, and
introduce the GMF. In Section 3, we introduce a range of regression-imputation,
weighting, ``hybrid,'' and multiply robust estimators for the GMF,
and present several techniques that could be used to improve the finite
sample performance of the multiply robust estimators. In Section 4,
we discuss how our results apply to a number of special cases such
as the NDE, NIE/TIE, nPSE, and cPSE. A simulation study and an empirical
example are given in Section 5 and Section 6 to illustrate the proposed
estimators. Proofs of Theorems 1-4 are given in Supplementary Materials
A, C, and D. Replication data and code for the simulation study and
the empirical example are available at \href{https://doi.org/10.7910/DVN/5TBUM3}{https://doi.org/10.7910/DVN/5TBUM3}.
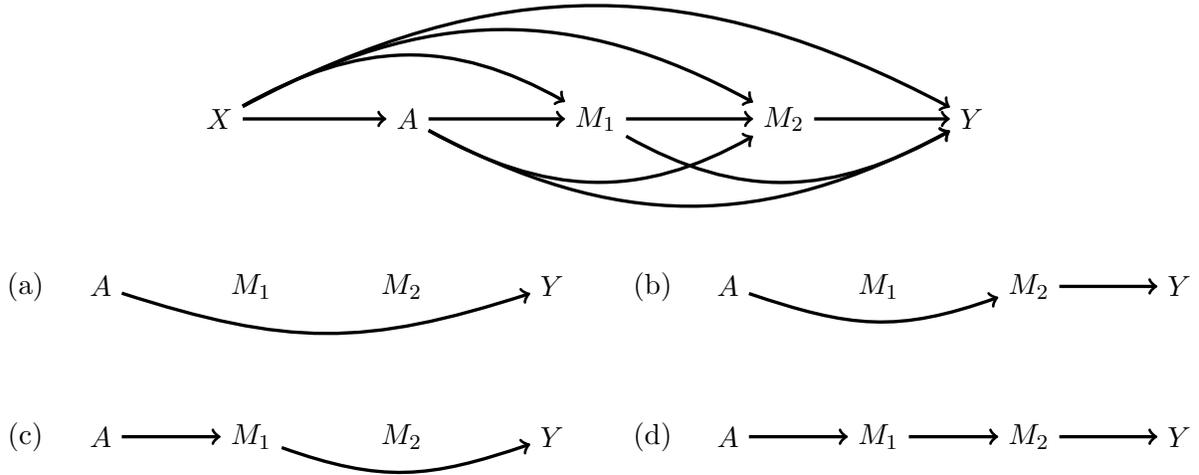
\begin{figure}[!th]
\noindent \begin{centering}
\begin{minipage}[t]{0.7\columnwidth}%
\noindent \begin{center}
\begin{tikzpicture}[yscale = 1.5, xscale = 2.5]
\node[text centered] at (-2,0) (x) {$X$};
\node[text centered] at (-1,0) (a) {$A$};
\node[text centered] at (1,0) (m) {$M_2$};
\node[text centered] at (0,0) (l) {$M_1$};
\node[text centered] at (2,0) (y) {$Y$};
\draw [->, line width= 1.25] (x) -- (a);
\draw [->, line width= 1.25] (x) to [out=40, in=140, looseness=1.2] (l);
\draw [->, line width= 1.25] (x) to [out=40, in=140, looseness=1.2] (y);
\draw [->, line width= 1.25] (x) to [out=40, in=140, looseness=1.2] (m);
\draw [->, line width= 1.25] (a) -- (l);
\draw [->, line width= 1.25] (a) to [out=-40, in=-140, looseness=1.2] (m);
\draw [->, line width= 1.25] (a) to [out=-40, in=-140, looseness=1.2] (y);
\draw [->, line width= 1.25] (l) -- (m);
\draw [->, line width= 1.25] (l) to [out=-40, in=-140, looseness=1.2] (y);
\draw [->, line width= 1.25] (m) -- (y); 
\end{tikzpicture}
\par\end{center}%
\end{minipage}\vfill{}
\begin{minipage}[t]{0.45\columnwidth}%
\noindent \begin{center}
\begin{tikzpicture}[yscale = 1, xscale = 2]
\node[text centered] at (-1.5,0) (p) {(a)};
\node[text centered] at (-1,0) (a) {$A$};
\node[text centered] at (1,0) (m) {$M_2$};
\node[text centered] at (0,0) (l) {$M_1$};
\node[text centered] at (2,0) (y) {$Y$};
\draw [->, line width= 1.25] (a) to [out=-30, in=-150, looseness=1.2] (y);

\node[text centered] at (-1.5,-2) (p) {(c)};
\node[text centered] at (-1,-2) (a) {$A$};
\node[text centered] at (1,-2) (m) {$M_2$};
\node[text centered] at (0,-2) (l) {$M_1$};
\node[text centered] at (2,-2) (y) {$Y$};
\draw [->, line width= 1.25] (a) -- (l);
\draw [->, line width= 1.25] (l) to [out=-30, in=-150, looseness=1.2] (y);
\end{tikzpicture}
\par\end{center}%
\end{minipage} \qquad{}%
\begin{minipage}[t]{0.45\columnwidth}%
\noindent \begin{center}
\begin{tikzpicture}[yscale = 1, xscale = 2]
\node[text centered] at (-1.5,0) (p) {(b)};
\node[text centered] at (-1,0) (a) {$A$};
\node[text centered] at (1,0) (m) {$M_2$};
\node[text centered] at (0,0) (l) {$M_1$};
\node[text centered] at (2,0) (y) {$Y$};
\draw [->, line width= 1.25] (a) to [out=-30, in=-150, looseness=1.2] (m);
\draw [->, line width= 1.25] (m) -- (y); 
\node[text centered] at (-1.5,-2) (p) {(d)};
\node[text centered] at (-1,-2) (a) {$A$};
\node[text centered] at (1,-2) (m) {$M_2$};
\node[text centered] at (0,-2) (l) {$M_1$};
\node[text centered] at (2,-2) (y) {$Y$};
\draw [->, line width= 1.25] (a) -- (l);
\draw [->, line width= 1.25] (l) -- (m);
\draw [->, line width= 1.25] (m) -- (y); 
\end{tikzpicture}
\par\end{center}%
\end{minipage} 
\par\end{centering}
\begin{centering}
\caption{Causal relationships with two causally ordered mediators. \label{fig:dag}}
\medskip{}
\par\end{centering}
Note: $A$ denotes the treatment, $Y$ denotes the outcome of interest,
$X$ denotes a vector of pretreatment covariates, and $M_{1}$ and
$M_{2}$ denote two causally ordered mediators.
\end{figure}

\section{Notation, Definitions, and Identification}

To ease exposition, we start with the case of two causally ordered
mediators before moving onto the general setting of $K$ mediators.

\subsection{The Case of Two Causally Ordered Mediators\label{subsec:Two Mediators}}

Let $A$ denote a binary treatment, $Y$ an outcome of interest, and
$X$ a vector of pretreatment covariates. In addition, let $M_{1}$
and $M_{2}$ denote two causally ordered mediators, and assume $M_{1}$
precedes $M_{2}$. We allow each of these mediators to be multivariate,
in which case the causal relationships among the component variables
are left unspecified. A directed acyclic graph (DAG) representing
the relationships between these variables is given in the top panel
of Figure~\ref{fig:dag}. In this DAG, four possible causal paths
exist from the treatment to the outcome, as shown in the lower panels:
(a) $A\to Y$; (b) $A\to M_{2}\to Y$; (c) $A\to M_{1}\to Y$; and
(d) $A\to M_{1}\to M_{2}\to Y$.

A formal definition of path-specific effects (PSEs) requires the potential-outcomes
notation for both the outcome and the mediators. Specifically, let
$Y(a,m_{1},m_{2})$ denote the potential outcome under treatment status
$a$ and mediator values $M_{1}=m_{1}$ and $M_{2}=m_{2}$, $M_{2}(a,m_{1})$
the potential value of the mediator $M_{2}$ under treatment status
$a$ and mediator value $M_{1}=m_{1}$, and $M_{1}(a)$ the potential
value of the mediator $M_{1}$ under treatment status $a$. This notation
allows us to define nested counterfactuals in the form of $Y\big(a,M_{1}(a_{1}),M_{2}(a_{2},M_{1}(a_{12}))\big)$,
where $a$, $a_{1}$, $a_{2}$, and $a_{12}$ can each take 0 or 1.
For example, $Y\big(1,M_{1}(0),M_{2}(0,M_{1}(0))\big)$ represents
the potential outcome in the hypothetical scenario where the subject
was treated but the mediators $M_{1}$ and $M_{2}$ were set to values
they would have taken if the subject had not been treated. Further,
if we let $Y(a)$ denote the potential outcome when treatment status
is set to $a$ and the mediators $M_{1}$ and $M_{2}$ take on their
``natural'' values under treatment status $a$ (i.e., $M_{1}(a)$
and $M_{2}(a,M_{1}(a))$), we have $Y(a)=Y\big(a,M_{1}(a),M_{2}(a,M_{1}(a))\big)$
by construction. This is sometimes referred to as the ``composition''
assumption (\citealt{vanderweele2009concerning}).

Under the above notation, for each of the causal paths shown in Figure
\ref{fig:dag}, its PSE can be defined in eight different ways, depending
on the reference levels chosen for $A$ for each of the other three
paths (\citealt{daniel2015causal}). For example, the average direct
effect of $A$ on $Y$, i.e., the portion of the treatment effect
that operates through the path $A\to Y$, can be defined as
\[
\tau_{A\to Y}(a_{1},a_{2},a_{12})=\mathbb{E}[Y\big(1,M_{1}(a_{1}),M_{2}(a_{2},M_{1}(a_{12}))\big)-Y\big(0,M_{1}(a_{1}),M_{2}(a_{2},M_{1}(a_{12}))\big)],
\]
where $a_{1}$, $a_{2}$, and $a_{12}$ can each take 0 or 1. In particular,
$\tau_{A\to Y}(0,0,0)$ corresponds to the natural direct effect (NDE;
\citealt{pearl2001direct}) or pure direct effect (PDE; \citealt{robins1992direct})
if the mediators $M_{1}$ and $M_{2}$ are considered as a whole.
In a similar vein, the PSEs via $A\to M_{2}\to Y$, $A\to M_{1}\to Y$,
and $A\to M_{1}\to M_{2}\to Y$ can be defined as
\begin{align*}
\tau_{A\to M_{2}\to Y}(a,a_{1},a_{12}) & =\mathbb{E}[Y\big(a,M_{1}(a_{1}),M_{2}(1,M_{1}(a_{12}))\big)-Y\big(a,M_{1}(a_{1}),M_{2}(0,M_{1}(a_{12}))\big)],\\
\tau_{A\to M_{1}\to Y}(a,a_{2},a_{12}) & =\mathbb{E}[Y\big(a,M_{1}(1),M_{2}(a_{2},M_{1}(a_{12}))\big)-Y\big(a,M_{1}(0),M_{2}(a_{2},M_{1}(a_{12}))\big)],\\
\tau_{A\to M_{1}\to M_{2}\to Y}(a,a_{1},a_{2}) & =\mathbb{E}[Y\big(a,M_{1}(a_{1}),M_{2}(a_{2},M_{1}(1))\big)-Y\big(a,M_{1}(a_{1}),M_{2}(a_{2},M_{1}(0))\big)].
\end{align*}
In addition, if we use $A\to M_{1}\rightsquigarrow Y$ to denote the
combination of the causal paths $A\to M_{1}\to Y$ and $A\to M_{1}\to M_{2}\to Y$,
the corresponding PSE for this ``composite path'' can be defined
as 
\[
\tau_{A\to M_{1}\rightsquigarrow Y}(a,a_{2})=\mathbb{E}[Y\big(a,M_{1}(1),M_{2}(a_{2},M_{1}(1))\big)-Y\big(a,M_{1}(0),M_{2}(a_{2},M_{1}(0))\big)].
\]
This quantity reflects the portion of the treatment effect that operates
through $M_{1}$, regardless of whether it further operates through
$M_{2}$ or not. In particular, $\tau_{A\to M_{1}\rightsquigarrow Y}(0,0)$
is often referred to as the natural indirect effect (NIE; \citealt{pearl2001direct})
or the pure indirect effect (PIE; \citealt{robins1992direct}) for
$M_{1}$, whereas $\tau_{A\to M_{1}\rightsquigarrow Y}(1,1)$ is sometimes
called the total indirect effect (TIE; \citealt{robins1992direct})
for $M_{1}$. Note, however, that the term NIE has also been used
to denote $\tau_{A\to M_{1}\rightsquigarrow Y}(1,1)$ (e.g., \citealt{tchetgen2012semiparametric}).
To avoid ambiguity, we use NIE and TIE to denote $\tau_{A\to M_{1}\rightsquigarrow Y}(0,0)$
and $\tau_{A\to M_{1}\rightsquigarrow Y}(1,1)$, respectively. By
definition, these PSEs are identified if the corresponding expected
potential outcomes, i.e., $\mathbb{E}[Y\big(a,M_{1}(a_{1}),M_{2}(a_{2},M_{1}(a_{12}))\big)]$,
are identified. Below, we review the assumptions under which these
expected potential outcomes are identified from observed data.

Following \citet{pearl2009}, we use a DAG to encode a nonparametric
structural equation model (NPSEM) with mutually independent errors.
In this framework, the top panel of Figure~\ref{fig:dag} implies
no unobserved confounding for any of the treatment-mediator, treatment-outcome,
mediator-mediator, and mediator-outcome relationships. Formally, we
invoke the following assumptions.
\begin{assumption}
Consistency of $A$ on $M_{1}$, $(A,M_{1})$ on $M_{2}$, and $(A,M_{1},M_{2})$
on $Y$: For any unit and any $a,m_{1},m_{2}$, $M_{1}=M_{1}(a)$
if $A=a$; $M_{2}=M_{2}(a,m_{1})$ if $A=a$ and $M_{1}=m_{1}$; and
$Y=Y(a,m_{1},m_{2})$ if $A=a$, $M_{1}=m_{1}$, and $M_{2}=m_{2}$.
\end{assumption}
\begin{assumption}
Conditional independence among treatment and potential outcomes: for
any $a,a_{1},a_{2},m_{1},m_{1}^{*},m_{2}$, $\big(M_{1}(a_{1}),M_{2}(a_{2},m_{1}),Y(a,m_{1},m_{2})\big)\ci A|X$;
$\big(M_{2}(a_{2},m_{1}),Y(a,m_{1},m_{2})\big)\ci M_{1}(a_{1})|X,A$,
and $Y(a,m_{1},m_{2})\ci M_{2}(a_{2},m_{1}^{*})|X,A,M_{1}$.
\end{assumption}
\begin{assumption}
Positivity: $p_{A|X}(a|x)>\epsilon>0$ whenever $p_{X}(x)>0$; $p_{A|X,M_{1}}(a|x,m_{1})>\epsilon>0$
whenever $p_{X,M_{1}}(x,m_{1})>0$, and $p_{A|X,M_{1},M_{2}}(a|x,m_{1},m_{2})>\epsilon>0$
whenever $p_{X,M,M_{2}}(x,m_{1},m_{2})>0$, where $p(\cdot)$ denotes
a probability density/mass function.
\end{assumption}
Note that Assumption 2 involves conditional independence relationships
between the so-called cross-world counterfactuals, such as $\big(M_{2}(a_{2},m_{1}),Y(a,m_{1},m_{2})\big)\ci M_{1}(a_{1})|X,A$.
This assumption is a direct consequence of Pearl's NPSEM with mutually
independent errors. It implies, but is not implied by, the sequential
ignorability assumption that \citet{robins2003semantics} invokes
in interpreting causal diagrams (see \citealt{robins2010alternative}
for an in-depth discussion). In addition, we note that Assumption
2 does not rule out all forms of unobserved confounding for the causal
effects of $X$ on its descendants. For example, unobserved variables
are permitted (although not shown) in Figure \ref{fig:dag} that affect
both $X$ and $Y$.

Under Assumptions 1-3, it can be shown that $\mathbb{E}[Y\big(a,M_{1}(a_{1}),M_{2}(a_{2},M_{1}(a_{12}))\big)]$
is identified if and only if $a_{12}=a_{1}$ (\citealt{avin2005identifiability,albert2011generalized,daniel2015causal}).
Consequently, none of the PSEs for the path $A\to M_{1}\to Y$ is
identified because given $a_{12}$, either $\mathbb{E}[Y\big(a,M_{1}(1),M_{2}(a_{2},M_{1}(a_{12}))\big)]$
or $\mathbb{E}[Y\big(a,M_{1}(0),M_{2}(a_{2},M_{1}(a_{12}))\big)]$
is unidentified. Similarly, none of the PSEs for the path $A\to M_{1}\to M_{2}\to Y$
is identified. Interestingly, the PSEs for the composite path $A\to M_{1}\rightsquigarrow Y$
are all identified, even if $a\neq a_{2}$. These results echo the
recanting witness criterion developed by \citet{avin2005identifiability},
which implies that the PSE for a (possibly composite) path from $A$
to $Y$ when $A$ is set to 0 (or 1) for all other paths is identified
if and only if the path of interest contains no ``recanting witness''
--- a variable $W$ that has an additional path to $Y$ that is not
contained in the path of interest. Thus the PSE $\tau_{A\to M_{1}\to Y}(0,0,0)$
is not identified because $M_{1}$ has an additional path to $Y$
($M_{1}\to M_{2}\to Y$) that is not contained in $A\to M_{1}\to Y$,
but the PSE $\tau_{A\to M_{1}\rightsquigarrow Y}(0,0)$ is identified
because all possible paths from $M_{1}$ to $Y$ is contained in $A\to M_{1}\rightsquigarrow Y$.

Because $\mathbb{E}[Y\big(a,M_{1}(a_{1}),M_{2}(a_{2},M_{1}(a_{12}))\big)]$
is identified if and only if $a_{1}=a_{12}$, we restrict our attention
to cases where $a_{1}=a_{12}$ and use the following notation:
\[
\psi_{a_{1},a_{2},a}\stackrel{\Delta}{=}\mathbb{E}\big[Y\big(a,M_{1}(a_{1}),M_{2}(a_{2},M_{1}(a_{1}))\big)\big].
\]
Under Assumptions 1-3, $\psi_{a_{1},a_{2},a}$ is identified via the
following formula:
\begin{align}
\psi_{a_{1},a_{2},a} & =\iiint\mathbb{E}[Y|x,a,m_{1},m_{2}]dP(m_{2}|x,a_{2},m_{1})dP(m_{1}|x,a_{1})dP(x).\label{eq:mFormula}
\end{align}
For a proof of the above formula, see \citet{daniel2015causal}. Equation
\eqref{eq:mFormula} can be seen as an extension of \citeauthor{pearl2001direct}'s
(2001) mediation formula to the case of two causally ordered mediators.

It should be noted that Assumptions 1-3 constitute a sufficient set
of conditions that allow us to identify $\psi_{a_{1},a_{2},a}$ for
arbitrary combinations of $a_{1}$, $a_{2}$, and $a$. For specific
combinations of $a_{1}$, $a_{2}$, and $a$, Assumption 2 can be
relaxed. For example, $\psi_{100}$ is still identified via equation
\eqref{eq:mFormula} when unobserved confounding exists for the $M_{2}$-$Y$
relationship, and $\psi_{010}$ is still identified via equation \eqref{eq:mFormula}
when unobserved confounding exists for the $M_{1}$-$Y$ relationship
(\citealt{shpitser2013counterfactual,miles2020semiparametric}).\renewcommand{\theassumption}{\arabic{assumption}$^*$}
\setcounter{assumption}{0} 

\subsection{The Case of $K(\protect\geq1)$ Causally Ordered Mediators\label{subsec:K Mediators}}

We now generalize the preceding results to the setting where the treatment
effect of $A$ on $Y$ operates through $K$ causally ordered, possibly
multivariate mediators, $M_{1},M_{2},\ldots M_{K}$. We assume that
for any $k<k'$, $M_{k}$ precedes $M_{k'}$, such that no component
of $M_{k'}$ causally affects any component of $M_{k}$. In a DAG
that is consistent with this setup, a directed path from the treatment
to the outcome can pass through any combination of the $K$ mediators,
resulting in $2^{K}$ possible paths. Among the $2^{K}$ paths, each
can be switched ``on'' or ``off,'' creating $2^{2^{K}}$ potential
outcomes. Also, for each of the $2^{K}$ paths, the corresponding
PSE can be defined in $2^{2^{K}-1}$ different ways, depending on
whether each of the other $2^{K}-1$ paths is switched ``on'' or
``off.'' For example, when $K=3$, for each causal path from $A$
to $Y$, its PSE can be defined in $2^{2^{3}-1}=128$ different ways.

As we will see, despite the exponential growth of possible causal
paths and the double exponential growth of possible PSEs, most of
these PSEs are not identified under the assumptions associated with
Pearl's NPSEM. To fix ideas, let an overbar denote a vector of variables,
so that $\overline{M}_{k}=(M_{1},M_{2},\ldots M_{k})$, $\overline{m}_{k}=(m_{1},m_{2},\ldots m_{k})$,
and $\overline{a}_{k}=(a_{1},a_{2},\ldots a_{k})$, where $\overline{M}_{l}=\overline{m}_{l}=\overline{a}_{l}=\varnothing$
if $l\leq0$. In addition, let $[K]$ denote the set $\{1,2,\ldots K\}$,
and let $a_{K+1}$, instead of $a$, denote the treatment status set
to the path $A\to Y$. Assumptions 1-3 can now be generalized as below.
\begin{assumption}
Consistency: For any unit, $M_{k}=M_{k}(a_{k},\overline{m}_{k-1})$
if $A=a_{k}$ and $\overline{M}_{k-1}=\overline{m}_{k-1}$, $\forall k\in[K]$;
and $Y=Y(a_{K+1},\overline{m}_{K})$ if $A=a_{K+1}$ and $\overline{M}_{K}=\overline{m}_{K}$.
\end{assumption}
\begin{assumption}
Conditional independence among treatment and potential outcomes: $\big(M_{1}(a_{1}),M_{2}(a_{2},\overline{m}_{1}),\ldots Y(a_{K+1},\overline{m}_{K})\big)\ci A|X$;
and $\big(M_{k+1}(a_{k+1},\overline{m}_{k}),\ldots M_{K}(a_{K},\overline{m}_{K-1}),Y(a_{K+1},\overline{m}_{K})\big)\ci M_{k}(a_{k},\overline{m}_{k-1}^{*})|X,A,\overline{M}_{k-1},\forall k\in[K]$.
\end{assumption}
\begin{assumption}
Positivity: $p_{A|X}(a|x)>\epsilon>0$ whenever $p_{X}(x)>0$; $p_{A|X,\overline{M}_{k}}(a|x,\overline{m}_{k})>\epsilon>0$
whenever $p_{X,\overline{M}_{k}}(x,\overline{m}_{k})>0$, $\forall k\in[K]$.
\end{assumption}
Before giving the identification results, we introduce the following
notational shorthands:
\begin{align*}
\overline{M}_{k}(\overline{a}_{k}) & \stackrel{\Delta}{=}\big(\overline{M}_{k-1}(\overline{a}_{k-1}),M_{k}(a_{k},\overline{M}_{k-1}(\overline{a}_{k-1}))\big),\forall k\in[K],\\
\psi_{\overline{a}} & \stackrel{\Delta}{=}\mathbb{E}[Y(a_{K+1},\overline{M}_{k}(\overline{a}_{k}))],
\end{align*}
where $\overline{M}_{k}(\overline{a}_{k})$ is defined iteratively,
with the assumption that $\overline{M}_{0}(\overline{a}_{0})=\varnothing$.
For example, when $K=3$,
\[
\psi_{\overline{a}}=\mathbb{E}\big[Y\big(a_{4},M_{1}(a_{1}),M_{2}(a_{2},M_{1}(a_{1})),M_{3}(a_{3},M_{1}(a_{1}),M_{2}(a_{2},M_{1}(a_{1})))\big)\big].
\]
Theorem 1 states that $\psi_{\overline{a}}$ is identified under Assumptions
1{*}-3{*}.
\begin{thm}
Under Assumptions 1{*}-3{*}, we have
\begin{equation}
\psi_{\overline{a}}=\int_{x}\int_{\overline{m}_{K}}\mathbb{E}[Y|x,a_{K+1},\overline{m}_{K}]\big[\prod\limits _{k=1}^{K}dP(m_{k}|x,a_{k},\overline{m}_{k-1})\big]dP(x).\label{eq:g-mformula}
\end{equation}
\end{thm}
The above equation extends \citeauthor{pearl2001direct}'s (2001)
and \citeauthor{daniel2015causal}'s (2015) mediation formula to the
case of $K$ causally ordered mediators. Following the terminology
of \citet{tchetgen2012semiparametric}, we refer to the right-hand
side of equation \eqref{eq:mFormula} as the generalized mediation
functional (GMF). Theorem 1 echoes \citeauthor{avin2005identifiability}'s
(2005) recanting witness criterion: a potential outcome is identified
(in expectation) if the value that a mediator $M_{k}$ takes, i.e.,
$M_{k}(a_{k})$, is carried over to all future mediators. This result
leads us to focus on the set of expected potential outcomes and PSEs
that are nonparametrically identified. For example, to assess the
mediating role of $M_{k}$, we focus on the composite causal path
$A\to M_{k}\rightsquigarrow Y$, where, as before, the squiggle arrow
encompasses all possible causal paths from $M_{k}$ to $Y$. An identifiable
PSE for this path can be expressed as 
\[
\tau_{A\to M_{k}\rightsquigarrow Y}(\overline{a}_{k-1},\cdot,\underline{a}_{k+1})=\psi_{\overline{a}_{k-1},1,\underline{a}_{k+1}}-\psi_{\overline{a}_{k-1},0,\underline{a}_{k+1}},
\]
where $\underline{a}_{k+1}\stackrel{\Delta}{=}(a_{k+1},\ldots a_{K+1})$.
The notation $\psi_{\overline{a}}$ makes it clear that the average
total effect (ATE) of $A$ on $Y$ can be decomposed into $K+1$ identifiable
PSEs corresponding to $A\to Y$ and $A\to M_{k}\rightsquigarrow Y$
($k\in[K]$):
\begin{equation}
\textup{ATE}=\psi_{\overline{1}}-\psi_{\overline{0}}=\underbrace{\psi_{\overline{0}_{K},1}-\psi_{\overline{0}_{K+1}}}_{A\to Y}+\sum_{k=1}^{K}\underbrace{\big(\psi_{\overline{0}_{k-1},\underline{1}_{k}}-\psi_{\overline{0}_{k},\underline{1}_{k+1}}\big)}_{A\to M_{k}\rightsquigarrow Y}.\label{eq:g-decomp1}
\end{equation}
To be sure, equation \eqref{eq:g-decomp1} is not the only way of
decomposing the ATE. Depending on the order in which the paths $A\to Y$
and $A\to M_{k}\rightsquigarrow Y$ ($k\in[K]$) are considered, there
are $(K+1)!$ different ways of decomposing the ATE. In the above
decomposition, $\psi_{\overline{0}_{K},1}-\psi_{\overline{0}_{K+1}}$
corresponds to the NDE if the mediators $\overline{M}_{K}$ are considered
as a whole.

\section{Estimation\label{sec:Estimation}}

In this section, we focus on the estimation of the GMF, i.e., the
right-hand side of equation \eqref{eq:g-mformula}. When Assumptions
1{*}-3{*} hold, the GMF is equal to the causal parameter $\psi_{\overline{a}}$,
but otherwise, it is still a well-defined statistical parameter of
potential scientific interest. To distinguish it from the causal parameter
$\psi_{\overline{a}}$, we henceforth denote the GMF by $\overline{\theta}_{a}$. 

\subsection{MLE, Regression-Imputation, and Weighting\label{subsec: Estimation}}

Equation \eqref{eq:g-mformula} suggests that $\theta_{\overline{a}}$
can be estimated via maximum likelihood (MLE) (\citealt{miles2017quantifying}).
Specifically, we can fit a parametric model for each $p(m_{k}|x,a_{k},\overline{m}_{k-1})$
($k\in[K]$) and for $\mathbb{E}[Y|x,a_{K+1},\overline{m}_{K}]$,
and then estimate the GMF via the following equation:
\begin{equation}
\hat{\theta}_{\overline{a}}^{\textup{mle}}=\mathbb{P}_{n}\big[\int_{\overline{m}_{K}}\hat{\mathbb{E}}[Y|X,a_{K+1},\overline{m}_{K}]\big(\prod\limits _{k=1}^{K}\hat{p}(m_{k}|x,a_{k},\overline{m}_{k-1})d\nu(m_{k})\big)\big],\label{eq:MLE}
\end{equation}
where $\mathbb{P}_{n}[\cdot]=n^{-1}\sum_{i}[\cdot]_{i}$ and $\nu(\cdot)$
is an appropriate dominating measure. This approach works best when
the mediators $M_{1},M_{2},\ldots M_{K}$ are all discrete and the
covariates $X$ are low-dimensional, in which case the working models
for $p(m_{k}|x,a_{k},\overline{m}_{k-1})$ are simply models for the
conditional probabilities of $M_{k}$ that can be reliably estimated.
When some of the mediators are continuous/multivariate or when the
covariates $X$ are high-dimensional, estimates of the corresponding
conditional density/probability functions can be unstable and sensitive
to model misspecification. This problem could be mitigated by imposing
highly constrained functional forms on the conditional means of the
mediators and the outcome. For example, when $\mathbb{E}[M_{k}|x,a_{k},\overline{m}_{k-1}]$
and $\mathbb{E}[Y|x,a_{K+1},\overline{m}_{K}]$ are all assumed to
be linear with no higher-order or interaction terms, $\hat{\theta}_{\overline{a}}^{\textup{mle}}$
will reduce to a simple function of regression coefficients (e.g.,
\citealt{alwin1975decomposition}). Yet, the assumptions of linearity
and additivity are unrealistic in many applications, which may lead
to biased estimates of $\theta_{\overline{a}}$. Below, we describe
several imputation- and weighting-based strategies for estimating
$\theta_{\overline{a}}$.

First, we observe that the GMF can be written as

\begin{equation}
\theta_{\overline{a}}=\mathbb{E}_{X}\big[\underbrace{\mathbb{E}_{M_{1}|X,a_{1}}\ldots\underbrace{\mathbb{E}_{M_{K}|X,a_{K},\overline{M}_{K-1}}\underbrace{\mathbb{E}[Y|X,a_{K+1},\overline{M}_{K}]}_{\stackrel{\Delta}{=}\mu_{K}(X,\overline{M}_{K})}}_{\stackrel{\Delta}{=}\mu_{K-1}(X,\overline{M}_{K-1})}}_{\stackrel{\Delta}{=}\mu_{0}(X)}\big].\label{eq:iterE}
\end{equation}
This expression suggests that $\theta_{\overline{a}}$ can be estimated
via an iterated regression-imputation (RI) approach (\citealt{zhou2020tracing}):
\begin{enumerate}
\item Estimate $\mu_{K}(X,\overline{M}_{K})$ by fitting a parametric model
for the conditional mean of $Y$ given $(X,A,\overline{M}_{K})$ and
then setting $A=a_{K+1}$ for all units;
\item For $k=K-1,\ldots0$, estimate $\mu_{k}(X,\overline{M}_{k})$ by fitting
a parametric model for the conditional mean of $\mu_{k+1}(X,\overline{M}_{k+1})$
and then setting $A=a_{k+1}$ for all units;
\item Estimate $\theta_{\overline{a}}$ by averaging the fitted values $\hat{\mu}_{0}(X)$
among all units: 
\begin{equation}
\hat{\theta}_{\overline{a}}^{\textup{ri}}=\mathbb{P}_{n}\big[\hat{\mu}_{0}(X)\big].\label{eq:RI}
\end{equation}
\end{enumerate}
The regression-imputation estimator can be seen as an extension of
the imputation strategy proposed by \citet{vansteelandt2012imputation}
for estimating the NDE and NIE in the one-mediator setting. Since
this approach requires modeling only the conditional means of observed/imputed
outcomes given different sets of mediators, it is more flexible to
use with continuous/multivariate mediators than MLE. Nonetheless,
because $\mu_{k}(x,\overline{m}_{k})$ is estimated iteratively, correct
specification of all of the outcome models is required for $\hat{\theta}_{\overline{a}}^{\textup{ri}}$
to be consistent. Thus, in practice, when parametric models are used
to estimate $\mu_{k}(x,\overline{m}_{k})$, care should be taken to
ensure that the outcome models used to estimate these functions are
mutually compatible. For example, if $\mu_{1}(X,M_{1})$ follows a
linear model that includes $X$ and $X^{2}$ as predictors, then the
model used to estimate $\mu_{0}(X)=\mathbb{E}[\mu_{1}(X,M_{1})|X,A=a_{1}]$
should also include $X$ and $X^{2}$ in the predictor set. 

The GMF can also be written as 
\begin{align*}
\theta_{\overline{a}} & =\mathbb{E}\big[\frac{\mathbb{I}(A=a_{K+1})}{p(a_{K+1}|X)}\big(\prod\limits _{k=1}^{K}\frac{p(M_{k}|X,a_{k},\overline{M}_{k-1})}{p(M_{k}|X,a_{K+1},\overline{M}_{k-1})}\big)Y\big].
\end{align*}
This expression suggests a weighting estimator of $\theta_{\overline{a}}$:
\begin{equation}
\hat{\theta}_{\overline{a}}^{\textup{w-m}}=\mathbb{P}_{n}\big[\frac{\mathbb{I}(A=a_{K+1})}{\hat{p}(a_{K+1}|X)}\big(\prod\limits _{k=1}^{K}\frac{\hat{p}(M_{k}|X,a_{k},\overline{M}_{k-1})}{\hat{p}(M_{k}|X,a_{K+1},\overline{M}_{k-1})}\big)Y\big].\label{eq:weighting-M}
\end{equation}
This estimator can be seen as an extension of the weighting estimator
proposed in \citet{vandwerweele2014effectdecomp} for the case of
two mediators. It shares a limitation of $\hat{\theta}_{\overline{a}}^{\textup{mle}}$
in that it requires estimates of the conditional densities/probabilities
of the mediators, which tend to be noisy if the mediators are continuous
or multivariate. This problem, however, can be sidestepped by recasting
the mediator density ratios, via Bayes' rule, as odds ratios in terms
of the treatment variable:
\begin{align*}
\frac{p(M_{k}|X,a_{k},\overline{M}_{k-1})}{p(M_{k}|X,a_{K+1},\overline{M}_{k-1})} & =\ensuremath{\frac{p(a_{k}|X,\overline{M}_{k})/p(a_{K+1}|X,\overline{M}_{k})}{p(a_{k}|X,\overline{M}_{k-1})/p(a_{K+1}|X,\overline{M}_{k-1})}}.
\end{align*}
This observation leads to an alternative weighting estimator based
on estimates of the conditional probabilities of treatment given different
sets of mediators:
\begin{equation}
\hat{\theta}_{\overline{a}}^{\textup{w-a}}=\mathbb{P}_{n}\big[\frac{\mathbb{I}(A=a_{K+1})}{\hat{p}(a_{1}|X)}\big(\prod\limits _{k=1}^{K}\frac{\hat{p}(a_{k}|X,\overline{M}_{k})}{\hat{p}(a_{k+1}|X,\overline{M}_{k})}\big)Y\big].\label{eq:weighting-A}
\end{equation}
In applications where the mediators are continuous/multivariate, $\hat{\theta}_{\overline{a}}^{\textup{w-a}}$
should be easier to work with than $\hat{\theta}_{\overline{a}}^{\textup{w-m}}$.
Yet, the parameters for $p(a|x,\overline{m}_{k})$ are not variationally
independent across different values of $k$. As in the case of the
regression-imputation estimator, care should be taken to ensure the
compatibility of the models specified for $p(a|x,\overline{m}_{k})$
(see \citealt{miles2020semiparametric} for some practical recommendations).

The regression-imputation approach and the weighting approach can
be combined to form various ``hybrid estimators'' of $\theta_{\overline{a}}$.
For example, in the case of $K=2$, one can use regression-imputation
to estimate $\mu_{2}(x,m_{1},m_{2})$, another regression-imputation
step to estimate $\mu_{1}(x,m_{1})$, and weighting to estimate $\theta_{\overline{a}}$,
yielding an ``RI-RI-W'' estimator:
\begin{equation}
\hat{\theta}_{\overline{a}}^{\textup{ri-ri-w}}=\mathbb{P}_{n}\big[\frac{\mathbb{I}(A=a_{1})}{\hat{p}(a_{1}|X)}\hat{\mu}_{1}(X,M_{1})\big].\label{eq:RI-RI-W}
\end{equation}
One can also use regression-imputation to estimate $\mu_{2}(x,m_{1},m_{2})$
and then employ appropriate weights to estimate $\theta_{\overline{a}}$,
which leads to an ``RI-W-W'' estimator:
\begin{equation}
\hat{\theta}_{\overline{a}}^{\textup{ri-w-w}}=\mathbb{P}_{n}\big[\frac{\mathbb{I}(A=a_{2})}{\hat{p}(a_{2}|X)}\frac{\hat{p}(M_{1}|X,a_{1})}{\hat{p}(M_{1}|X,a_{2})}\hat{\mu}_{2}(X,M_{1},M_{2})\big].\label{eq:RI-W-W}
\end{equation}
In fact, with $K$ mediators, there are $2^{K+1}$ different ways
to combine regression-imputation and weighting, each of which involves
estimating $K+1$ nuisance functions, which entail a choice between
$p(a|x)$ and $\mu_{0}(x)$ and a choice between $p(m_{k}|x,a,\overline{m}_{k-1})$
and $\mu_{k}(x,\overline{m}_{k})$ for each $k\in[K]$ (see Supplementary
Material \ref{sec:Hybrid-Estimators} for detailed expressions of
these hybrid estimators in the case of $K=2$). As with $\hat{\theta}_{\overline{a}}^{\textup{mle}}$,
$\hat{\theta}_{\overline{a}}^{\textup{ri}}$, $\hat{\theta}_{\overline{a}}^{\textup{w-m}}$,
and $\hat{\theta}_{\overline{a}}^{\textup{w-a}}$, each of these hybrid
estimators will be consistent only if the corresponding nuisance functions
are all correctly specified and consistently estimated. In applications
where the pretreatment covariates $X$ and/or the mediators have many
components, all of the above estimators will be prone to model misspecification
bias.

\subsection{Multiply Robust and Semiparametric Efficient Estimation \label{subsec:Multiply-Robust-and}}

Henceforth, let $O=(X,A,\overline{M}_{K},Y)$ denote the observed
data, and $\mathcal{P}_{\textup{np}}$ a nonparametric model over
$O$ wherein all laws $P$ satisfy the positivity assumption described
in Section \ref{subsec:K Mediators}. In addition, define $\mu_{k}(X,\overline{M}_{k})$
iteratively as in equation \eqref{eq:iterE}:
\begin{align*}
\mu_{K}(X,\overline{M}_{K}) & \stackrel{\Delta}{=}\mathbb{E}[Y|X,a_{K+1},\overline{M}_{K}]\\
\mu_{k}(X,\overline{M}_{k}) & \stackrel{\Delta}{=}\mathbb{E}[\mu_{k+1}(X,\overline{M}_{k+1})|X,a_{k+1},\overline{M}_{k}],\quad k=K-1,\ldots,0.
\end{align*}

\begin{thm}
The efficient influence function (EIF) of $\theta_{\overline{a}}$
in $\mathcal{P}_{\textup{np}}$ is given by
\begin{equation}
\varphi_{\overline{a}}(O)=\sum_{k=0}^{K+1}\varphi_{k}(O),\label{eq:EIF}
\end{equation}
where 
\begin{align*}
\varphi_{0}(O) & =\mu_{0}(X)-\theta_{\overline{a}},\\
\varphi_{k}(O) & =\frac{\mathbb{I}(A=a_{k})}{p(a_{k}|X)}\Big(\prod_{j=1}^{k-1}\frac{p(M_{j}|X,a_{j},\overline{M}_{j-1})}{p(M_{j}|X,a_{k},\overline{M}_{j-1})}\Big)\big(\mu_{k}(X,\overline{M}_{k})-\mu_{k-1}(X,\overline{M}_{k-1})\big),\quad k\in[K],\\
\varphi_{K+1}(O) & =\frac{\mathbb{I}(A=a_{K+1})}{p(a_{K+1}|X)}\Big(\prod_{j=1}^{K}\frac{p(M_{j}|X,a_{j},\overline{M}_{j-1})}{p(M_{j}|X,a_{K+1},\overline{M}_{j-1})}\Big)\big(Y-\mu_{K}(X,\overline{M}_{K})\big).
\end{align*}
The semiparametric efficiency bound for any regular and asymptotically
linear estimator of $\theta_{\overline{a}}$ in $\mathcal{P}_{\textup{np}}$
is therefore $\mathbb{E}\big[\big(\varphi_{\overline{a}}(O)\big)^{2}\big]$.
\end{thm}
We now present two estimators of $\theta_{\overline{a}}$ based on
the EIF. First, consider the factorized likelihood of $O$: $p(O)=p(X)p(A|X)\Big(\prod_{k=1}^{K}p(M_{k}|X,A,\overline{M}_{k-1})\big)p(Y|X,A,\overline{M}_{K}).$
Suppose we have estimated $K+2$ nuisance functions, each of which
corresponds to a component of $p(O)$: $\hat{\pi}_{0}(a|x)$ for $p(a|x)$,
$\hat{f}_{k}(m_{k}|x,a,\overline{m}_{k-1})$ for $p(m_{k}|x,a,\overline{m}_{k-1})$,
and $\hat{\mu}_{K}(x,\overline{m}_{K})$ for $\mathbb{E}[Y|x,a_{K+1},\overline{m}_{K}]$.
The GMF can now be estimated as
\begin{align}
\hat{\theta}_{\overline{a}}^{\textup{eif}_{1}}= & \mathbb{P}_{n}\big[\frac{\mathbb{I}(A=a_{K+1})}{\hat{\pi}_{0}(a_{K+1}|X)}\Big(\prod_{j=1}^{K}\frac{\hat{f}_{j}(M_{j}|X,a_{j},\overline{M}_{j-1})}{\hat{f}_{j}(M_{j}|X,a_{K+1},\overline{M}_{j-1})}\Big)\big(Y-\hat{\mu}_{K}(X,\overline{M}_{K})\big)\nonumber \\
 & +\sum_{k=1}^{K}\frac{\mathbb{I}(A=a_{k})}{\hat{\pi}_{0}(a_{k}|X)}\Big(\prod_{j=1}^{k-1}\frac{\hat{f}_{j}(M_{j}|X,a_{j},\overline{M}_{j-1})}{\hat{f}_{j}(M_{j}|X,a_{k},\overline{M}_{j-1})}\Big)\big(\hat{\mu}_{k}^{\textup{mle}}(X,\overline{M}_{k})-\hat{\mu}_{k-1}^{\textup{mle}}(X,\overline{M}_{k-1})\big)\nonumber \\
 & +\hat{\mu}_{0}^{\textup{mle}}(X)\big],\label{eq:EIF1}
\end{align}
where $\hat{\mu}_{K}^{\textup{mle}}(X,\overline{M}_{K})=\hat{\mu}_{K}(X,\overline{M}_{K})$
and $\hat{\mu}_{k}^{\textup{mle}}(X,\overline{M}_{k})$ is iteratively
constructed as 
\begin{equation}
\hat{\mu}_{k}^{\textup{mle}}(X,\overline{M}_{k})=\int\hat{\mu}_{k+1}^{\textup{mle}}(X,\overline{M}_{k},m_{k+1})\hat{f}_{k+1}(m_{k+1}|X,a_{k+1},\overline{M}_{k})d\nu(m_{k+1}),\quad k=K-1,\ldots0.\label{eq:mu_k^mle}
\end{equation}
When $M_{k+1}$ involves continuous components, equation \eqref{eq:mu_k^mle}
can be evaluated via Monte Carlo simulation.

When some of the mediators are continuous/multivariate, it can be
difficult to estimate the conditional distributions $p(m_{k}|x,a,\overline{m}_{k-1})$.
In such cases, it is often preferable to estimate the mediator density
ratios using the corresponding odds ratios of the treatment variable,
and estimate the functions $\mu_{k}(x,\overline{m}_{k})$ using the
regression-imputation approach. Specifically, suppose we have estimated
$2(K+1)$ nuisance functions: $\hat{\pi}_{0}(a|x)$ for $p(a|x)$,
$\hat{\pi}_{k}(a|x,\overline{m}_{k})$ for $p(a|x,\overline{m}_{k})$
($k\in[K]$), and $\hat{\mu}_{k}(x,\overline{m}_{k})$ for $\mu_{k}(x,\overline{m}_{k})$
($k\in\{0,1,\ldots K\}$), where for $k<K$, $\mu_{k}(x,\overline{m}_{k})$
is estimated iteratively by fitting a model for the conditional mean
of $\hat{\mu}_{k+1}(X,\overline{M}_{k+1})$ given $(X,A,\overline{M}_{k})$
and then setting $A=a_{k+1}$ for all units. The GMF can then be estimated
as
\begin{align}
\hat{\theta}_{\overline{a}}^{\textup{eif}_{2}}= & \mathbb{P}_{n}\big[\frac{\mathbb{I}(A=a_{K+1})}{\hat{\pi}_{0}(a_{1}|X)}\Big(\prod_{j=1}^{K}\frac{\hat{\pi}_{j}(a_{j}|X,\overline{M}_{j})}{\hat{\pi}_{j}(a_{j+1}|X,\overline{M}_{j})}\Big)\big(Y-\hat{\mu}_{K}(X,\overline{M}_{K})\big)\nonumber \\
 & +\sum_{k=1}^{K}\frac{\mathbb{I}(A=a_{k})}{\hat{\pi}_{0}(a_{1}|X)}\Big(\prod_{j=1}^{k-1}\frac{\hat{\pi}_{j}(a_{j}|X,\overline{M}_{j})}{\hat{\pi}_{j}(a_{j+1}|X,\overline{M}_{j})}\Big)\big(\hat{\mu}_{k}(X,\overline{M}_{k})-\hat{\mu}_{k-1}(X,\overline{M}_{k-1})\big)\nonumber \\
 & +\hat{\mu}_{0}(X)\big].\label{eq:EIF2}
\end{align}
The multiple robustness and semiparametric efficiency of $\hat{\theta}_{\overline{a}}^{\textup{eif}_{1}}$
and $\hat{\theta}_{\overline{a}}^{\textup{eif}_{2}}$ are given below.
\begin{thm}
Let $\eta_{1}=\{\pi_{0},f_{1},\ldots f_{K},\mu_{K}\}$ denote the
$K+2$ nuisance functions involved in $\hat{\theta}_{\overline{a}}^{\textup{eif}_{1}}$,
and $\eta_{2}=\{\pi_{0},\ldots\pi_{K},\mu_{0},\ldots\mu_{K}\}$ denote
the $2(K+1)$ nuisance functions involved in $\hat{\theta}_{\overline{a}}^{\textup{eif}_{2}}$.
Suppose that Assumption 3{*} (positivity) and suitable regularity
conditions for estimating equations (e.g., \citealt[p. 2148]{newey1994large})
hold. In addition, suppose that $\mu_{K}(x,\overline{m}_{K})$ is
bounded over the support of $(X,\overline{M}_{K})$. Then, when the
elements of $\eta_{1}$ and $\eta_{2}$ are estimated via parametric
models,
\begin{enumerate}
\item $\hat{\theta}_{\overline{a}}^{\textup{eif}_{1}}$ is consistent and
asymptotically normal (CAN) if $K+1$ of the $K+2$ nuisance functions
in $\eta_{1}$ are correctly specified and their parameter estimates
are $\sqrt{n}$-consistent; it is semiparametric efficient if all
of the $K+2$ nuisance functions in $\mathcal{\eta}_{1}$ are correctly
specified and their parameter estimates are $\sqrt{n}$-consistent.
\item $\hat{\theta}_{\overline{a}}^{\textup{eif}_{2}}$ is CAN if $\exists k\in\{0,\ldots K+1\}$,
the first $k$ treatment models $\pi_{0},\ldots\pi_{k-1}$ and the
last $K+1-k$ outcome models $\mu_{k},\ldots\mu_{K}$ in $\eta_{2}$
are correctly specified and their parameter estimates are $\sqrt{n}$-consistent;
it is semiparametric efficient if all of the treatment and outcome
models in $\eta_{2}$ are correctly specified and their parameter
estimates are $\sqrt{n}$-consistent.
\end{enumerate}
\end{thm}
Both $\hat{\theta}_{\overline{a}}^{\textup{eif}_{1}}$ and $\hat{\theta}_{\overline{a}}^{\textup{eif}_{2}}$
are $K+2$-robust in the sense that they are CAN provided that one
of $K+2$ sets of nuisance functions is correctly specified and the
corresponding parameter estimates are $\sqrt{n}$-consistent. Several
special cases are worth noting. First, in the degenerate case where
$K=0$, it is clear that both $\hat{\theta}_{\overline{a}}^{\textup{eif}_{1}}$
and $\hat{\theta}_{\overline{a}}^{\textup{eif}_{2}}$ reduce to the
standard doubly robust estimator for $\mathbb{E}[Y(a)]$ (\citealt{scharfstein1999adjusting}).
Second, when $K=1$, $\hat{\theta}_{01}^{\textup{eif}_{1}}$ coincides
with \citeauthor{tchetgen2012semiparametric}'s (2012) triply robust
estimator for $\mathbb{E}[Y\big(1,M(0)\big)]$. Finally, when $K=2$,
$\hat{\theta}_{010}^{\textup{eif}_{1}}$ is identical to \citeauthor{miles2020semiparametric}'s
(2020) estimator for $\theta_{010}$. For this case, however, \citeauthor{miles2020semiparametric}
provide a slightly weaker condition than that implied by Theorem 3
for $\hat{\theta}_{010}^{\textup{eif}_{1}}$ to be CAN. Specifically,
they showed that $\hat{\theta}_{010}^{\textup{eif}_{1}}$ remains
CAN even if both $f_{1}$ and $\mu_{2}$ are misspecified. In Section
\ref{sec:Special-Cases}, we show that the conditions for $\hat{\theta}_{\overline{a}}^{\textup{eif}_{2}}$
to be CAN can also be relaxed for several particular types of PSEs,
including the natural path-specific effect (nPSE), of which $\psi_{010}-\psi_{000}$
is a special case. For the $K=2$ case, \citeauthor{miles2020semiparametric}
also noted that the mediator density ratios in $\hat{\theta}_{010}^{\textup{eif}_{1}}$
can be indirectly estimated through models for $\pi_{1}$ and $\pi_{2}$.
Clearly, this approach will result in $\hat{\theta}_{010}^{\textup{eif}_{2}}$
if the $\mu_{k}(x,\overline{m}_{k})$ functions are in the meanwhile
estimated through regression-imputation. The $K+2$-robustness of
$\hat{\theta}_{\overline{a}}^{\textup{eif}_{1}}$ and $\hat{\theta}_{\overline{a}}^{\textup{eif}_{2}}$,
interestingly, resembles the multiple robustness of the Bang-Robins
\citeyearpar{bang2005doubly} estimator for the mean of a potential
outcome with time-varying treatments and time-varying confounders
(\citealt{luedtke2017sequential,molina2017multiple,rotnitzky2017multiply}).

To gain some intuition as to why $\hat{\theta}_{\overline{a}}^{\textup{eif}_{1}}$
is $K+2$-robust, consider cases in which only one nuisance function
in $\eta_{1}$ is misspecified. When only $\pi_{0}$ is misspecified,
all terms inside $\mathbb{P}_{n}[\cdot]$ but $\hat{\mu}_{0}^{\textup{mle}}(X)$
will have a zero mean (asymptotically), leaving only $\mathbb{P}_{n}[\hat{\mu}_{0}^{\textup{mle}}(X)]$
(i.e., the MLE estimator \eqref{eq:MLE}), which is consistent because
the corresponding nuisance functions $\{f_{1},\ldots f_{K},\mu_{K}\}$
are all correctly specified. When only $\mu_{K}$ is misspecified,
all terms involving $\hat{\mu}_{K}(X,\overline{M}_{K})$ and $\hat{\mu}_{k}^{\textup{mle}}(X,\overline{M}_{k})$
($k=0,1,\ldots K-1$) inside $\mathbb{P}_{n}[\cdot]$ will have a
zero mean (asymptotically), leaving only a weighted average of $Y$
(i.e., the weighting estimator \eqref{eq:weighting-M}), which is
consistent because the corresponding nuisance functions $\{\pi_{0},f_{1},\ldots f_{K}\}$
are all correctly specified. Finally, when only $f_{k'}$ is misspecified
(for some $k'\in[K]$), it can be shown that all terms involving $\hat{f}_{k'}$
and $\hat{\mu}_{k}^{\textup{mle}}(X,\overline{M}_{k})$ ($\forall k<k'$)
inside $\mathbb{P}_{n}[\cdot]$ will have a zero mean (asymptotically),
leaving only a weighted average of $\hat{\mu}_{k'}^{\textup{mle}}(X,\overline{M}_{k'})$.
The latter constitutes a ``hybrid'' estimator similar to those mentioned
in the previous section, and it is consistent in this case because
its nuisance functions $\{\pi_{0},f_{1},\ldots f_{k'-1},f_{k'+1},\ldots f_{K},\mu_{K}\}$
are all correctly specified.

The $K+2$-robustness of $\hat{\theta}_{\overline{a}}^{\textup{eif}_{2}}$
is due to a similar logic to that of $\hat{\theta}_{\overline{a}}^{\textup{eif}_{1}}$.
Yet, different from $\hat{\theta}_{\overline{a}}^{\textup{eif}_{1}}$,
$\hat{\theta}_{\overline{a}}^{\textup{eif}_{2}}$ involves estimating
$2(K+1)$ nuisance functions, $K+1$ for the conditional probabilities
of treatment and $K+1$ for the conditional means of observed/imputed
outcomes. Also, unlike $\hat{\theta}_{\overline{a}}^{\textup{eif}_{1}}$,
the treatment models involved in $\hat{\theta}_{\overline{a}}^{\textup{eif}_{2}}$
are not variationally independent; neither are the outcome models.
For example, when $M_{K}\ci A|X,\overline{M}_{K-1}$, $\pi_{K}(A|X,\overline{M}_{K})$
should be identical to $\pi_{K-1}(A|X,\overline{M}_{K-1})$; similarly,
when $M_{K}\ci Y|X,A,\overline{M}_{K-1}$, $\mu_{K}(X,\overline{M}_{K})$
should be identical to $\mu_{K-1}(X,\overline{M}_{K-1})$. Thus, in
practice, both the treatment and outcome models should be specified
in a mutually compatible way, otherwise some of the conditions in
Theorem 3 may fail by design.

The local efficiency of $\hat{\theta}_{\overline{a}}^{\textup{eif}_{1}}$
and $\hat{\theta}_{\overline{a}}^{\textup{eif}_{2}}$ is due to the
fact that both of the EIF-based estimating equations \eqref{eq:EIF1}
and \eqref{eq:EIF2} have a zero derivative with respect to the nuisance
functions at the truth. This property, referred to as ``Neyman orthogonality''
by \citet{chernozhukov2018double}, implies that first step estimation
of the nuisance functions has no first order effect on the influence
functions of $\hat{\theta}_{\overline{a}}^{\textup{eif}_{1}}$ and
$\hat{\theta}_{\overline{a}}^{\textup{eif}_{2}}$. This property suggests
that the nuisance functions can be estimated using data-adaptive/machine
learning methods or their ensembles. In this case, these estimators
will still be consistent as long as the nuisance functions associated
with one of the $K+2$ conditions in Theorem 3 are consistently estimated.
For $\hat{\theta}_{\overline{a}}^{\textup{eif}_{2}}$, an added advantage
of employing data-adaptive methods to estimate the nuisance functions
is that, by exploring a larger space within $\mathcal{P}_{\textup{np}}$,
the risk of model incompatibility is reduced.

When data-adaptive/machine learning methods are used to estimate the
nuisance functions, it is advisable to use sample splitting to render
the empirical process term asymptotically negligible (\citealt{zheng2011cross,chernozhukov2018double,newey2018cross}).
For example, \citet{chernozhukov2018double} suggest the method of
``cross-fitting,'' which involves the following steps: (a) randomly
partition the sample $S$ into $J$ folds: $S_{1},S_{2}\ldots S_{J}$;
(b) for each $j$, obtain a fold-specific estimate of the target parameter
using only data from $S_{j}$ (``main sample''), but with nuisance
functions learned from the remainder of the sample (i.e., $S\backslash S_{j}$;
``auxiliary sample''); (c) average these fold-specific estimates
to form a final estimate of the target parameter.

When cross-fitting is used, $\hat{\theta}_{\overline{a}}^{\textup{eif}_{1}}$
and $\hat{\theta}_{\overline{a}}^{\textup{eif}_{2}}$ will be semiparametric
efficient if the corresponding nuisance function estimates are all
consistent and converge at sufficiently fast rates. For example, a
sufficient (but not necessary) condition for $\hat{\theta}_{\overline{a}}^{\textup{eif}_{1}}$
and $\hat{\theta}_{\overline{a}}^{\textup{eif}_{2}}$ to attain the
semiparametric efficiency bound is when all of the nuisance function
estimates converge at faster-than-$n^{-1/4}$ rates. More precise
conditions are given in Theorem 4.
\begin{thm}
Let $\hat{\eta}_{1}=\{\hat{\pi}_{0},\hat{f}_{1},\ldots\hat{f}_{K},\hat{\mu}_{K}\}$
and $\hat{\eta}_{2}=\{\hat{\pi}_{0},\ldots\hat{\pi}_{K},\hat{\mu}_{0},\ldots\hat{\mu}_{K}\}$
denote estimates of the nuisance functions involved in $\hat{\theta}_{\overline{a}}^{\textup{eif}_{1}}$
and $\hat{\theta}_{\overline{a}}^{\textup{eif}_{2}}$, respectively.
Let $r_{n}(\cdot)$ denote a mapping from a nuisance function estimator
to its $L_{2}(P)$ convergence rate where $P$ represents the true
distribution of $O=(X,A,\overline{M}_{K},Y)$. Suppose that Assumption
3{*} (positivity) holds for both the true distribution $P$ and its
estimates implied by $\hat{\eta}_{1}$ and $\hat{\eta}_{2}$, and
that all other assumptions required for Theorem 3 hold. Then, when
the nuisance functions are estimated via data-adaptive methods and
cross-fitting,
\begin{enumerate}
\item $\hat{\theta}_{\overline{a}}^{\textup{eif}_{1}}$ is consistent if
$K+1$ of the $K+2$ elements in $\hat{\eta}_{1}$ are consistent
in the $L_{2}$-norm; it is CAN and semiparametric efficient if all
elements in $\hat{\eta}_{1}$ are consistent in the $L_{2}$-norm
and $\sum\limits _{u,v\in\hat{\eta}_{1};u\neq v}r_{n}(u)r_{n}(v)=o(n^{-1/2})$;
\item $\hat{\theta}_{\overline{a}}^{\textup{eif}_{2}}$ is consistent if
$\exists k\in\{0,\ldots K+1\}$, $\hat{\pi}_{0},\ldots\hat{\pi}_{k-1},\hat{\mu}_{k},\ldots\hat{\mu}_{K}$
are all consistent in the $L_{2}$-norm; it is CAN and semiparametric
efficient if all elements in $\hat{\eta}_{2}$ are consistent in the
$L_{2}$-norm and $\sum\limits _{j=0}^{K}r_{n}(\hat{\pi}_{j})r_{n}(\hat{\mu}_{j})=o(n^{-1/2})$.
\end{enumerate}
\end{thm}
The multiple robustness result for $\hat{\theta}_{\overline{a}}^{\textup{eif}_{1}}$
echoes Theorem 3. Moreover, the first part of Theorem 4 states that
$\hat{\theta}_{\overline{a}}^{\textup{eif}_{1}}$ is CAN and semiparametric
efficient if all nuisance functions in $\eta_{1}$ are consistently
estimated and, for every two nuisance functions in $\eta_{1}$, the
product of their convergence rates is $o(n^{-1/2})$. Thus $\hat{\theta}_{\overline{a}}^{\textup{eif}_{1}}$
is CAN and semiparametric efficient if all of the $K+2$ nuisance
function estimates are consistent and converge at faster-than-$n^{-1/4}$
rates, but it will also attain semiparametric efficiency under alternative
conditions. For example, when estimates of the treatment and mediator
models $\{\hat{\pi}_{0}$, $\hat{f}_{1},\ldots\hat{f}_{K}\}$ all
converge to the truth at a rate of $n^{-1/3}$ and estimates of the
outcome model $\hat{\mu}_{K}$ converge to the truth at a rate of
$n^{-1/5}$, the product of the convergence rates of any two elements
in $\hat{\eta}_{1}$ is either $O(n^{-1/3})O(n^{-1/3})=O(n^{-2/3})$
or $O(n^{-1/3})O(n^{-1/5})=O(n^{-8/15})$, both faster than $O(n^{-1/2})$.

The second part of Theorem 4 states that $\hat{\theta}_{\overline{a}}^{\textup{eif}_{2}}$
is consistent if there exists a $k$ such that the first $k$ treatment
models and the last $K+1-k$ outcome models in $\eta_{2}$ are consistently
estimated, echoing Theorem 3. As with $\hat{\theta}_{\overline{a}}^{\textup{eif}_{1}}$,
$\hat{\theta}_{\overline{a}}^{\textup{eif}_{2}}$ will be CAN and
semiparametric efficient if all of the required nuisance functions
are consistently estimated and converge at faster-than-$n^{-1/4}$
rates. The rate condition $\sum\limits _{j=0}^{K}r_{n}(\hat{\pi}_{j})r_{n}(\hat{\mu}_{j})=o(n^{-1/2})$
appears to be weaker than that for $\hat{\theta}_{\overline{a}}^{\textup{eif}_{1}}$
as it involves the sum of only $K+1$, rather than ${K+2 \choose 2}$,
product terms. Because the outcome models are estimated iteratively,
the convergence rate of $\hat{\mu}_{k}$ will in general depend on
the convergence rates of $\{\hat{\mu}_{k+1},\ldots\hat{\mu}_{K}\}$.
That is, if $r_{n}(\hat{\mu}_{k+1})=O(n^{\delta})$, $r_{n}(\hat{\mu}_{k})$
is unlikely to be faster than $O(n^{\delta})$. Nonetheless, $\hat{\theta}_{\overline{a}}^{\textup{eif}_{2}}$
will be CAN and semiparametric efficient under relatively weak conditions
--- for example, when estimates of the treatment models all converge
to the truth at a rate of $n^{-1/3}$ and estimates of the outcome
models all converge to the truth at a rate of $n^{-1/5}$, in which
case $\sum\limits _{j=0}^{K}r_{n}(\hat{\pi}_{j})r_{n}(\hat{\mu}_{j})=\sum\limits _{j=0}^{K}O(n^{-8/15})=o(n^{-1/2})$.

For inference on $\hat{\theta}_{\overline{a}}^{\textup{eif}_{1}}$
and $\hat{\theta}_{\overline{a}}^{\textup{eif}_{2}}$, a simple variance
estimator can be constructed from the empirical analog of the EIF,
i.e., $\mathbb{P}_{n}[\hat{\varphi}_{\overline{a}}^{2}(O)]/n$. However,
unlike $\hat{\theta}_{\overline{a}}^{\textup{eif}_{1}}$ and $\hat{\theta}_{\overline{a}}^{\textup{eif}_{2}}$,
this variance estimator is not multiply robust --- it will be consistent
only if the conditions for semiparametric efficiency in Theorem 3
or Theorem 4 are satisfied. Thus, when the nuisance functions are
estimated using parametric models, the variance estimator constructed
from the empirical EIF may be inconsistent even when the corresponding
estimator for $\theta_{\overline{a}}$ is CAN --- for example, when
only $K+1$ of the $K+2$ nuisance functions involved in $\hat{\theta}_{\overline{a}}^{\textup{eif}_{1}}$
are correctly specified. In this case, the nonparametric bootstrap
is a convenient approach to more robust inference. When the nuisance
functions are estimated using data-adaptive/machine learning methods,
however, the nonparametric bootstrap is not theoretically justified,
and the EIF-based variance estimator may still be preferred.

\subsection{Multiply Robust Regression-Imputation Estimators \label{subsec:Stable-Estimation}}

Both of the multiply robust estimators described above involve inverse
probability weights. When the positivity assumption is nearly violated,
the inverse probability weights tend to be highly variable, which
may lead to poor finite sample performance (\citealt{kang2007demystifying,petersen2012diagnosing}).
A variety of methods have been proposed to reduce the influence of
highly variable weights on doubly robust and multiply robust estimators
in similar settings (e.g., \citealt{robins2007comment,tchetgen2012semiparametric,seaman2018introduction}).
Among them, a common strategy is to tailor the estimating equation
of the outcome model(s) such that the terms involving inverse probability
weights will equal zero, leaving only a regression-imputation or ``substitution''
estimator that typically resides in the parameter space of the estimand.
Below, we briefly describe how this approach can be adapted to $\hat{\theta}_{\overline{a}}^{\textup{eif}_{1}}$
and $\hat{\theta}_{\overline{a}}^{\textup{eif}_{2}}$.

Let us start with $\hat{\theta}_{\overline{a}}^{\textup{eif}_{2}}$,
which can be written as
\begin{align}
\hat{\theta}_{\overline{a}}^{\textup{eif}_{2}}= & \mathbb{P}_{n}\big[\hat{w}_{K}(X,A,\overline{M}_{K})\big(Y-\hat{\mu}_{K}(X,\overline{M}_{K})\big)\nonumber \\
 & +\sum_{k=1}^{K}\hat{w}_{k-1}(X,A,\overline{M}_{k-1})\big(\hat{\mu}_{k}(X,\overline{M}_{k})-\hat{\mu}_{k-1}(X,\overline{M}_{k-1})\big)\nonumber \\
 & +\hat{\mu}_{0}(X)\big],\label{eq:EIF2-1}
\end{align}
where $\hat{w}_{k}(A,X,\overline{M}_{k})$ ($0\leq k\leq K$) are
estimates of the corresponding inverse probability weights as displayed
in equation \eqref{eq:EIF2}. Note that the nuisance functions $\hat{\mu}_{k}(X,\overline{M}_{k})$
($0\leq k\leq K$) here are all estimated via the regression-imputation
approach. When the corresponding outcome models are fitted via generalized
linear models (GLM) with canonical links, one can either (a) fit weighted
GLMs (with an intercept term) for $\hat{\mu}_{k}(X,\overline{M}_{k})$
using $\hat{w}_{k}(A,X,\overline{M}_{k})$ as weights, or (b) add
the corresponding inverse probability weight as an additional covariate
in these regressions (\citealt{robins2007comment}). Either way, the
score equations for GLMs will ensure that all terms inside $\mathbb{P}_{n}[\cdot]$
but $\hat{\mu}_{0}(X)$ have a sample mean of zero, leaving only $\mathbb{P}_{n}[\hat{\mu}_{0}(X)]$,
which will reside in the parameter space of $\theta_{\overline{a}}$
if the latter equals the range of the GLM specified for $\mu_{0}(x)$.

Alternatively, one can use the method of targeted maximum likelihood
estimation (TMLE; \citealt{van2006targeted,zheng2012targeted}), which,
by fitting each of the outcome models in two steps, will also ensure
a zero sample mean for all terms inside $\mathbb{P}_{n}[\cdot]$ but
$\hat{\mu}_{0}(X)$. This approach does not require the first-step
models to be GLM and thus can be used with a wider range of outcome
models. In our case, it involves the following steps:
\begin{enumerate}
\item For $k=K,\ldots0$
\begin{enumerate}
\item Using $\hat{\mu}_{k+1}^{\textup{tmle}}(X,\overline{M}_{k+1})$ (or,
in the case $k=K$, the observed outcome $Y$) as the response variable,
obtain a first-step regression-imputation estimate of $\mu_{k}(X,\overline{M}_{k})$;
\item Fit a one-parameter GLM for the conditional mean of $\hat{\mu}_{k+1}^{\textup{tmle}}(X,\overline{M}_{k+1})$
(or, in the case $k=K$, the observed outcome $Y$), using $g(\hat{\mu}_{k}(X,\overline{M}_{k}))$
as an offset term and $\hat{w}_{k}(A,X,\overline{M}_{k})$ as the
only covariate (without an intercept term), and obtain an updated
estimate $\hat{\mu}_{k}^{\textup{tmle}}(X,\overline{M}_{k})=g^{-1}\big(g(\hat{\mu}_{k}(X,\overline{M}_{k}))+\text{\ensuremath{\hat{\beta}}}_{k}\hat{w}_{k}(A,X,\overline{M}_{k})\big)$,
where $g(\cdot)$ is the link function for the GLM and $\hat{\beta}_{k}$
is the estimated coefficient on $\hat{w}_{k}(A,X,\overline{M}_{k})$;
\end{enumerate}
\item Obtain the final estimate $\hat{\theta}_{\overline{a}}^{\textup{tmle}}=\mathbb{P}_{n}[\hat{\mu}_{0}^{\textup{tmle}}(X)]$.
\end{enumerate}
In the one-mediator case, the above estimator is similar to the TMLE
estimator proposed by \citet{zheng2012targeted} for the NDE, i.e.,
$\psi_{01}-\psi_{00}$. Since Zheng and van der Laan's estimand is
the NDE instead of the mediation functional, their TMLE procedure
involves fitting a model for the ``mediated mean outcome difference''
(p. 6), i.e., $\mathbb{E}[\mathbb{E}[Y|X,A=1,M]-\mathbb{E}[Y|X,A=0,M]|X,A=0]$,
instead of the conditional mean of the imputed outcome itself, i.e.,
$\mu_{0}(X)$.

As with the GLM-based adjustments, the TMLE approach also yields a
regression-imputation estimator that resides in the parameter space
of $\theta_{\overline{a}}$ if the latter equals the range of the
model specified for $\mu_{0}(x)$. It should be noted that when data-adaptive
methods are used to obtain first-step estimates of the nuisance functions,
sample splitting should be employed so that steps 1(a) and steps 1(b)
are implemented on different subsamples. In cross-fitting, for example,
steps 1(a) should be implemented in the auxiliary sample ($S\backslash S_{j}$)
and steps 1(b) implemented in the main sample $S_{j}$. The method
of TMLE can also be used to adjust $\hat{\theta}_{\overline{a}}^{\textup{eif}_{1}}$,
in which case the first step estimates of $\mu_{k}(X,\overline{M}_{k})$
($0\leq k\leq K-1$) are based on equation \eqref{eq:mu_k^mle}, and
the weights $\hat{w}_{k}(A,X,\overline{M}_{k})$ ($0\leq k\leq K$)
reflect the corresponding terms in equation \eqref{eq:EIF1}.

\section{Special Cases\label{sec:Special-Cases}}

We have so far considered $\theta_{\overline{a}}$ for the unconstrained
case where $a_{1},\ldots a_{K+1}$ can each take 0 or 1. In many applications,
the researcher may be interested in particular causal estimands such
as the natural direct effect (NDE), the natural/total indirect effect
(NIE/TIE), and natural path-specific effects (nPSE; \citealt{daniel2015causal}).
Below, we discuss how the multiply robust semiparametric estimators
of $\theta_{\overline{a}}$ apply to these estimands. In addition,
we discuss a set of cumulative path-specific effects (cPSEs) that
together compose the ATE. In Supplementary Material \ref{sec:Decomposition-of-Between-group},
we connect these cPSEs to noncausal decompositions of between-group
disparities that are widely used in the social sciences. For illustrative
purposes, we focus on estimators based on $\hat{\theta}_{\overline{a}}^{\textup{eif}_{2}}$,
although similar results hold for those based on $\hat{\theta}_{\overline{a}}^{\textup{eif}_{1}}$.
Throughout this section, we maintain Assumptions 1{*}-3{*} so that
$\theta_{\overline{a}}=\psi_{\overline{a}}$.
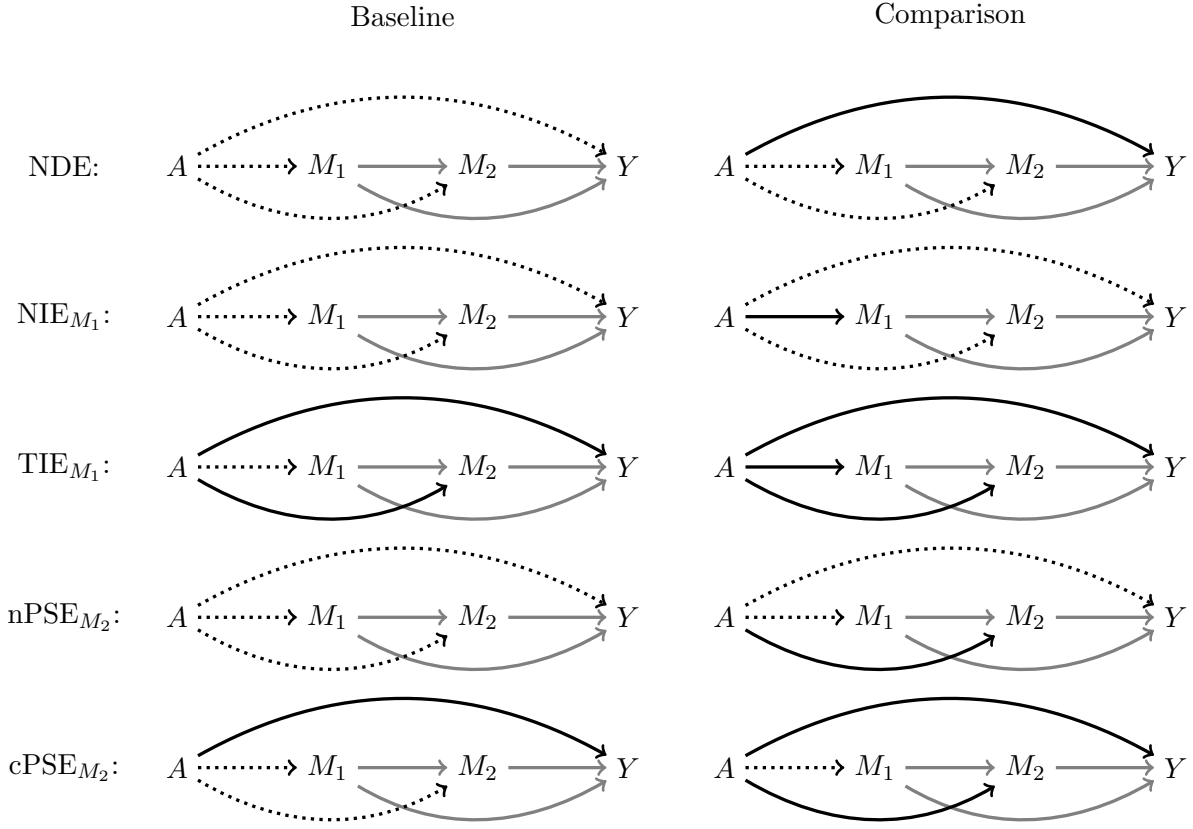
\begin{figure}[!th]
\noindent \begin{raggedright}
\begin{minipage}[t]{0.45\columnwidth}%
\noindent \begin{center}
\begin{tikzpicture}[yscale = 1, xscale = 2]
\node[text centered] at (0.5,2) (ref) {Baseline};
\node[text centered] at (-1.75,0) (p) {$\textup{NDE}$:};
\node[text centered] at (-1,0) (a) {$A$};
\node[text centered] at (1,0) (m) {$M_2$};
\node[text centered] at (0,0) (l) {$M_1$};
\node[text centered] at (2,0) (y) {$Y$};
\draw [->, line width= 1.25, gray] (l) -- (m);
\draw [->, line width= 1.25, gray] (l) to [out=-45, in=-135, looseness=1.2] (y);
\draw [->, line width= 1.25, gray] (m) -- (y); 
\draw [->, line width= 1.25, dotted] (a) to [out=-45, in=-135, looseness=1.2] (m);
\draw [->, line width= 1.25, dotted] (a) to [out=45, in=135, looseness=1.2] (y);
\draw [->, line width= 1.25, dotted] (a) -- (l);

\node[text centered] at (-1.75,-2) (p) {$\textup{NIE}_{M_1}$:};
\node[text centered] at (-1,-2) (a) {$A$};
\node[text centered] at (1,-2) (m) {$M_2$};
\node[text centered] at (0,-2) (l) {$M_1$};
\node[text centered] at (2,-2) (y) {$Y$};
\draw [->, line width= 1.25, gray] (l) -- (m);
\draw [->, line width= 1.25, gray] (l) to [out=-45, in=-135, looseness=1.2] (y);
\draw [->, line width= 1.25, gray] (m) -- (y); 
\draw [->, line width= 1.25, dotted] (a) to [out=-45, in=-135, looseness=1.2] (m);
\draw [->, line width= 1.25, dotted] (a) to [out=45, in=135, looseness=1.2] (y);
\draw [->, line width= 1.25, dotted] (a) -- (l);

\node[text centered] at (-1.75,-4) (p) {$\textup{TIE}_{M_1}$:};
\node[text centered] at (-1,-4) (a) {$A$};
\node[text centered] at (1,-4) (m) {$M_2$};
\node[text centered] at (0,-4) (l) {$M_1$};
\node[text centered] at (2,-4) (y) {$Y$};
\draw [->, line width= 1.25, gray] (l) -- (m);
\draw [->, line width= 1.25, gray] (l) to [out=-45, in=-135, looseness=1.2] (y);
\draw [->, line width= 1.25, gray] (m) -- (y); 
\draw [->, line width= 1.25] (a) to [out=-45, in=-135, looseness=1.2] (m);
\draw [->, line width= 1.25] (a) to [out=45, in=135, looseness=1.2] (y);
\draw [->, line width= 1.25, dotted] (a) -- (l);

\node[text centered] at (-1.75,-6) (p) {$\textup{nPSE}_{M_2}$:};
\node[text centered] at (-1,-6) (a) {$A$};
\node[text centered] at (1,-6) (m) {$M_2$};
\node[text centered] at (0,-6) (l) {$M_1$};
\node[text centered] at (2,-6) (y) {$Y$};
\draw [->, line width= 1.25, gray] (l) -- (m);
\draw [->, line width= 1.25, gray] (l) to [out=-45, in=-135, looseness=1.2] (y);
\draw [->, line width= 1.25, gray] (m) -- (y); 
\draw [->, line width= 1.25, dotted] (a) to [out=-45, in=-135, looseness=1.2] (m);
\draw [->, line width= 1.25, dotted] (a) to [out=45, in=135, looseness=1.2] (y);
\draw [->, line width= 1.25, dotted] (a) -- (l);

\node[text centered] at (-1.75,-8) (p) {$\textup{cPSE}_{M_2}$:};
\node[text centered] at (-1,-8) (a) {$A$};
\node[text centered] at (1,-8) (m) {$M_2$};
\node[text centered] at (0,-8) (l) {$M_1$};
\node[text centered] at (2,-8) (y) {$Y$};
\draw [->, line width= 1.25, gray] (l) -- (m);
\draw [->, line width= 1.25, gray] (l) to [out=-45, in=-135, looseness=1.2] (y);
\draw [->, line width= 1.25, gray] (m) -- (y); 
\draw [->, line width= 1.25, dotted] (a) to [out=-45, in=-135, looseness=1.2] (m);
\draw [->, line width= 1.25] (a) to [out=45, in=135, looseness=1.2] (y);
\draw [->, line width= 1.25, dotted] (a) -- (l);
\end{tikzpicture}
\par\end{center}%
\end{minipage}\hfill{}%
\begin{minipage}[t]{0.45\columnwidth}%
\noindent \begin{center}
\begin{tikzpicture}[yscale = 1, xscale = 2]
\node[text centered] at (0.5,2) (ref){Comparison};
\node[text centered] at (-1,0) (a) {$A$};
\node[text centered] at (1,0) (m) {$M_2$};
\node[text centered] at (0,0) (l) {$M_1$};
\node[text centered] at (2,0) (y) {$Y$};
\draw [->, line width= 1.25, gray] (l) -- (m);
\draw [->, line width= 1.25, gray] (l) to [out=-45, in=-135, looseness=1.2] (y);
\draw [->, line width= 1.25, gray] (m) -- (y); 
\draw [->, line width= 1.25, dotted] (a) to [out=-45, in=-135, looseness=1.2] (m);
\draw [->, line width= 1.25] (a) to [out=45, in=135, looseness=1.2] (y);
\draw [->, line width= 1.25, dotted] (a) -- (l);

\node[text centered] at (-1,-2) (a) {$A$};
\node[text centered] at (1,-2) (m) {$M_2$};
\node[text centered] at (0,-2) (l) {$M_1$};
\node[text centered] at (2,-2) (y) {$Y$};
\draw [->, line width= 1.25, gray] (l) -- (m);
\draw [->, line width= 1.25, gray] (l) to [out=-45, in=-135, looseness=1.2] (y);
\draw [->, line width= 1.25, gray] (m) -- (y); 
\draw [->, line width= 1.25, dotted] (a) to [out=-45, in=-135, looseness=1.2] (m);
\draw [->, line width= 1.25, dotted] (a) to [out=45, in=135, looseness=1.2] (y);
\draw [->, line width= 1.25] (a) -- (l);

\node[text centered] at (-1,-4) (a) {$A$};
\node[text centered] at (1,-4) (m) {$M_2$};
\node[text centered] at (0,-4) (l) {$M_1$};
\node[text centered] at (2,-4) (y) {$Y$};
\draw [->, line width= 1.25, gray] (l) -- (m);
\draw [->, line width= 1.25, gray] (l) to [out=-45, in=-135, looseness=1.2] (y);
\draw [->, line width= 1.25, gray] (m) -- (y); 
\draw [->, line width= 1.25] (a) to [out=-45, in=-135, looseness=1.2] (m);
\draw [->, line width= 1.25] (a) to [out=45, in=135, looseness=1.2] (y);
\draw [->, line width= 1.25] (a) -- (l);

\node[text centered] at (-1,-6) (a) {$A$};
\node[text centered] at (1,-6) (m) {$M_2$};
\node[text centered] at (0,-6) (l) {$M_1$};
\node[text centered] at (2,-6) (y) {$Y$};
\draw [->, line width= 1.25, gray] (l) -- (m);
\draw [->, line width= 1.25, gray] (l) to [out=-45, in=-135, looseness=1.2] (y);
\draw [->, line width= 1.25, gray] (m) -- (y); 
\draw [->, line width= 1.25] (a) to [out=-45, in=-135, looseness=1.2] (m);
\draw [->, line width= 1.25, dotted] (a) to [out=45, in=135, looseness=1.2] (y);
\draw [->, line width= 1.25, dotted] (a) -- (l);

\node[text centered] at (-1,-8) (a) {$A$};
\node[text centered] at (1,-8) (m) {$M_2$};
\node[text centered] at (0,-8) (l) {$M_1$};
\node[text centered] at (2,-8) (y) {$Y$};
\draw [->, line width= 1.25, gray] (l) -- (m);
\draw [->, line width= 1.25, gray] (l) to [out=-45, in=-135, looseness=1.2] (y);
\draw [->, line width= 1.25, gray] (m) -- (y); 
\draw [->, line width= 1.25] (a) to [out=-45, in=-135, looseness=1.2] (m);
\draw [->, line width= 1.25] (a) to [out=45, in=135, looseness=1.2] (y);
\draw [->, line width= 1.25, dotted] (a) -- (l);
\end{tikzpicture}
\par\end{center}%
\end{minipage}
\par\end{raggedright}
\begin{centering}
\caption{Illustrations of NDE, NIE, TIE, nPSE, and cPSE in the case of two
mediators.\label{fig:dag2}}
\medskip{}
\par\end{centering}
Note: $A$ denotes the treatment, $Y$ denotes the outcome of interest,
and $M_{1}$ and $M_{2}$ denote two causally ordered mediators. Solid
and dashed arrows for $A\to M_{1}$, $A\to M_{2}$, and $A\to Y$
denote activated ($A=1$) and unactivated ($A=0$) paths, respectively.
Gray arrows $M_{1}\to M_{2}$, $M_{1}\to Y$, and $M_{2}\to Y$ signify
that the mediators $M_{1}$ and $M_{2}$ are not under direct intervention.
\end{figure}

\subsection{Natural Direct Effect (NDE)}

The NDE measures the effect of switching treatment status from $0$
to $1$ in a hypothetical world where the mediators $(M_{1},\ldots M_{K})$
were all set to values they would have \textquotedblleft naturally\textquotedblright{}
taken for each unit under treatment status $A=0$. It is thus given
by $\psi_{\overline{0}_{K},1}-\psi_{\overline{0}_{K+1}}$. The first
row of Figure \ref{fig:dag2} illustrates the baseline and comparison
interventions associated with the NDE for the case of $K=2$, where
the black solid and dashed arrows for $A\to M_{1}$, $A\to M_{2}$,
and $A\to Y$ denote activated ($A=1$) and unactivated ($A=0$) paths,
respectively. A semiparametric efficient estimator for the NDE can
be constructed as
\begin{equation}
\widehat{\textup{NDE}}^{\textup{eif}_{2}}=\hat{\theta}_{\overline{0}_{K},1}^{\textup{eif}_{2}}-\hat{\theta}_{\overline{0}_{K+1}}^{\textup{eif}_{2}}.\label{eq:NDE}
\end{equation}
If we treat $\overline{M}_{K}=(M_{1},\ldots M_{K})$ as a whole, $\psi_{\overline{0}_{K},1}-\psi_{\overline{0}_{K+1}}$
coincides with the NDE defined in the single mediator setting. In
fact, $\widehat{\textup{NDE}}^{\textup{eif}_{2}}$ is akin to the
semiparametric estimator of the NDE given in \citet{zheng2012targeted}.
By contrast, if we use $\hat{\theta}_{\overline{a}}^{\textup{eif}_{1}}$
instead of $\hat{\theta}_{\overline{a}}^{\textup{eif}_{2}}$ in equation
\eqref{eq:NDE}, we obtain \citeauthor{tchetgen2012semiparametric}'s
(2012) estimator of the NDE.

Setting $a_{1}=\ldots a_{K+1}=0$ in equation \eqref{eq:EIF2}, we
have
\begin{equation}
\hat{\theta}_{\overline{0}_{K+1}}^{\textup{eif}_{2}}=\mathbb{P}_{n}\big[\frac{\mathbb{I}(A=0)}{\hat{\pi}_{0}(0|X)}\big(Y-\hat{\mu}_{0}(X)\big)+\hat{\mu}_{0}(X)\big],\label{eq:DR}
\end{equation}
where $\mu_{0}(X)=\mathbb{E}[Y|X,A=0]$. Not surprisingly, $\hat{\theta}_{\overline{0}_{K+1}}^{\textup{eif}_{2}}$
is the standard doubly robust estimator for $\mathbb{E}[Y(0)]$, which
is consistent if either $\hat{\pi}_{0}(0|X)$ or $\hat{\mu}_{0}(X)$
is consistent. Similarly, by setting $a_{1}=\ldots a_{K}=0$ and $a_{K+1}=1$
in equation \eqref{eq:EIF2}, we have
\[
\hat{\theta}_{\overline{0}_{K},1}^{\textup{eif}_{2}}=\mathbb{P}_{n}\big[\frac{\mathbb{I}(A=1)}{\hat{\pi}_{0}(0|X)}\frac{\hat{\pi}_{K}(0|X,\overline{M}_{K})}{\hat{\pi}_{K}(1|X,\overline{M}_{K})}\big(Y-\hat{\mu}_{K}(X,\overline{M}_{K})\big)+\frac{\mathbb{I}(A=0)}{\hat{\pi}_{0}(0|X)}\big(\hat{\mu}_{K}(X,\overline{M}_{K})-\hat{\mu}_{0,K}(X)\big)+\hat{\mu}_{0,K}(X)\big].
\]
In contrast to the general case where $\overline{a}_{K}$ is unconstrained,
$\hat{\theta}_{\overline{0}_{K},1}^{\textup{eif}_{2}}$ involves estimating
only four nuisance functions: $\pi_{0}(a|x)$, $\pi_{K}(a|x,\overline{m}_{K})$,
$\mu_{0,K}(x)$, and $\mu_{K}(x,\overline{m}_{K})$, where $\mu_{K}(x,\overline{m}_{K})=\mathbb{E}[Y|x,A=1,\overline{m}_{K}]$
and $\mu_{0,K}(x)=\mathbb{E}[\mu_{K}(X,\overline{M}_{K})|x,A=0]$.
Hence $\mu_{0,K}(x)$ can be estimated by fitting a model for the
conditional mean of $\hat{\mu}_{K}(X,\overline{M}_{K})$ given $(X,A)$
and then setting $A=0$ for all units. It follows from Theorem 3 that
$\hat{\theta}_{\overline{0}_{K},1}^{\textup{eif}_{2}}$ is triply
robust in that it is consistent if one of the following three conditions
holds: (a) $\hat{\pi}_{0}$ and $\hat{\pi}_{K}$ are consistent; (b)
$\hat{\pi}_{0}$ and $\hat{\mu}_{K}$ are consistent; and (c) $\hat{\mu}_{0,K}$
and $\hat{\mu}_{K}$ are consistent. In the meantime, we know that
$\hat{\theta}_{\overline{0}_{K+1}}^{\textup{eif}_{2}}$ is consistent
if either $\hat{\pi}_{0}$ or $\hat{\mu}_{0}$ is consistent. By taking
the intersection of the multiple robustness conditions for $\hat{\theta}_{\overline{0}_{K},1}^{\textup{eif}_{2}}$
and $\hat{\theta}_{\overline{0}_{K+1}}^{\textup{eif}_{2}}$, we deduce
that $\widehat{\textup{NDE}}^{\textup{eif}_{2}}$ is also triply robust,
as detailed in Corollary 1.
\begin{cor}
Suppose all assumptions required for Theorem 4 hold. When the nuisance
functions are estimated via parametric models, $\widehat{\textup{NDE}}^{\textup{eif}_{2}}$
is CAN provided that one of the following three sets of nuisance functions
is correctly specified and its parameter estimates are $\sqrt{n}$-consistent:
$\{\pi_{0},\pi_{K}\}$, $\{\pi_{0},\mu_{K}\}$, $\{\mu_{0},\mu_{0,K},\mu_{K}\}$.
$\widehat{\textup{NDE}}^{\textup{eif}_{2}}$ is semiparametric efficient
if all of the above nuisance functions are correctly specified and
their parameter estimates $\sqrt{n}$-consistent. When the nuisance
functions are estimated via data-adaptive methods and cross-fitting,
$\widehat{\textup{NDE}}^{\textup{eif}_{2}}$ is CAN and semiparametric
efficient if all of the nuisance functions are consistently estimated
and $r_{n}(\hat{\pi}_{0})r_{n}(\hat{\mu}_{0,K})+r_{n}(\hat{\pi}_{K})r_{n}(\hat{\mu}_{K})+r_{n}(\hat{\pi}_{0})r_{n}(\hat{\mu}_{0})=o(n^{-1/2})$.
\end{cor}

\subsection{Natural and Total Indirect Effects for $M_{1}$}

In Section \ref{subsec:Two Mediators}, we noted that $\psi_{100}-\psi_{000}$
and $\psi_{111}-\psi_{011}$ correspond to the NIE and TIE for the
first mediator $M_{1}$ (illustrated in the second and third rows
of Figure \ref{fig:dag2}). This correspondence extends naturally
to the case of $K$ mediators, where the NIE and TIE for $M_{1}$
are given by
\[
\textup{NIE}_{M_{1}}=\psi_{1,\underline{0}_{2}}-\psi_{\overline{0}_{K+1}},\quad\textup{TIE}_{M_{1}}=\psi_{\overline{1}_{K+1}}-\psi_{0,\underline{1}_{2}},
\]
where $\underline{0}_{2}=(0,\ldots0)$ and $\underline{1}_{2}=(1,\ldots1)$
are vectors of length $K$ representing the fact that $a_{2}=\ldots=a_{K+1}=0$
in $\textup{NIE}_{M_{1}}$ and $a_{2}=\ldots=a_{K+1}=1$ in $\textup{TIE}_{M_{1}}$.
Since $\textup{TIE}_{M_{1}}$ can be obtained by switching the 0s
and 1s in $\textup{NIE}_{M_{1}}$ and then flipping the sign, we focus
on $\textup{NIE}_{M_{1}}$ below, noting that analogous results hold
for $\textup{TIE}_{M_{1}}$.

A semiparametric efficient estimator of $\textup{NIE}_{M_{1}}$ can
be constructed as 
\[
\widehat{\textup{NIE}}_{M_{1}}^{\textup{eif}_{2}}=\hat{\theta}_{1,\underline{0}_{2}}^{\textup{eif}_{2}}-\hat{\theta}_{\overline{0}_{K+1}}^{\textup{eif}_{2}}.
\]
As shown previously, $\hat{\theta}_{\overline{0}_{K+1}}^{\textup{eif}_{2}}$
is given by the doubly robust estimator \eqref{eq:DR}. Setting $a_{1}=1$
and $a_{2}=\ldots a_{K+1}=0$ in equation \eqref{eq:EIF2}, we obtain
\begin{align*}
\hat{\theta}_{1,\underline{0}_{2}}^{\textup{eif}_{2}}= & \mathbb{P}_{n}\big[\frac{\mathbb{I}(A=0)}{\hat{\pi}_{0}(1|X)}\frac{\hat{\pi}_{1}(1|X,M_{1})}{\hat{\pi}_{1}(0|X,M_{1})}\big(Y-\hat{\mu}_{1}(X,M_{1})\big)+\frac{\mathbb{I}(A=1)}{\hat{\pi}_{0}(1|X)}\big(\hat{\mu}_{1}(X,M_{1})-\hat{\mu}_{0,1}(X)\big)+\hat{\mu}_{0,1}(X)\big].
\end{align*}
Like $\hat{\theta}_{\overline{0}_{K},1}^{\textup{eif}_{2}}$, $\hat{\theta}_{1,\underline{0}_{2}}^{\textup{eif}_{2}}$
also involves estimating four nuisance functions: $\pi_{0}(a|x)$,
$\pi_{1}(a|x,m_{1})$, $\mu_{0,1}(x)$, and $\mu_{1}(x,m_{1})$, where
$\mu_{1}(x,m_{1})=\mathbb{E}[Y|x,A=0,m_{1}]$ and $\mu_{0,1}(x)=\mathbb{E}[\mu_{1}(X,M_{1})|x,A=1]$.
It follows from Theorem 3 that $\hat{\theta}_{1,\underline{0}_{2}}^{\textup{eif}_{2}}$
is triply robust in that it is consistent if one of the following
three conditions holds: (a) $\hat{\pi}_{0}$ and $\hat{\pi}_{1}$
are consistent; (b) $\hat{\pi}_{0}$ and $\hat{\mu}_{1}$ are consistent;
and (c) $\hat{\mu}_{0,1}$ and $\hat{\mu}_{1}$ are consistent. By
taking the intersection of the multiple robustness conditions for
$\hat{\theta}_{1,\underline{0}_{2}}^{\textup{eif}_{2}}$ and $\hat{\theta}_{\overline{0}_{K+1}}^{\textup{eif}_{2}}$,
we deduce that $\widehat{\textup{NIE}}_{M_{1}}^{\textup{eif}_{2}}$
is also triply robust, as detailed in Corollary 2.
\begin{cor}
Suppose all assumptions required for Theorem 4 hold. When the nuisance
functions are estimated via parametric models, $\widehat{\textup{NIE}}_{M_{1}}^{\textup{eif}_{2}}$
is CAN provided that one of the following three sets of nuisance functions
is correctly specified and its parameter estimates are $\sqrt{n}$-consistent:
$\{\pi_{0},\pi_{1}\}$, $\{\pi_{0},\mu_{1}\}$, $\{\mu_{0},\mu_{0,1},\mu_{1}\}$.
$\widehat{\textup{NIE}}_{M_{1}}^{\textup{eif}_{2}}$ is semiparametric
efficient if all of the above nuisance functions are correctly specified
and their parameter estimates $\sqrt{n}$-consistent. When the nuisance
functions are estimated via data-adaptive methods and cross-fitting,
$\widehat{\textup{NIE}}_{M_{1}}^{\textup{eif}_{2}}$ is semiparametric
efficient if all of the nuisance functions are consistently estimated
and $r_{n}(\hat{\pi}_{0})r_{n}(\hat{\mu}_{0,1})+r_{n}(\hat{\pi}_{1})r_{n}(\hat{\mu}_{1})+r_{n}(\hat{\pi}_{0})r_{n}(\hat{\mu}_{0})=o(n^{-1/2})$.
\end{cor}

\subsection{Natural Path-Specific Effects (nPSE) for $M_{k}$ ($k\protect\geq2$)\label{subsec:nPSEs}}

In the same spirit of the NIE for $M_{1}$, the natural path-specific
effect (nPSE; \citealt{daniel2015causal}) for mediator $M_{k}$ ($k\geq2$)
is defined as
\[
\textup{nPSE}_{M_{k}}=\psi_{\overline{0}_{k-1},1,\underline{0}_{k+1}}-\psi_{\overline{0}_{K+1}}.
\]
It can be interpreted as the effect of activating the path $A\to M_{k}\rightsquigarrow Y$
while all other causal paths are ``switched off,'' as shown in the
fourth row of Figure \ref{fig:dag2}. A semiparametric efficient estimator
of $\textup{nPSE}_{M_{k}}$ can be constructed as
\[
\widehat{\textup{nPSE}}_{M_{k}}^{\textup{eif}_{2}}=\hat{\theta}_{\overline{0}_{k-1},1,\underline{0}_{k+1}}^{\textup{eif}_{2}}-\hat{\theta}_{\overline{0}_{K+1}}^{\textup{eif}_{2}}.
\]
If, instead, we use $\hat{\theta}_{\overline{a}}^{\textup{eif}_{1}}$
in the above equation, the resulting estimator $\widehat{\textup{nPSE}}_{M_{k}}^{\textup{eif}_{1}}$
can be seen as \citeauthor{miles2020semiparametric}'s (2020) estimator
of $\theta_{010}-\theta_{000}$ applied to $\widetilde{M}_{1}=(M_{1},M_{2},\ldots M_{k-1})$
and $\widetilde{M}_{2}=M_{k}$.

Again, $\hat{\theta}_{\overline{0}_{K+1}}^{\textup{eif}_{2}}$ is
given by the doubly robust estimator \eqref{eq:DR}. Setting $a_{1}=\ldots a_{k-1}=a_{k+1}=\ldots a_{K+1}=0$
and $a_{k}=1$ in equation \eqref{eq:EIF2}, we obtain
\begin{align*}
\hat{\theta}_{\overline{0}_{k-1},1,\underline{0}_{k+1}}^{\textup{eif}_{2}}= & \mathbb{P}_{n}\big[\frac{\mathbb{I}(A=0)}{\hat{\pi}_{0}(0|X)}\frac{\hat{\pi}_{k-1}(0|X,\overline{M}_{k-1})}{\hat{\pi}_{k-1}(1|X,\overline{M}_{k-1})}\frac{\hat{\pi}_{k}(1|X,\overline{M}_{k})}{\hat{\pi}_{k}(0|X,\overline{M}_{k})}\big(Y-\hat{\mu}_{k}(X,\overline{M}_{k})\big).\\
 & +\frac{\mathbb{I}(A=1)}{\hat{\pi}_{0}(0|X)}\frac{\hat{\pi}_{k-1}(0|X,\overline{M}_{k-1})}{\hat{\pi}_{k-1}(1|X,\overline{M}_{k-1})}\big(\hat{\mu}_{k}(X,\overline{M}_{k})-\hat{\mu}_{k-1,k}(X,\overline{M}_{k-1})\big)\\
 & +\frac{\mathbb{I}(A=0)}{\hat{\pi}_{0}(0|X)}\big(\hat{\mu}_{k-1,k}(X,\overline{M}_{k-1})-\hat{\mu}_{0,k-1,k}(X)\big)+\hat{\mu}_{0,k-1,k}(X)\big].
\end{align*}
We can see that $\hat{\theta}_{\overline{0}_{k-1},1,\underline{0}_{k+1}}^{\textup{eif}_{2}}$involves
estimating six nuisance functions: $\pi_{0}(a|x)$, $\pi_{k-1}(a|x,\overline{m}_{k-1})$,
$\pi_{k}(a|x,\overline{m}_{k})$, $\mu_{0,k-1,k}(x)$, $\mu_{k-1,k}(x,\overline{m}_{k-1})$,
and $\mu_{k}(x,\overline{m}_{k})$, where $\mu_{k}(X,\overline{M}_{k})=\mathbb{E}[Y|X,A=0,\overline{M}_{k}]$,
$\mu_{k-1,k}(X,\overline{M}_{k-1})=\mathbb{E}[\mu_{k}(X,\overline{M}_{k})|X,A=1,\overline{M}_{k-1}]$,
and $\mu_{0,k-1,k}(X)=\mathbb{E}[\mu_{k-1,k}(X,\overline{M}_{k-1})|X,A=0]$.
Hence $\mu_{k-1,k}(x)$ can be estimated by fitting a model for the
conditional mean of $\hat{\mu}_{k}(X,\overline{M}_{k})$ given ($X,A,\overline{M}_{k-1})$
and then setting $A=1$ for all units, and $\mu_{0,k-1,k}(x)$ can
be estimated by fitting a model for the conditional mean of $\hat{\mu}_{k-1,k}(X,\overline{M}_{k-1})$
given ($X,A)$ and then setting $A=0$ for all units. It follows from
Theorem 3 that $\hat{\theta}_{\overline{0}_{k-1},1,\underline{0}_{k+1}}^{\textup{eif}_{2}}$
is quadruply robust in that it is consistent if one of the following
four conditions holds: (a) $\hat{\pi}_{0}$, $\hat{\pi}_{k-1}$, and
$\hat{\pi}_{k}$ are consistent; (b) $\hat{\pi}_{0}$, $\hat{\pi}_{k-1}$,
and $\hat{\mu}_{k}$ are consistent; (c) $\hat{\pi}_{0}$, $\hat{\mu}_{k-1,k}$,
and $\hat{\mu}_{k}$ are consistent; and (d) $\hat{\mu}_{0,k-1,k}$,
$\hat{\mu}_{k-1,k}$, and $\hat{\mu}_{k}$ are consistent. By taking
the intersection of the multiple robustness conditions for $\hat{\theta}_{1,\underline{0}_{2}}^{\textup{eif}_{2}}$
and $\hat{\theta}_{\overline{0}_{K+1}}^{\textup{eif}_{2}}$, we deduce
that $\widehat{\textup{nPSE}}_{M_{k}}^{\textup{eif}_{2}}$ is also
quadruply robust, as detailed in Corollary 3.
\begin{cor}
\noindent Suppose all assumptions required for Theorem 4 hold. When
the nuisance functions are estimated via parametric models, $\widehat{\textup{nPSE}}_{M_{k}}^{\textup{eif}_{2}}$
is CAN provided that one of the following four sets of nuisance functions
is correctly specified and its parameter estimates are $\sqrt{n}$-consistent:
$\{\pi_{0},\pi_{k-1},\pi_{k}\}$, $\{\pi_{0},\pi_{k-1},\mu_{k}\}$,
$\{\pi_{0},\mu_{k-1,k},\mu_{k}\}$, $\{\mu_{0},\mu_{0,k-1,k},\mu_{k-1,k},\mu_{k}\}$.
$\widehat{\textup{nPSE}}_{M_{k}}^{\textup{eif}_{2}}$ is semiparametric
efficient if all of the above nuisance functions are correctly specified
and their parameter estimates $\sqrt{n}$-consistent. When the nuisance
functions are estimated via data-adaptive methods and cross-fitting,
$\widehat{\textup{nPSE}}_{M_{k}}^{\textup{eif}_{2}}$ is semiparametric
efficient if all of the nuisance functions are consistently estimated
and $r_{n}(\hat{\pi}_{0})r_{n}(\hat{\mu}_{0,k-1,k})+r_{n}(\hat{\pi}_{k-1})r_{n}(\hat{\mu}_{k-1,k})+r_{n}(\hat{\pi}_{k})r_{n}(\hat{\mu}_{k})+r_{n}(\hat{\pi}_{0})r_{n}(\hat{\mu}_{0})=o(n^{-1/2})$.
\end{cor}

\subsection{Cumulative Path-Specific Effects (cPSE) for $M_{k}$ ($k\protect\geq2$)
\label{subsec:cPSEs}}

The NDE, NIE, and nPSE are all defined as the effect of activating
one causal path while keeping all other causal paths ``switched off.''
By contrast, in equation \eqref{eq:g-decomp1}, the ATE is decomposed
into $K+1$ components, each of which reflects the \textit{cumulative}
contribution of a specific mediator to the ATE. Specifically, the
component $\psi_{\overline{0}_{K},1}-\psi_{\overline{0}_{K+1}}$equals
the NDE, the component $\psi_{\underline{1}_{1}}-\psi_{0,\underline{1}_{2}}$
equals $\textup{TIE}_{M_{1}}$, and the component $\psi_{\overline{0}_{k-1},\underline{1}_{k}}-\psi_{\overline{0}_{k},\underline{1}_{k+1}}$
gauges the additional contribution of the causal path $A\to M_{k}\rightsquigarrow Y$
after the causal paths $A\to M_{k+1}\rightsquigarrow Y,\ldots A\to M_{K}\rightsquigarrow Y,A\to Y$
are ``switched on.'' Such a decomposition will be useful in applications
where the investigator aims to partition the ATE into its path-specific
components.

We define the cumulative path-specific effect (cPSE) for mediator
$M_{k}$ ($k\ge2$) as 
\[
\textup{cPSE}_{M_{k}}=\psi_{\overline{0}_{k-1},\underline{1}_{k}}-\psi_{\overline{0}_{k},\underline{1}_{k+1}}.
\]
The last row of Figure \ref{fig:dag2} gives the baseline and comparison
interventions associated with $\textup{cPSE}_{M_{2}}$ in the case
of $K=2$. A semiparametric efficient estimator for $\textup{cPSE}_{M_{k}}$
can be constructed as 
\[
\widehat{\textup{cPSE}}_{M_{k}}^{\textup{eif}_{2}}=\hat{\theta}_{\overline{0}_{k-1},\underline{1}_{k}}^{\textup{eif}_{2}}-\hat{\theta}_{\overline{0}_{k},\underline{1}_{k+1}}^{\textup{eif}_{2}}.
\]
Setting $a_{1}=\ldots a_{k}=0$ and $a_{k+1}=\ldots=a_{K+1}=1$ in
equation \eqref{eq:EIF2}, we obtain
\begin{align}
\hat{\theta}_{\overline{0}_{k},\underline{1}_{k+1}}^{\textup{eif}_{2}} & =\mathbb{P}_{n}\big[\frac{\mathbb{I}(A=1)}{\hat{\pi}_{0}(0|X)}\frac{\hat{\pi}_{k}(0|X,\overline{M}_{k})}{\hat{\pi}_{k}(1|X,\overline{M}_{k})}\big(Y-\hat{\mu}_{k}(X,\overline{M}_{k})\big)+\frac{\mathbb{I}(A=0)}{\hat{\pi}_{0}(0|X)}\big(\hat{\mu}_{k}(X,\overline{M}_{k})-\hat{\mu}_{0,k}(X)\big)+\hat{\mu}_{0,k}(X)\big],\label{eq:cPSE_k}
\end{align}
where $\mu_{k}(X,\overline{M}_{k})=\mathbb{E}[Y|X,A=1,\overline{M}_{k}]$
and $\mu_{0,k}(X)=\mathbb{E}[\mu_{k}(X,\overline{M}_{k})|X,A=0]$.
It follows from Theorem 3 that $\hat{\theta}_{\overline{0}_{k},\underline{1}_{k+1}}^{\textup{eif}_{2}}$
is triply robust in that it is consistent if one of the following
three conditions holds: (a) $\hat{\pi}_{0}$ and $\hat{\pi}_{k}$
are consistent; (b) $\hat{\pi}_{0}$ and $\hat{\mu}_{k}$ are consistent;
and (c) $\hat{\mu}_{0,k}$ and $\hat{\mu}_{k}$ are consistent. By
replacing $k$ with $k-1$ in equation \eqref{eq:cPSE_k}, we obtain
a similar expression for $\hat{\theta}_{\overline{0}_{k-1},\underline{1}_{k}}^{\textup{eif}_{2}}$,
which is also triply robust in that it is consistent if one of the
following three conditions holds: (a) $\hat{\pi}_{0}$ and $\hat{\pi}_{k-1}$
are consistent; (b) $\hat{\pi}_{0}$ and $\hat{\mu}_{k-1}$ are consistent;
and (c) $\hat{\mu}_{0,k-1}$ and $\hat{\mu}_{k-1}$ are consistent.
As a result, $\widehat{\textup{cPSE}}_{M_{k}}^{\textup{eif}_{2}}$
involves fitting seven working models --- for $\pi_{0}(a|x)$, $\pi_{k-1}(a|x,\overline{m}_{k-1})$,
$\pi_{k}(a|x,\overline{m}_{k})$, $\mu_{k-1}(x,\overline{m}_{k-1})$,
$\mu_{0,k-1}(x)$, $\mu_{k}(x,\overline{m}_{k})$, and $\mu_{0,k}(x)$.
By taking the intersection of the multiple robustness conditions for
$\hat{\theta}_{\overline{1}_{k},\underline{0}_{k+1}}^{\textup{eif}_{2}}$and
$\hat{\theta}_{\overline{1}_{k-1},\underline{0}_{k}}^{\textup{eif}_{2}}$,
we deduce that $\widehat{\textup{cPSE}}_{M_{k}}^{\textup{eif}_{2}}$
is quintuply robust in that it is consistent if one of five sets of
nuisance functions is correctly specified and consistently estimated,
as detailed in Corollary 4.
\begin{cor}
Suppose all assumptions required for Theorem 4 hold. When the nuisance
functions are estimated via parametric models, $\widehat{\textup{cPSE}}_{M_{k}}^{\textup{eif}_{2}}$
is CAN provided that one of the following five sets of nuisance functions
is correctly specified and its parameter estimates are $\sqrt{n}$-consistent:
$\{\pi_{0},\pi_{k-1},\pi_{k}\}$;$\{\pi_{0},\pi_{k-1},\mu_{k}\}$;$\{\pi_{0},\mu_{k-1},\pi_{k}\}$;$\{\pi_{0},\mu_{k-1},\mu_{k}\}$;$\{\mu_{0,k-1},\mu_{0,k},\mu_{k-1},\mu_{k}\}$.
$\widehat{\textup{cPSE}}_{M_{k}}^{\textup{eif}_{2}}$ is semiparametric
efficient if all of the above nuisance functions are correctly specified
and their parameter estimates $\sqrt{n}$-consistent. When the nuisance
functions are estimated via data-adaptive methods and cross-fitting,
$\widehat{\textup{cPSE}}_{M_{k}}^{\textup{eif}_{2}}$ is semiparametric
efficient if all of the nuisance functions are consistently estimated
and $r_{n}(\hat{\pi}_{0})r_{n}(\hat{\mu}_{0,k-1})+r_{n}(\hat{\pi}_{0})r_{n}(\hat{\mu}_{0,k})+r_{n}(\hat{\pi}_{k-1})r_{n}(\hat{\mu}_{k-1})+r_{n}(\hat{\pi}_{k})r_{n}(\hat{\mu}_{k})=o(n^{-1/2})$.
\end{cor}

\section{A Simulation Study\label{sec:A-Simulation-Study}}

In this section, we conduct a simulation study to demonstrate the
robustness of various estimators under different forms of model misspecification.
Specifically, we consider a binary treatment $A$, a continuous outcome
$Y$, two causally ordered mediators $M_{1}$ and $M_{2}$, and four
pretreatment covariates $X_{1},X_{2},X_{3},X_{4}$ generated from
the following model:
\begin{align*}
(U_{1},U_{2},U_{3},U_{XY}) & \sim N(0,I_{4}),\\
X_{j} & \sim N((U_{1},U_{2},U_{3},U_{XY})\beta_{X_{j}},1),\quad j=1,2,3,4,\\
A & \sim\textup{Bernoulli}\big(\textup{logit}^{-1}[(1,X_{1},X_{2},X_{3},X_{4})\beta_{A}]\big),\\
M_{1} & \sim N\big((1,X_{1},X_{2},X_{3},X_{4},A)\beta_{M_{1}},1\big),\\
M_{2} & \sim N\big((1,X_{1},X_{2},X_{3},X_{4},A,M_{1})\beta_{M_{2}},1\big),\\
Y & \sim N\big((1,U_{XY},X_{1},X_{2},X_{3},X_{4},A,M_{1},M_{2})\beta_{Y},1\big).
\end{align*}
The coefficients $\beta_{X_{j}}(\text{1\ensuremath{\leq j\leq4}})$,
$\beta_{A}$, $\beta_{M_{1}}$, $\beta_{M_{2}},$ $\beta_{Y}$ are
produced from a set of uniform distributions (see Supplementary Material
\ref{sec:Simulation-Details} for more details). Given the coefficients,
we generate 1,000 Monte Carlo samples of size 2,000. Note that in
the above model, the unobserved variable $U_{XY}$ confounds the $X$-$Y$
relationship but does not pose an identification threat for $\psi_{\overline{a}}$
and the associated PSEs (i.e., Assumption 2 still holds). 

Without loss of generality, we focus on the estimand $\textup{cPSE}_{M_{2}}$,
which we estimate by $\hat{\theta}_{011}-\hat{\theta}_{001}$. To
highlight the general results stated in Theorem 3, we use only estimators
for the generic $\theta_{\overline{a}}$ (i.e., those described in
Section \ref{sec:Estimation}). First, we consider the weighting estimator
$\hat{\theta}_{\overline{a}}^{\textup{w-a}}$, the regression-imputation
estimator $\hat{\theta}_{\overline{a}}^{\textup{ri}}$, and the hybrid
estimators $\hat{\theta}_{\overline{a}}^{\textup{ri-w-w}}$ and $\hat{\theta}_{\overline{a}}^{\textup{ri-ri-w}}$,
where the mediator density ratio involved in $\hat{\theta}_{\overline{a}}^{\textup{ri-w-w}}$
is estimated via the corresponding odds ratio of the treatment variable.
We then consider four EIF-based estimators $\hat{\theta}_{\textup{\ensuremath{\overline{a}}}}^{\textup{par,eif}_{2}}$,
$\hat{\theta}_{\textup{\ensuremath{\overline{a}}}}^{\textup{par}_{2},\textup{eif}_{2}}$,
$\hat{\theta}_{\textup{\ensuremath{\overline{a}}}}^{\textup{np,eif}_{2}}$,
and $\hat{\theta}_{\textup{\ensuremath{\overline{a}}}}^{\textup{tmle,eif}_{2}}$.
For $\hat{\theta}_{\textup{\ensuremath{\overline{a}}}}^{\textup{par,eif}_{2}}$
and $\hat{\theta}_{\textup{\ensuremath{\overline{a}}}}^{\textup{par}_{2},\textup{eif}_{2}}$,
the nuisance functions are estimated via GLMs. $\hat{\theta}_{\textup{\ensuremath{\overline{a}}}}^{\textup{par}_{2},\textup{eif}_{2}}$
differs from $\hat{\theta}_{\textup{\ensuremath{\overline{a}}}}^{\textup{par,eif}_{2}}$
in that the outcome models $\mu_{2}(x,m_{1},m_{2})$, $\mu_{1}(x,m_{1})$,
and $\mu_{0}(x)$ are fitted using a set of weighted GLMs such that
in equation \eqref{eq:EIF2-1}, all terms inside $\mathbb{P}_{n}[\cdot]$
but $\hat{\mu}_{0}(X)$ have a zero sample mean, yielding a regression-imputation
estimator that may perform better in finite samples.

All of the above estimators are constructed using estimates of six
nuisance functions: $\pi_{0}(a|x)$, $\pi_{1}(a|x,m_{1})$, $\pi_{2}(a|x,m_{1},m_{2})$,
$\mu_{0}(x)$, $\mu_{1}(x,m_{1})$, and $\mu_{2}(x,m_{1},m_{2})$.
To demonstrate the consequences of model misspecification and the
multiple robustness of $\hat{\theta}_{\textup{\ensuremath{\overline{a}}}}^{\textup{par,eif}_{2}}$
and $\hat{\theta}_{\textup{\ensuremath{\overline{a}}}}^{\textup{par}_{2},\textup{eif}_{2}}$,
we generate a set of ``false covariates'' $Z=\big(X_{1},e^{X_{2}/2},(X_{3}/X_{1})^{1/3},X_{4}/(e^{X_{1}/2}+1)\big)$
and use them to fit a misspecified GLM for each of the nuisance functions
(with only the main effects of $Z_{1},Z_{2},Z_{3},Z_{4}$). We evaluate
each of the parametric estimators under five different cases: (a)
only $\pi_{0}$, $\pi_{1}$, $\pi_{2}$ are correctly specified; (b)
only $\pi_{0}$, $\pi_{1}$, $\mu_{2}$ are correctly specified; (c)
only $\pi_{0}$, $\mu_{1}$, $\mu_{2}$ are correctly specified; (d)
only $\mu_{0}$, $\mu_{1}$, $\mu_{2}$ are correctly specified; and
(e) all of the six nuisance functions are misspecified. In theory,
$\hat{\theta}_{\overline{a}}^{\textup{w-a}}$ is consistent in case
(a), $\hat{\theta}_{\overline{a}}^{\textup{ri-w-w}}$ is consistent
in case (b), $\hat{\theta}_{\overline{a}}^{\textup{ri-ri-w}}$ is
consistent in case (c), $\hat{\theta}_{\overline{a}}^{\textup{ri}}$
is consistent in case (d), and $\hat{\theta}_{\textup{\ensuremath{\overline{a}}}}^{\textup{par,eif}_{2}}$
and $\hat{\theta}_{\textup{\ensuremath{\overline{a}}}}^{\textup{par}_{2},\textup{eif}_{2}}$
are consistent in cases (a)-(d). The corresponding estimators of $\textup{cPSE}_{M_{2}}$
should follow the same properties.

For the two nonparametric estimators, $\hat{\theta}_{\textup{\ensuremath{\overline{a}}}}^{\textup{np,eif}_{2}}$
is based on estimating equation \eqref{eq:EIF2}, and $\hat{\theta}_{\textup{\ensuremath{\overline{a}}}}^{\textup{tmle,eif}_{2}}$
is based on the method of TMLE. Like $\hat{\theta}_{\textup{\ensuremath{\overline{a}}}}^{\textup{par}_{2},\textup{eif}_{2}}$,
$\hat{\theta}_{\textup{\ensuremath{\overline{a}}}}^{\textup{tmle,eif}_{2}}$
is a regression-imputation estimator, which may have better finite-sample
performance than $\hat{\theta}_{\textup{\ensuremath{\overline{a}}}}^{\textup{np,eif}_{2}}$.
For both $\hat{\theta}_{\textup{\ensuremath{\overline{a}}}}^{\textup{np,eif}_{2}}$
and $\hat{\theta}_{\textup{\ensuremath{\overline{a}}}}^{\textup{tmle,eif}_{2}}$,
the nuisance functions are estimated via a super learner (\citealt{van2007super})
composed of Lasso and random forest, where the feature matrix consists
of first-order, second-order, and interaction terms of \textit{the
false covariates $Z$}. The super learner is more flexible than a
misspecified GLM consisting of only the main effects of $Z$, but
it remains agnostic about the true nuisance functions, which are either
logit or linear models that depend on $X=(Z_{1},2\log(Z_{2}),Z_{1}Z_{3}^{3},(1+e^{Z_{1}/2})Z_{4})$.
We obtain nonparametric estimates of $\textup{cPSE}_{M_{2}}$ using
both five-fold cross-fitting and no cross-fitting.

\begin{figure}[!th]
\centering{}\includegraphics[width=1\textwidth]{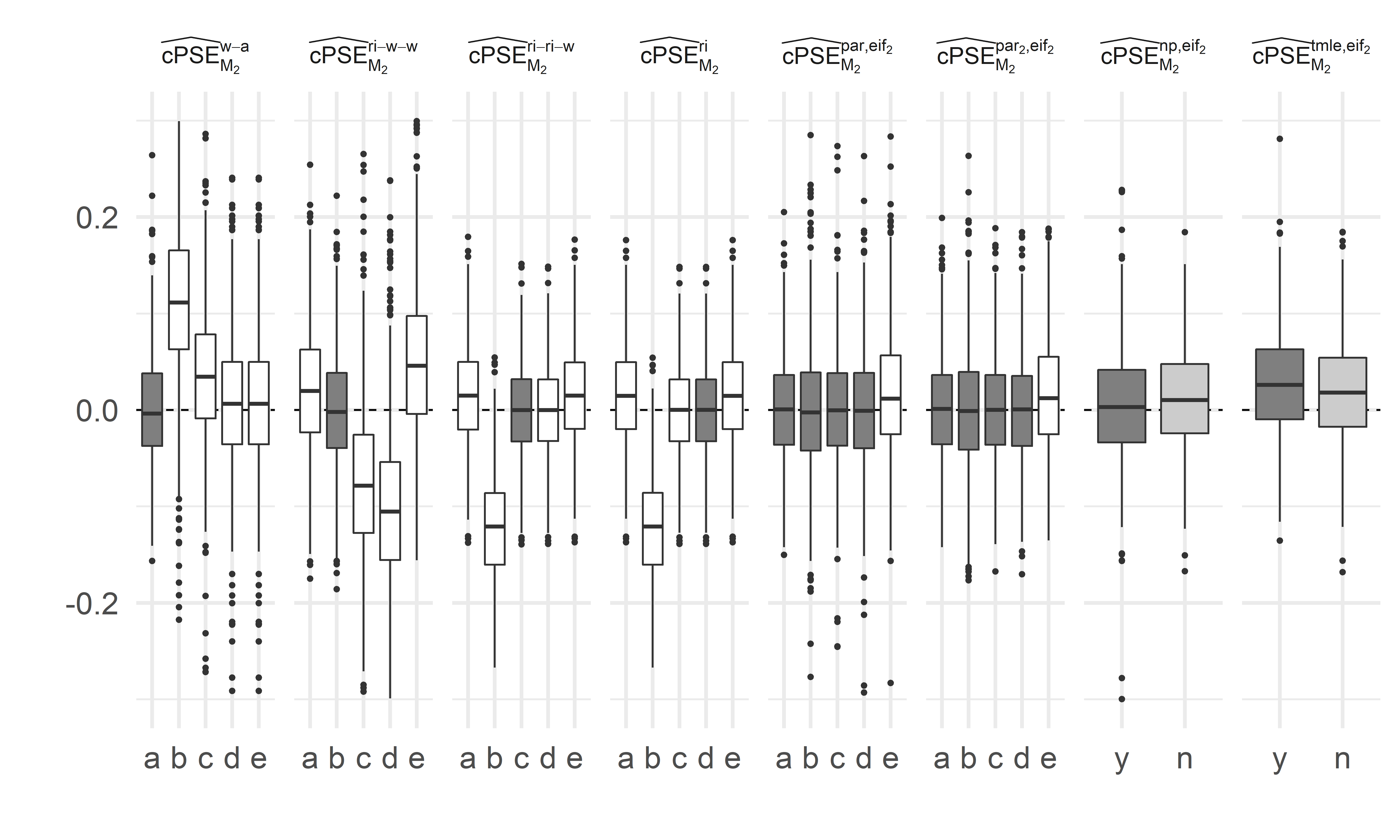}\caption{Sampling distributions of eight different estimators for $n=2,000$.
Cases (a)-(e) are described in the main text. The symbols\textsf{
y }and \textsf{n} denote whether cross-fitting is used to implement
the nonparametric estimators \texttt{(}\textsf{y} = yes, \textsf{n}
= no). \label{fig:simulation}}
\end{figure}

Results from the simulation study are shown in Figure \ref{fig:simulation},
where each panel corresponds to an estimator, and the \textit{y} axis
is recentered at the true value of $\textup{cPSE}_{M_{2}}$. The shaded
box plots highlight cases under which a given estimator should be
consistent, and the box plots with a lighter shade in the last two
panels denote nonparametric estimators obtained without cross-fitting.
From the first four panels, we can see that the weighting, regression-imputation,
and hybrid estimators all behave as expected. They center around the
true value if the requisite nuisance functions are all correctly specified,
and deviate from the truth in most other cases. The next four panels
show the box plots of the EIF-based estimators. As expected, both
of the parametric EIF-based estimators are quadruply robust, as their
sampling distributions roughly concentrate around the true value in
all of the four cases from (a) to (d). Moreover, it is reassuring
to see that when all of the nuisance functions are misspecified (case
(e)), the multiply robust estimators do not show a larger amount of
bias than those of the other parametric estimators. Finally, both
of the nonparametric EIF-based estimators perform reasonably well.
When cross-fitting is used, the estimating equation estimator $\widehat{\textup{cPSE}}_{M_{2}}^{\textup{np},\textup{eif}_{2}}$
appears to have a smaller bias than the TMLE estimator $\widehat{\textup{cPSE}}_{M_{2}}^{\textup{tmle},\textup{eif}_{2}}$,
but it occasionally gives rise to extreme estimates. Their 95\% Wald
confidence intervals, constructed using the estimated variance $\hat{\mathbb{E}}\big[\big(\hat{\varphi}_{011}-\hat{\varphi}_{001}\big)^{2}\big]/n$,
have close-to-nominal coverage rates --- 95.5\% for $\widehat{\textup{cPSE}}_{M_{2}}^{\textup{np},\textup{eif}_{2}}$
and 90.9\% for $\widehat{\textup{cPSE}}_{M_{2}}^{\textup{tmle},\textup{eif}_{2}}$.
Without cross-fitting, the point estimates exhibit similar distributions,
but the coverage rates of the corresponding 95\% confidence intervals
are somewhat lower --- 87.3\% for $\widehat{\textup{cPSE}}_{\textup{\ensuremath{M_{2}}}}^{\textup{np},\textup{eif}_{2}}$
and 85.8\% for $\widehat{\textup{cPSE}}_{M_{2}}^{\textup{tmle},\textup{eif}_{2}}$.

\section{An Empirical Application}

In this section, we illustrate semiparametric estimation of PSEs by
analyzing the causal pathways through which higher education affects
political participation. Prior research suggests that college attendance
has a substantial positive effect on political participation in the
United States (e.g., \citealt{dee2004there,milligan2004does}). Yet,
the mechanisms underlying this causal link remain unclear. The effect
of college on political participation may operate through the development
of civic and political interest (e.g., \citealt{hillygus2005missing}),
through an increase in economic status (e.g., \citealt{kingston2003education}),
or through other pathways such as social and occupational networks
(e.g., \citealt{rolfe2012voter}). To examine these direct and indirect
effects, we consider a causal structure akin to the top panel of Figure
\ref{fig:dag}, where $A$ denotes college attendance, $Y$ denotes
political participation, and $M_{1}$ and $M_{2}$ denote two causally
ordered mediators that reflect (a) economic status, and (b) civic
and political interest, respectively.

In this model, economic status is allowed to affect civic and political
interest but not vice versa, which we consider to be a reasonable
approximation to reality. Nonetheless, the conditional independence
assumption (Assumption 2) is still strong in this context, as it rules
out unobserved confounding for any of the pairwise relationships between
college attendance, economic status, civic and political interest,
and political participation. Thus, the following analyses should be
viewed as an illustration of the proposed methodology rather than
a definitive assessment of the PSEs of interest.

We use data from $n=2,969$ individuals in the National Longitudinal
Survey of Youth 1997 (NLSY97) who were age 15-17 in 1997 and had completed
high school by age 20. The treatment $A$ is a binary indicator for
whether the individual had attended a two-year or four-year college
by age 20. The outcome $Y$ is a binary indicator for whether the
individual voted in the 2010 general election. We measure economic
status ($M_{1}$) using the respondent's average annual earnings from
2006 to 2009. To gauge civic and political interest ($M_{2}$), we
use a set of variables that reflect the respondent's interest in government
and public affairs and involvement in volunteering, donation, community
group activities between 2007 and 2010. The overlap of the periods
in which $M_{1}$ and $M_{2}$ were measured is a limitation of this
analysis, and it makes our earlier assumption that $M_{2}$ does not
affect $M_{1}$ essential for identifying the direct and path-specific
effects.

To minimize potential bias due to unobserved confounding, we include
a rich set of pre-college individual and contextual characteristics
in the vector of pretreatment covariates $X$. They include gender,
race, ethnicity, age at 1997, parental education, parental income,
parental assets, presence of a father figure, co-residence with both
biological parents, percentile score on the Armed Services Vocational
Aptitude Battery (ASVAB), high school GPA, an index of substance use
(ranging from 0 to 3), an index of delinquency (ranging from 0 to
10), whether the respondent had any children by age 18, college expectation
among the respondent's peers, and a number of school-level characteristics.
Descriptive statistics on these pre-college characteristics as well
as the mediators and the outcome are given in Supplementary Material
\ref{sec:NLSY97}. Some components of $X$, $M_{1}$, and $M_{2}$
contain a small fraction of missing values. They are imputed via a
random-forest-based multiple imputation procedure (with ten imputed
data sets). The standard errors of our parameter estimates are adjusted
using Rubin's \citeyearpar{rubin1987imputation} method.

Under Assumptions 1-3 given in Section \ref{subsec:Two Mediators},
a set of PSEs reflecting the causal paths $A\to Y$, $A\to M_{1}\rightsquigarrow Y$,
and $A\to M_{2}\to Y$ are identified. For illustrative purposes,
we focus on the cumulative PSEs (cPSEs) defined in Section \ref{subsec:cPSEs}:
\begin{equation}
\textup{ATE}=\underbrace{\psi_{001}-\psi_{000}}_{A\to Y}+\underbrace{\psi_{011}-\psi_{001}}_{A\to M_{2}\to Y}+\underbrace{\psi_{111}-\psi_{011}}_{A\to M_{1}\rightsquigarrow Y}.\label{eq:decomp2}
\end{equation}
Here, the first component is the NDE of college attendance, and the
second and third components reflect the amounts of treatment effect
that are additionally mediated by civic/political interest and economic
status, respectively. Since $M_{2}$ is multivariate, it would be
difficult to model its conditional distributions directly. We thus
estimate the PSEs using the estimator $\hat{\theta}_{a_{1},a_{2},a}^{\textup{eif}_{2}}$.
Each of the nuisance functions is estimated using a super learner
composed with Lasso and random forest. For computational reasons,
the feature matrix supplied to the super learner consists of only
first-order terms of the corresponding variables. As in our simulation
study, we implement two versions of this EIF-based estimator, one
based on the original estimating equation ($\hat{\theta}_{\textup{\ensuremath{\overline{a}}}}^{\textup{np,eif}_{2}}$),
and one based on the method of TMLE ($\hat{\theta}_{\textup{\ensuremath{\overline{a}}}}^{\textup{tmle,eif}_{2}}$).
Five-fold cross-fitting is used to obtain the final estimates.

\begin{table}
\caption{Estimates of total and path-specific effects of college attendance
on voting.\label{tab:Estimates}}
\smallskip{}

\begin{centering}
\begin{tabular}{lcc}
\hline 
 & Estimating equation ($\hat{\theta}_{\textup{\ensuremath{\overline{a}}}}^{\textup{np,eif}_{2}}$) & TMLE ($\hat{\theta}_{\textup{\ensuremath{\overline{a}}}}^{\textup{tmle,eif}_{2}}$)\tabularnewline
\hline 
Average total effect & 0.152 (0.022) & 0.156 (0.023)\tabularnewline
Through economic status ($A\to M_{1}\rightsquigarrow Y$) & 0.007 (0.005) & 0.002 (0.005)\tabularnewline
Through civic/political interest ($A\to M_{2}\to Y$) & 0.042 (0.008) & 0.049 (0.008)\tabularnewline
Direct effect ($A\to Y$) & 0.103 (0.021) & 0.105 (0.021)\tabularnewline
\hline 
\end{tabular}\smallskip{}
\par\end{centering}
Note: Numbers in parentheses are estimated standard errors, which
are constructed using sample variances of the estimated efficient
influence functions and adjusted for multiple imputation via Rubin's
\citeyearpar{rubin1987imputation} method.
\end{table}

The results are shown in Table \ref{tab:Estimates}. We can see that
the two estimators yield similar estimates of the total and path-specific
effects. By $\hat{\theta}_{\textup{\ensuremath{\overline{a}}}}^{\textup{np,eif}_{2}}$,
for example, the estimated total effect of college attendance on voting
is 0.152, meaning that, on average, college attendance increases the
likelihood of voting in 2010 by about 15 percentage points. The estimated
PSE via $M_{2}$ is 0.042, suggesting that a small fraction of the
college effect operates through the development of civic and political
interest. By contrast, the estimated PSE via economic status is substantively
negligible and statistically insignificant. A large portion of the
college effect appears to be ``direct,'' i.e., operating neither
through increased economic status nor through increased civic and
political interest.

\section{Concluding Remarks}

By considering the general case of $K$($\geq1$) causally ordered
mediators, this paper offers several new insights into the identification
and estimation of PSEs. First, under the assumptions associated with
Pearl's NPSEM with mutually independent errors, we have defined a
set of PSEs as contrasts between the expectations of $2^{K+1}$ potential
outcomes, which are identified via what we call the generalized mediation
functional (GMF). Second, building on its efficient influence function,
we have developed two $K+2$-robust and semiparametric efficient estimators
for the GMF. By virtue of their multiple robustness, these estimators
are well suited to the use of data-adaptive methods for estimating
their nuisance functions. For such cases, we have established the
rate conditions required of the nuisance functions for consistency
and semiparametric efficiency.

As we have seen, our proposed methodology is general in that the GMF
encompasses a variety of causal estimands such as the NDE, NIE/TIE,
nPSE, cPSE. Nonetheless, it does not accommodate PSEs that are not
identified under Pearl's NPSEM, some of which may be scientifically
important. For example, social and biomedical scientists are often
interested in testing hypotheses about ``serial mediation,'' i.e.,
the degree to which the effect of a treatment operates through multiple
mediators sequentially, such as that reflected in the causal path
$A\to M_{1}\to M_{2}\to Y$ (e.g., \citealt{jones2015health}). Given
that the corresponding PSEs are not nonparametrically identified under
Pearl's NPSEM, previous research has proposed strategies that involve
either additional assumptions (\citealt{albert2011generalized}) or
alternative estimands (\citealt{lin2017interventional}). We consider
semiparametric estimation and inference for these alternative approaches
a promising direction for future research.

\noindent \onehalfspacing

\noindent \bibliographystyle{rss}
\bibliography{causality_ref}

\clearpage{}

\appendix
\noindent \onehalfspacing

\part*{Supplementary Materials}

\section{Proof of Theorem 1}

Assumption 2{*} implies that for any $k\in\{2,\ldots K\}$ and any
$j\in\{1,\ldots k-1\}$, 
\begin{align}
 & \big(M_{k}(a_{k},\overline{m}_{k-1}),\ldots M_{K}(a_{K},\overline{m}_{K-1}),Y(a_{K+1},\overline{m}_{K})\big)\ci M_{k-j}(a_{k-j},\overline{m}_{k-j-1}^{*})|X,A,\overline{M}_{k-j-1}\nonumber \\
\Rightarrow & \big(M_{k}(a_{k},\overline{m}_{k-1}),\ldots M_{K}(a_{K},\overline{m}_{K-1}),Y(a_{K+1},\overline{m}_{K})\big)\ci M_{k-j}(a_{k-j},\overline{m}_{k-j-1}^{*})|X,A=a_{k-j},\overline{M}_{k-j-1}=\overline{m}_{k-j-1}^{*}\nonumber \\
\Rightarrow & \big(M_{k}(a_{k},\overline{m}_{k-1}),\ldots M_{K}(a_{K},\overline{m}_{K-1}),Y(a_{K+1},\overline{m}_{K})\big)\ci M_{k-j}|X,A=a_{k-j},\overline{M}_{k-j-1}=\overline{m}_{k-j-1}^{*}\nonumber \\
\Rightarrow & \big(M_{k}(a_{k},\overline{m}_{k-1}),\ldots M_{K}(a_{K},\overline{m}_{K-1}),Y(a_{K+1},\overline{m}_{K})\big)\ci M_{k-j}|X,A,\overline{M}_{k-j-1}.\label{eq:ct_indep_obs}
\end{align}
Setting $\overline{m}_{k-1}^{*}=\overline{m}_{k-1}$, Assumption 2{*}
also implies that for any $k\in[K]$,
\begin{equation}
\big(M_{k+1}(a_{k+1},\overline{m}_{k}),\ldots M_{K}(a_{K},\overline{m}_{K-1}),Y(a_{K+1},\overline{m}_{K})\big)\ci M_{k}(a_{k},\overline{m}_{k-1})|X,A,\overline{M}_{k-1}.\label{eq:ct_indep_ct}
\end{equation}
Now suppose that for some $j\in\{1,\ldots k-1\}$,
\begin{align}
\big(M_{k+1}(a_{k+1},\overline{m}_{k}),\ldots M_{K}(a_{K},\overline{m}_{K-1}),Y(a_{K+1},\overline{m}_{K})\big) & \ci M_{k}(a_{k},\overline{m}_{k-1})|X,A,\overline{M}_{k-j}.\label{eq:ct_indep_ct_j}
\end{align}
By the contraction rule of conditional independence, the relationships
\eqref{eq:ct_indep_obs} and \eqref{eq:ct_indep_ct_j} imply
\[
\big(M_{k+1}(a_{k+1},\overline{m}_{k}),\ldots M_{K}(a_{K},\overline{m}_{K-1}),Y(a_{K+1},\overline{m}_{K})\big)\ci M_{k}(a_{k},\overline{m}_{k-1})|X,A,\overline{M}_{k-j-1}.
\]
Hence, by the initial relationship (\ref{eq:ct_indep_ct}) and mathematical
induction, we have
\begin{equation}
\big(M_{k+1}(a_{k+1},\overline{m}_{k}),\ldots M_{K}(a_{K},\overline{m}_{K-1}),Y(a_{K+1},\overline{m}_{K})\big)\ci M_{k}(a_{k},\overline{m}_{k-1})|X,A,\quad\forall k\in[K].\label{eq:ct_indep_ct_given_A}
\end{equation}
In the meantime, because $\big(M_{k+1}(a_{k+1},\overline{m}_{k}),\ldots M_{K}(a_{K},\overline{m}_{K-1}),Y(a_{K+1},\overline{m}_{K})\big)\ci A|X$,
we have (by the contraction rule)
\[
\big(M_{k+1}(a_{k+1},\overline{m}_{k}),\ldots M_{K}(a_{K},\overline{m}_{K-1}),Y(a_{K+1},\overline{m}_{K})\big)\ci\big(A,M_{k}(a_{k},\overline{m}_{k-1})\big)|X,\quad\forall k\in[K].
\]
Thus the components in $\big(A,M_{1}(a_{1}),\ldots M_{K}(a_{K},\overline{m}_{K-1}),Y(a_{K+1},\overline{m}_{K})\big)$
are mutually independent given $X$. Therefore, 
\begin{align*}
\psi_{\overline{a}} & =\mathbb{E}[Y(a_{K+1},\overline{M}_{K}(\overline{a}_{K}))]\\
 & =\int_{x}\int_{\overline{m}_{K}}\mathbb{E}[Y(a_{K+1},\overline{m}_{K})|X=x,A=a_{K+1},M_{1}(a_{1})=m_{1},\ldots M_{K}(a_{K},\overline{m}_{K-1})=m_{K}]\\
 & \big(\prod_{k=1}^{K}dP_{M_{k}(a_{k},\overline{m}_{k-1})|X,A,M_{1}(a_{1}),\ldots M_{k-1}(a_{k-1},\overline{m}_{k-2})}(m_{k}|x,a_{K+1},\overline{m}_{k-1})\big)dP_{X}(x)\\
 & =\int_{x}\int_{\overline{m}_{K}}\mathbb{E}[Y(a_{K+1},\overline{m}_{K})|X=x,A=a_{K+1}]\big(\prod_{k=1}^{K}dP_{M_{k}(a_{k},\overline{m}_{k-1})|X}(m_{k}|x)\big)dP_{X}(x)\\
 & =\int_{x}\int_{\overline{m}_{K}}\mathbb{E}[Y(a_{K+1},\overline{m}_{K})|x,a_{K+1},\overline{M}_{K}=\overline{m}_{K}]\big(\prod_{k=1}^{K}dP_{M_{k}(a_{k},\overline{m}_{k-1})|X,A,\overline{M}_{k-1}}(m_{k}|x,a_{k},\overline{m}_{k-1})\big)dP_{X}(x)\\
 & =\int_{x}\int_{\overline{m}_{K}}\mathbb{E}[Y|x,a_{K+1},\overline{m}_{K}]\big(\prod_{k=1}^{K}dP_{M_{k}|X,A,\overline{M}_{k-1}}(m_{k}|x,a_{k},\overline{m}_{k-1})\big)dP_{X}(x).
\end{align*}

\section{Hybrid Estimators of $\theta_{a_{1},a_{2},a}$\label{sec:Hybrid-Estimators}}

For notational brevity, let us use the following shorthands:
\begin{align*}
\lambda_{0}^{j}(A|X) & \stackrel{\Delta}{=}\frac{\mathbb{I}(A=a_{j})}{p(a_{j}|X)}\\
\lambda_{1}^{j}(M_{1}|X) & \stackrel{\Delta}{=}\frac{p(M_{1}|X,a_{1})}{p(M_{1}|X,a_{j})}\\
\lambda_{2}^{j}(M_{2}|X,M_{1}) & \stackrel{\Delta}{=}\frac{p(M_{2}|X,a_{2},M_{1})}{p(M_{2}|X,a_{j},M_{1})},
\end{align*}
In addition, define $\lambda_{0}(A|X)=\mathbb{I}(A=a)/p(a|X)$, $\lambda_{1}(M_{1}|X)=p(M_{1}|X,a_{1})/p(M_{1}|X,a)$
and $\lambda_{2}(M_{2}|X,M_{1})=p(M_{2}|X,a_{2},M_{1})/p(M_{2}|X,a,M_{1})$.
With the above notation, the iterated conditional means $\mu_{1}(X,M_{1})$,
$\mu_{0}(X)$, and $\theta_{a_{1},a_{2},a}$ can each be written in
several different forms:
\begin{align*}
\mu_{1}(X,M_{1}) & =\begin{cases}
\mathbb{E}[\mu_{2}(X,M_{1},M_{2})|X,a_{2},M_{1}]\\
\mathbb{E}[\lambda_{2}(M_{2}|X,M_{1})Y|X,a,M_{1}]
\end{cases}\\
\mu_{0}(X) & =\begin{cases}
\mathbb{E}[\mu_{1}(X,M_{1})|X,a_{1}]=\begin{cases}
\mathbb{E}\big[\mathbb{E}[\mu_{2}(X,M_{1},M_{2})|X,a_{2},M_{1}]|X,a_{1}\big]\\
\mathbb{E}\big[\mathbb{E}[\lambda_{2}(M_{2}|X,M_{1})Y|X,a,M_{1}]|X,a_{1}\big]
\end{cases}\\
\mathbb{E}[\lambda_{1}^{2}(M_{1}|X)\mu_{2}(X,M_{1},M_{2})|X,a_{2}]\\
\mathbb{E}[\lambda_{1}(M_{1}|X)\lambda_{2}(X,M_{1},M_{2})Y|X,a]
\end{cases}\\
\theta_{a_{1},a_{2},a} & =\begin{cases}
\mathbb{E}[\mu_{0}(X)]=\begin{cases}
\mathbb{E}\Big[\mathbb{E}\big[\mathbb{E}[\mu_{2}(X,M_{1},M_{2})|X,a_{2},M_{1}]|X,a_{1}\big]\Big] & \textup{(RI-RI-RI)}\\
\mathbb{E}\Big[\mathbb{E}\big[\mathbb{E}[\lambda_{2}(M_{2}|X,M_{1})Y|X,a,M_{1}]|X,a_{1}\big]\Big] & \textup{(W-RI-RI)}\\
\mathbb{E}\Big[\mathbb{E}[\lambda_{1}^{2}(M_{1}|X)\mu_{2}(X,M_{1},M_{2})|X,a_{2}]\Big] & \textup{(RI-W-RI)}\\
\mathbb{E}\Big[\mathbb{E}[\lambda_{1}(M_{1}|X)\lambda_{2}(X,M_{1},M_{2})Y|X,a]\Big] & \textup{(W-W-RI)}
\end{cases}\\
\mathbb{E}[\lambda_{0}^{1}(A|X)\mu_{1}(X,M_{1})]=\begin{cases}
\mathbb{E}\big[\lambda_{0}^{1}(A|X)\mathbb{E}[\mu_{2}(X,M_{1},M_{2})|X,a_{2},M_{1}]\big] & \textup{(RI-RI-W)}\\
\mathbb{E}\big[\lambda_{0}^{1}(A|X)\mathbb{E}[\lambda_{2}(M_{2}|X,M_{1})Y|X,a,M_{1}]\big] & \textup{(W-RI-W)}
\end{cases}\\
\mathbb{E}[\lambda_{0}^{2}(A|X)\lambda_{1}^{2}(M_{1}|X)\mu_{2}(X,M_{1},M_{2})]\quad\;\;\textup{(RI-W-W)}\\
\mathbb{E}[\lambda_{0}(A|X)\lambda_{1}(M_{1}|X)\lambda_{2}(M_{2}|X,M_{1})Y]\quad\textup{(W-W-W)}
\end{cases}
\end{align*}
The first set of equations suggest two different ways of estimating
$\mu_{1}(x,m_{1})$: (a) fit a model for the conditional mean of $\hat{\mu}_{2}(X,M_{1},M_{2})$
given $X$, $A$, $M_{1}$ and then set $A=a_{2}$ for all units;
(b) fit a model for the conditional mean of $\hat{\lambda}_{2}(M_{2}|X,M_{1})Y$
given $X$, $A$, and $M_{1}$ and then set $A=a$ for all units.
Similarly, the second set of equations suggest four different ways
of estimating $\mu_{0}(x)$, and the last set of equations point to
eight different ways of estimating $\theta_{a_{1},a_{2},a}$. Each
of these eight estimators corresponds to a unique combination of regression-imputation
and weighting.

\section{Proof of Theorem 2}

To show that equation \eqref{eq:EIF} is the EIF of $\theta_{\overline{a}}$
in $\mathcal{P}_{\textup{np}}$, it suffices to show 
\begin{equation}
\frac{\partial\theta_{\overline{a}}(t)}{\partial t}\biggl\vert_{t=0}=\mathbb{E}[\varphi_{\overline{a}}(O)S_{0}(O)],\label{eq:proof}
\end{equation}
where $S_{0}(O)$ is the score function for any one-dimensional submodel
$P_{t}(O)$ evaluated at $t=0$. We first note that $S_{t}(O)$ can
be written as $S_{t}(O)=S_{t}(X)+S_{t}(A|X)+\sum_{k=1}^{K}S_{t}(M_{k}|X,A,\overline{M}_{k-1})+S_{t}(Y|X,A,\overline{M}_{K})$,
where $S_{t}(u|v)=\partial\log p_{t}(u|v)/\partial t$ and $p_{t}(u|v)$
is the conditional probability density/mass function of $U$ given
$V$. Using equation \eqref{eq:g-mformula} and the product rule,
the left-hand side of equation \eqref{eq:proof} can be written as
\begin{align*}
\frac{\partial\theta_{\overline{a}}(t)}{\partial t}\biggl\vert_{t=0} & =\frac{\partial\iiint ydP_{t}(y|x,a_{K+1},\overline{m}_{K})\big[\prod_{k=1}^{K}dP_{t}(m_{k}|x,a_{k},\overline{m}_{k-1})\big]dP_{t}(x)}{\partial t}\biggl\vert_{t=0}\\
 & =\underbrace{\iiint yS_{0}(x)dP_{0}(y|x,a_{K+1},\overline{m}_{K})\big[\prod_{k=1}^{K}dP_{0}(m_{k}|x,a_{k},\overline{m}_{k-1})\big]dP_{0}(x)}_{=:\textup{\ensuremath{\phi_{0}}}}\\
 & +\sum_{k=1}^{K}\underbrace{\iiint yS_{0}(m_{k}|x,a_{k},\overline{m}_{k-1})dP_{0}(y|x,a_{K+1},\overline{m}_{K})\big[\prod_{k=1}^{K}dP_{0}(m_{k}|x,a_{k},\overline{m}_{k-1})\big]dP_{0}(x)}_{=:\phi_{k}}\\
 & +\underbrace{\iiint yS_{0}(y|x,a_{K+1},\overline{m}_{K})dP_{0}(y|x,a_{K+1},\overline{m}_{K})\big[\prod_{k=1}^{K}dP_{0}(m_{k}|x,a_{k},\overline{m}_{k-1})\big]dP_{0}(x)}_{=:\phi_{K+1}}\\
 & =\sum_{k=0}^{K+1}\phi_{k}
\end{align*}
where the second equality follows from the fact that $\partial dP_{t}(u|v)/\partial t=S_{t}(u|v)dP_{t}(u|v).$
Below, we verify that $\phi_{k}=\mathbb{E}[\varphi_{k}(O)S_{0}(O)]$
for all $k\in\{0,\ldots K+1\}$, where $\varphi_{k}(O)$ is defined
in Theorem 2. First,
\begin{align*}
 & \mathbb{E}[\varphi_{0}(O)S_{0}(O)]\\
= & \mathbb{E}[\big(\mu_{0}(X)-\theta_{\overline{a}}\big)S_{0}(O)]\\
= & \mathbb{E}[\mu_{0}(X)S_{0}(O)]\\
= & \mathbb{E}\big[\mu_{0}(X)\big(S_{0}(X)+S_{0}(A|X)+\sum_{k=1}^{K}S_{0}(M_{k}|X,A,\overline{M}_{k-1})+S_{0}(Y|X,A,\overline{M}_{K})\big)\big]\\
= & \mathbb{E}\big[\mu_{0}(X)S_{0}(X)\big]+\mathbb{E}\Big[\mu_{0}(X)\underbrace{\mathbb{E}\big[S_{0}(A|X)|X\big]}_{=0}\Big]+\sum_{k=1}^{K}\mathbb{E}\Big[\mu_{0}(X)\underbrace{\mathbb{E}\big[S_{0}(M_{k}|X,A,\overline{M}_{k-1})|X,A,\overline{M}_{k-1}\big]}_{=0}\Big]\\
 & +\mathbb{E}\Big[\mu_{0}(X)\underbrace{\mathbb{E}\big[S_{0}(Y|X,A,\overline{M}_{K})|X,A,\overline{M}_{K}\big]}_{=0}\Big]\\
= & \int\mu_{0}(x)S_{0}(x)dP_{0}(x)\\
= & \iiint yS_{0}(x)dP_{0}(y|x,a_{K+1},\overline{m}_{K})\big[\prod_{k=1}^{K}dP_{0}(m_{k}|x,a_{k},\overline{m}_{k-1})\big]dP_{0}(x)\\
= & \phi_{0}.
\end{align*}
Second, for $k\in[K]$,
\begin{align*}
 & \mathbb{E}[\varphi_{k}(O)S_{0}(O)]\\
= & \mathbb{E}\big[\varphi_{k}(O)\big(S_{0}(X)+S_{0}(A|X)+\sum_{j=1}^{K}S_{0}(M_{j}|X,A,\overline{M}_{j-1})+S_{0}(Y|X,A,\overline{M}_{K})\big)\big]\\
= & \mathbb{E}\Big[\mathbb{E}\big[\varphi_{k}(O)\big(S_{0}(X)+S_{0}(A|X)+\sum_{j=1}^{k-1}S_{0}(M_{j}|X,A,\overline{M}_{j-1})\big)|X,A,\overline{M}_{k-1}\big]\Big]+\mathbb{E}\big[\varphi_{k}(O)S_{0}(M_{k}|X,A,\overline{M}_{k-1})\big)\big]\\
 & +\sum_{j=k+1}^{K}\mathbb{E}\Big[\varphi_{k}(O)\underbrace{\mathbb{E}\big[S_{0}(M_{j}|X,A,\overline{M}_{j-1})|X,A,\overline{M}_{j-1}\big]}_{=0}\Big]+\mathbb{E}\Big[\varphi_{k}(O)\underbrace{\mathbb{E}\big[S_{0}(Y|X,A,\overline{M}_{K})|X,A,\overline{M}_{K}\big]}_{=0}\Big]\\
= & \mathbb{E}\Big[\big(S_{0}(X)+S_{0}(A|X)+\sum_{j=1}^{k-1}S_{0}(M_{j}|X,A,\overline{M}_{j-1})\big)\underbrace{\mathbb{E}\big[\varphi_{k}(O)|X,A,\overline{M}_{k-1}\big]}_{=0}\Big]+\mathbb{E}\big[\varphi_{k}(O)S_{0}(M_{k}|X,A,\overline{M}_{k-1})\big)\big]\\
= & \mathbb{E}\big[\varphi_{k}(O)S_{0}(M_{k}|X,A,\overline{M}_{k-1})\big)\big]\\
= & \mathbb{E}\Big[\mathbb{E}\big[\frac{\mathbb{I}(A=a_{k})}{p(a_{k}|X)}\Big(\prod_{j=1}^{k-1}\frac{p(M_{j}|X,a_{j},\overline{M}_{j-1})}{p(M_{j}|X,a_{k},\overline{M}_{j-1})}\Big)\big(\mu_{k}(X,\overline{M}_{k})-\mu_{k-1}(X,\overline{M}_{k-1})\big)S_{0}(M_{k}|X,A,\overline{M}_{k-1})\big)\big|X,A,\overline{M}_{k-1}\big]\Big]\\
= & \mathbb{E}\big[\frac{\mathbb{I}(A=a_{k})}{p(a_{k}|X)}\Big(\prod_{j=1}^{k-1}\frac{p(M_{j}|X,a_{j},\overline{M}_{j-1})}{p(M_{j}|X,a_{k},\overline{M}_{j-1})}\Big)\mu_{k}(X,\overline{M}_{k})S_{0}(M_{k}|X,A,\overline{M}_{k-1})\big)\big]\\
= & \mathbb{E}_{X}\mathbb{E}\big[\Big(\prod_{j=1}^{k-1}\frac{p(M_{j}|X,a_{j},\overline{M}_{j-1})}{p(M_{j}|X,a_{k},\overline{M}_{j-1})}\Big)\mu_{k}(X,\overline{M}_{k})S_{0}(M_{k}|X,A,\overline{M}_{k-1})\big)|X,A=a_{k}\big]\\
= & \iiint S_{0}(m_{k}|x,a_{k},\overline{m}_{k-1})\Big(\int_{y}\int_{\overline{m}_{K}}ydP_{0}(y|x,a_{K+1},\overline{m}_{K})\prod_{j=k+1}^{K}dP_{0}(m_{j}|x,a_{j},\overline{m}_{j-1})\Big)\\
\cdot & dP_{0}(m_{k}|x,a_{k},\overline{m}_{k-1})\Big(\prod_{j=1}^{k-1}\frac{p(m_{j}|x,a_{j},\overline{m}_{j-1})}{p(m_{j}|x,a_{k},\overline{m}_{j-1})}\Big)\Big(\prod_{j=1}^{k-1}dP_{0}(m_{j}|x,a_{k},\overline{m}_{j-1})\Big)dP_{0}(x)\\
= & \iiint yS_{0}(m_{k}|x,a_{k},\overline{m}_{k-1})dP_{0}(y|x,a_{K+1},\overline{m}_{K})\big(\prod_{j=1}^{K}dP_{0}(m_{j}|x,a_{j},\overline{m}_{j-1})\big)dP_{0}(x)\\
= & \phi_{k},
\end{align*}
where the fourth equality is due to the fact that
\begin{align*}
 & \mathbb{E}\big[\varphi_{k}(O)|X,A,\overline{M}_{k-1}\big]\\
= & \mathbb{E}\big[\frac{\mathbb{I}(A=a_{k})}{p(a_{k}|X)}\Big(\prod_{j=1}^{k-1}\frac{p(M_{j}|X,a_{j},\overline{M}_{j-1})}{p(M_{j}|X,a_{k},\overline{M}_{j-1})}\Big)\big(\mu_{k}(X,\overline{M}_{k})-\mu_{k-1}(X,\overline{M}_{k-1})\big)|X,A,\overline{M}_{k-1}\big]\\
= & \mathbb{E}\big[\Big(\prod_{j=1}^{k-1}\frac{p(M_{j}|X,a_{j},\overline{M}_{j-1})}{p(M_{j}|X,a_{k},\overline{M}_{j-1})}\Big)\big(\mu_{k}(X,\overline{M}_{k})-\mu_{k-1}(X,\overline{M}_{k-1})\big)|X,A=a_{k},\overline{M}_{k-1}\big]\\
= & \Big(\prod_{j=1}^{k-1}\frac{p(M_{j}|X,a_{j},\overline{M}_{j-1})}{p(M_{j}|X,a_{k},\overline{M}_{j-1})}\Big)\underbrace{\mathbb{E}\big[\mu_{k}(X,\overline{M}_{k})-\mu_{k-1}(X,\overline{M}_{k-1})|X,A=a_{k},\overline{M}_{k-1}\big]}_{=0}\\
= & 0.
\end{align*}
Finally, 
\begin{align*}
 & \mathbb{E}[\varphi_{K+1}(O)S_{0}(O)]\\
= & \mathbb{E}\big[\varphi_{K+1}(O)\big(S_{0}(X)+S_{0}(A|X)+\sum_{j=1}^{K}S_{0}(M_{j}|X,A,\overline{M}_{j-1})+S_{0}(Y|X,A,\overline{M}_{K})\big)\big]\\
= & \mathbb{E}\Big[\mathbb{E}\big[\varphi_{K+1}(O)\big(S_{0}(X)+S_{0}(A|X)+\sum_{j=1}^{K}S_{0}(M_{j}|X,A,\overline{M}_{j-1})\big)|X,A,\overline{M}_{K}\big]\Big]+\mathbb{E}\big[\varphi_{K+1}(O)S_{0}(Y|X,A,\overline{M}_{K})\big)\big]\\
= & \mathbb{E}\Big[\big(S_{0}(X)+S_{0}(A|X)+\sum_{j=1}^{K}S_{0}(M_{j}|X,A,\overline{M}_{j-1})\big)\underbrace{\mathbb{E}\big[\varphi_{K+1}(O)|X,A,\overline{M}_{K}\big]}_{=0}\Big]+\mathbb{E}\big[\varphi_{K+1}(O)S_{0}(Y|X,A,\overline{M}_{K})\big)\big]\\
= & \mathbb{E}\big[\varphi_{K+1}(O)S_{0}(Y|X,A,\overline{M}_{K})\big)\big]\\
= & \mathbb{E}\Big[\mathbb{E}\big[\frac{\mathbb{I}(A=a_{K+1})}{p(a_{K+1}|X)}\Big(\prod_{j=1}^{K}\frac{p(M_{j}|X,a_{j},\overline{M}_{j-1})}{p(M_{j}|X,a_{K+1},\overline{M}_{j-1})}\Big)\big(Y-\mu_{K}(X,\overline{M}_{K})\big)S_{0}(Y|X,A,\overline{M}_{K})\big)\big|X,A,\overline{M}_{K}\big]\Big]\\
= & \mathbb{E}\Big[\mathbb{E}\big[\frac{\mathbb{I}(A=a_{K+1})}{p(a_{K+1}|X)}\Big(\prod_{j=1}^{K}\frac{p(M_{j}|X,a_{j},\overline{M}_{j-1})}{p(M_{j}|X,a_{K+1},\overline{M}_{j-1})}\Big)YS_{0}(Y|X,A,\overline{M}_{K})\big)\big|X,A,\overline{M}_{K}\big]\Big]\\
= & \mathbb{E}\Big[\mathbb{E}\big[\frac{\mathbb{I}(A=a_{K+1})}{p(a_{K+1}|X)}\Big(\prod_{j=1}^{K}\frac{p(M_{j}|X,a_{j},\overline{M}_{j-1})}{p(M_{j}|X,a_{K+1},\overline{M}_{j-1})}\Big)YS_{0}(Y|X,A,\overline{M}_{K})\big)\big|X,A\big]\Big]\\
= & \mathbb{E}_{X}\Big[\mathbb{E}\big[\Big(\prod_{j=1}^{K}\frac{p(M_{j}|X,a_{j},\overline{M}_{j-1})}{p(M_{j}|X,a_{K+1},\overline{M}_{j-1})}\Big)YS_{0}(Y|X,A,\overline{M}_{K})\big)\big|X,A=a_{K+1}\big]\Big]\\
= & \iiint yS_{0}(y|x,a_{K+1},\overline{m}_{K})dP_{0}(y|x,a_{K+1},\overline{m}_{K})\Big(\prod_{j=1}^{K}\frac{p(m_{j}|x,a_{j},\overline{m}_{j-1})}{p(m_{j}|x,a_{K+1},\overline{m}_{j-1})}\Big)\big(\prod_{j=1}^{K}dP_{0}(m_{j}|x,a_{K+1},\overline{m}_{j-1})\big)dP_{0}(x)\\
= & \iiint yS_{0}(y|x,a_{K+1},\overline{m}_{K})dP_{0}(y|x,a_{K+1},\overline{m}_{K})\big(\prod_{j=1}^{K}dP_{0}(m_{j}|x,a_{j},\overline{m}_{j-1})\big)dP_{0}(x)\\
= & \phi_{K+1},
\end{align*}
where the third equality is due to the fact that
\begin{align*}
 & \mathbb{E}\big[\varphi_{K+1}(O)|X,A,\overline{M}_{K}\big]\\
= & \mathbb{E}\big[\frac{\mathbb{I}(A=a_{K+1})}{p(a_{K+1}|X)}\Big(\prod_{j=1}^{K}\frac{p(M_{j}|X,a_{j},\overline{M}_{j-1})}{p(M_{j}|X,a_{K+1},\overline{M}_{j-1})}\Big)\big(Y-\mu_{K}(X,\overline{M}_{K})\big)|X,A,\overline{M}_{K}\big]\\
= & \frac{p(a_{K+1}|X,\overline{M}_{K})}{p(a_{K+1}|X)}\Big(\prod_{j=1}^{K}\frac{p(M_{j}|X,a_{j},\overline{M}_{j-1})}{p(M_{j}|X,a_{K+1},\overline{M}_{j-1})}\Big)\underbrace{\mathbb{E}\big[Y-\mu_{K}(X,\overline{M}_{K})|X,A=a_{K+1},\overline{M}_{K}\big]}_{=0}\\
= & 0.
\end{align*}
Since $\phi_{k}=\mathbb{E}[\varphi_{k}(O)S_{0}(O)]$ for all $k\in\{0,\ldots K+1\}$,
we have 
\begin{align*}
\frac{\partial\theta_{\overline{a}}(t)}{\partial t}\biggl\vert_{t=0} & =\sum_{k=0}^{K+1}\phi_{k}=\mathbb{E}\big[\big(\sum_{k=0}^{K+1}\varphi_{k}(O)\big)S_{0}(O)\big]=\mathbb{E}\big[\varphi_{\overline{a}}(O)S_{0}(O)\big].
\end{align*}

\section{Proof of Theorems 3 and 4}

\subsection{Parametric Estimation of Nuisance Parameters\label{subsec:D1_Parametric-Estimation}}

In this subsection, we prove the multiple robustness of $\hat{\theta}_{\overline{a}}^{\textup{eif}_{1}}$and
$\hat{\theta}_{\overline{a}}^{\textup{eif}_{2}}$ for the case where
parametric models are used to estimate the corresponding nuisance
functions. The local efficiency of these estimators is implied by
our proof in Section \ref{subsec:D2_Data-Adaptive-Estimation}, which
considers the case where data-adaptive methods and cross-fitting are
used to estimate the nuisance functions.

Let us start with $\hat{\theta}_{\overline{a}}^{\textup{eif}_{1}}=\text{\ensuremath{\mathbb{P}_{n}[m_{1}(O;\hat{\eta}_{1})]}}$,
where $m_{1}(O;\hat{\eta}_{1})$ denotes the quantity inside $\mathbb{P}_{n}[\cdot]$
in equation \eqref{eq:EIF1}, and $\hat{\eta}_{1}=(\hat{\pi}_{0},\hat{f}_{1},\ldots\hat{f}_{K},\hat{\mu}_{K})$.
In the meantime, let $\eta_{1}=(\pi_{0},f_{1},\ldots f_{K},\mu_{K})$
denote the truth and $\eta_{1}^{*}=(\pi_{0}^{*},f_{1}^{*},\ldots f_{K}^{*},\mu_{K}^{*})$
the probability limit of $\hat{\eta}_{1}$. A first-order Taylor expansion
of $\hat{\theta}_{\overline{a}}^{\textup{eif}_{1}}$ yields
\[
\hat{\theta}_{\overline{a}}^{\textup{eif}_{1}}=\mathbb{P}_{n}\big[m_{1}(O;\eta_{1}^{*})\big]+o_{p}(1).
\]
Hence it suffices to show $\mathbb{E}[m_{1}(O;\eta_{1}^{*})]=\theta_{\overline{a}}$
whenever all but one elements in $\eta_{1}^{*}$ equal the truth.
Consistency follows from the law of large numbers. By treating $\hat{\theta}_{\overline{a}}^{\textup{eif}_{1}}=\text{\ensuremath{\mathbb{P}_{n}[m_{1}(O;\hat{\eta}_{1})]}}$
as a two-stage M-estimator, asymptotic normality follows from standard
regularity conditions for estimating equations (e.g., \citealt[p. 2148]{newey1994large}).

First, if $\eta_{1}^{*}=(\pi_{0}^{*},f_{1},\ldots f_{K},\mu_{K})$,
the MLE of $\mu_{k}$ ($0\leq k\leq K-1$) will also be consistent.
Thus,
\begin{align*}
 & \mathbb{E}[m_{1}(O;\eta_{1}^{*})]\\
= & \mathbb{E}\big[\frac{\mathbb{I}(A=a_{K+1})}{\pi_{0}^{*}(a_{K+1}|X)}\Big(\prod_{j=1}^{K}\frac{f_{j}(M_{j}|X,a_{j},\overline{M}_{j-1})}{f_{j}(M_{j}|X,a_{K+1},\overline{M}_{j-1})}\Big)\big(Y-\mu_{K}(X,\overline{M}_{K})\big)\\
 & +\sum_{k=1}^{K}\frac{\mathbb{I}(A=a_{k})}{\pi_{0}^{*}(a_{k}|X)}\Big(\prod_{j=1}^{k-1}\frac{f_{j}(M_{j}|X,a_{j},\overline{M}_{j-1})}{f_{j}(M_{j}|X,a_{k},\overline{M}_{j-1})}\Big)\big(\mu_{k}(X,\overline{M}_{k})-\mu_{k-1}(X,\overline{M}_{k-1})\big)\\
 & +\mu_{0}(X)\big]\\
= & \mathbb{E}\Big[\frac{\pi_{0}(a_{K+1}|X,\overline{M}_{K})}{\pi_{0}^{*}(a_{K+1}|X)}\Big(\prod_{j=1}^{K}\frac{f_{j}(M_{j}|X,a_{j},\overline{M}_{j-1})}{f_{j}(M_{j}|X,a_{K+1},\overline{M}_{j-1})}\Big)\underbrace{\mathbb{E}\big[Y-\mu_{K}(X,\overline{M}_{K})\big|X,A=a_{K+1},\overline{M}_{K}\big]}_{=0}\\
 & +\sum_{k=1}^{K}\frac{\pi_{0}(a_{k}|X,\overline{M}_{k-1})}{\pi_{0}^{*}(a_{k}|X)}\Big(\prod_{j=1}^{k-1}\frac{f_{j}(M_{j}|X,a_{j},\overline{M}_{j-1})}{f_{j}(M_{j}|X,a_{k},\overline{M}_{j-1})}\Big)\underbrace{\mathbb{E}\big[\mu_{k}(X,\overline{M}_{k})-\mu_{k-1}(X,\overline{M}_{k-1})\big|X,A=a_{k},\overline{M}_{k-1}\big]}_{=0}\\
 & +\mu_{0}(X)\Big]\\
= & \mathbb{E}[\mu_{0}(X)]\\
= & \theta_{\overline{a}}.
\end{align*}
Second, if $\eta_{1}^{*}=(\pi_{0},f_{1},\ldots f_{k'-1},f_{k'}^{*},f_{k'+1},\ldots f_{K},\mu_{K})$,
the MLE of $\mu_{k}$ for any $k\geq k'$ will also be consistent.
Thus,
\begin{align*}
 & \mathbb{E}[m_{1}(O;\eta_{1}^{*})]\\
= & \mathbb{E}\big[\frac{\mathbb{I}(A=a_{K+1})}{\pi_{0}(a_{K+1}|X)}\Big(\prod_{j=1}^{K}\frac{f_{j}^{*}(M_{j}|X,a_{j},\overline{M}_{j-1})}{f_{j}^{*}(M_{j}|X,a_{K+1},\overline{M}_{j-1})}\Big)\big(Y-\mu_{K}(X,\overline{M}_{K})\big)\\
 & +\sum_{k=k'+1}^{K}\frac{\mathbb{I}(A=a_{k})}{\pi_{0}(a_{k}|X)}\Big(\prod_{j=1}^{k-1}\frac{f_{j}^{*}(M_{j}|X,a_{j},\overline{M}_{j-1})}{f_{j}^{*}(M_{j}|X,a_{k},\overline{M}_{j-1})}\Big)\big(\mu_{k}(X,\overline{M}_{k})-\mu_{k-1}(X,\overline{M}_{k-1})\big)\\
 & +\frac{\mathbb{I}(A=a_{k'})}{\pi_{0}(a_{k'}|X)}\Big(\prod_{j=1}^{k'-1}\frac{f_{j}(M_{j}|X,a_{j},\overline{M}_{j-1})}{f_{j}(M_{j}|X,a_{k'},\overline{M}_{j-1})}\Big)\big(\mu_{k'}(X,\overline{M}_{k'})-\mu_{k'-1}^{*}(X,\overline{M}_{k'-1})\big)\\
 & +\sum_{k=1}^{k'-1}\frac{\mathbb{I}(A=a_{k})}{\pi_{0}(a_{k}|X)}\Big(\prod_{j=1}^{k-1}\frac{f_{j}(M_{j}|X,a_{j},\overline{M}_{j-1})}{f_{j}(M_{j}|X,a_{k},\overline{M}_{j-1})}\Big)\big(\mu_{k}^{*}(X,\overline{M}_{k})-\mu_{k-1}^{*}(X,\overline{M}_{k-1})\big)\\
 & +\mu_{0}^{*}(X)\big]\\
= & \mathbb{E}\Big[\frac{\pi_{0}(a_{K+1}|X,\overline{M}_{K})}{\pi_{0}(a_{K+1}|X)}\Big(\prod_{j=1}^{K}\frac{f_{j}^{*}(M_{j}|X,a_{j},\overline{M}_{j-1})}{f_{j}^{*}(M_{j}|X,a_{K+1},\overline{M}_{j-1})}\Big)\underbrace{\mathbb{E}\big[Y-\mu_{K}(X,\overline{M}_{K})\big|X,A=a_{K+1},\overline{M}_{K}\big]}_{=0}\\
 & +\sum_{k=k'+1}^{K}\frac{\pi_{0}(a_{k}|X,\overline{M}_{k-1})}{\pi_{0}(a_{k}|X)}\Big(\prod_{j=1}^{k-1}\frac{f_{j}^{*}(M_{j}|X,a_{j},\overline{M}_{j-1})}{f_{j}^{*}(M_{j}|X,a_{k},\overline{M}_{j-1})}\Big)\underbrace{\mathbb{E}\big[\mu_{k}(X,\overline{M}_{k})-\mu_{k-1}(X,\overline{M}_{k-1})\big|X,A=a_{k},\overline{M}_{k-1}\big]}_{=0}\\
 & +\frac{\mathbb{I}(A=a_{k'})}{\pi_{0}(a_{k'}|X)}\Big(\prod_{j=1}^{k'-1}\frac{f_{j}(M_{j}|X,a_{j},\overline{M}_{j-1})}{f_{j}(M_{j}|X,a_{k'},\overline{M}_{j-1})}\Big)\mu_{k'}(X,\overline{M}_{k'})\\
 & +\sum_{k=1}^{k'-1}\mu_{k}^{*}(X,\overline{M}_{k})\mathbb{E}\big[\Big(\frac{\mathbb{I}(A=a_{k})}{\pi_{0}(a_{k}|X)}\prod_{j=1}^{k-1}\frac{f_{j}(M_{j}|X,a_{j},\overline{M}_{j-1})}{f_{j}(M_{j}|X,a_{k},\overline{M}_{j-1})}-\frac{\mathbb{I}(A=a_{k+1})}{\pi_{0}(a_{k+1}|X)}\prod_{j=1}^{k}\frac{f_{j}(M_{j}|X,a_{j},\overline{M}_{j-1})}{f_{j}(M_{j}|X,a_{k+1},\overline{M}_{j-1})}\Big)|X,\overline{M}_{k}\big]\\
 & +\mu_{0}^{*}(X)\underbrace{\mathbb{E}\big[1-\frac{\mathbb{I}(A=a_{1})}{\pi_{0}(a_{1}|X)}|X\big]}_{=0}\Big]\\
= & \mathbb{E}\Big[\frac{\mathbb{I}(A=a_{k'})}{\pi_{0}(a_{k'}|X)}\Big(\prod_{j=1}^{k'-1}\frac{f_{j}(M_{j}|X,a_{j},\overline{M}_{j-1})}{f_{j}(M_{j}|X,a_{k'},\overline{M}_{j-1})}\Big)\mu_{k'}(X,\overline{M}_{k'})\Big]\\
 & +\mathbb{E}\Big[\sum_{k=1}^{k'-1}\mu_{k}^{*}(X,\overline{M}_{k})\Big(\frac{\pi_{k}(a_{k}|X,\overline{M}_{k})}{\pi_{0}(a_{k}|X)}\prod_{j=1}^{k-1}\frac{f_{j}(M_{j}|X,a_{j},\overline{M}_{j-1})}{f_{j}(M_{j}|X,a_{k},\overline{M}_{j-1})}-\frac{\pi_{k}(a_{k+1}|X,\overline{M}_{k})}{\pi_{0}(a_{k+1}|X)}\prod_{j=1}^{k}\frac{f_{j}(M_{j}|X,a_{j},\overline{M}_{j-1})}{f_{j}(M_{j}|X,a_{k+1},\overline{M}_{j-1})}\Big)\Big]\\
= & \underbrace{\mathbb{E}\Big[\frac{\mathbb{I}(A=a_{k'})}{\pi_{0}(a_{k'}|X)}\Big(\prod_{j=1}^{k'-1}\frac{f_{j}(M_{j}|X,a_{j},\overline{M}_{j-1})}{f_{j}(M_{j}|X,a_{k'},\overline{M}_{j-1})}\Big)\mu_{k'}(X,\overline{M}_{k'})\Big]}_{=\theta_{\overline{a}}}\\
 & +\mathbb{E}\Big[\sum_{k=1}^{k'-1}\mu_{k}^{*}(X,\overline{M}_{k})\underbrace{\Big(\prod_{j=1}^{k}\frac{\pi_{j}(a_{j}|X,\overline{M}_{j})}{\pi_{j-1}(a_{j}|X,\overline{M}_{j-1})}-\prod_{j=1}^{k}\frac{\pi_{j}(a_{j}|X,\overline{M}_{j})}{\pi_{j-1}(a_{j}|X,\overline{M}_{j-1})}\Big)}_{=0}\Big]\\
= & \theta_{\overline{a}},
\end{align*}
where the penultimate equality is due to the fact that 
\begin{align*}
 & \frac{\pi_{k}(a_{k}|X,\overline{M}_{k})}{\pi_{0}(a_{k}|X)}\prod_{j=1}^{k-1}\frac{f_{j}(M_{j}|X,a_{j},\overline{M}_{j-1})}{f_{j}(M_{j}|X,a_{k},\overline{M}_{j-1})}\\
= & \frac{\pi_{k}(a_{k}|X,\overline{M}_{k})}{\pi_{0}(a_{k}|X)}\prod_{j=1}^{k-1}\Big(\frac{\pi_{j}(a_{j}|X,\overline{M}_{j})}{\pi_{j}(a_{k}|X,\overline{M}_{j})}\cdot\frac{\pi_{j-1}(a_{k}|X,\overline{M}_{j-1})}{\pi_{j-1}(a_{j}|X,\overline{M}_{j-1})}\Big)\\
= & \frac{\pi_{k}(a_{k}|X,\overline{M}_{k})}{\pi_{0}(a_{k}|X)}\prod_{j=1}^{k-1}\Big(\frac{\pi_{j-1}(a_{k}|X,\overline{M}_{j-1})}{\pi_{j}(a_{k}|X,\overline{M}_{j})}\Big)\prod_{j=1}^{k-1}\Big(\frac{\pi_{j}(a_{j}|X,\overline{M}_{j})}{\pi_{j-1}(a_{j}|X,\overline{M}_{j-1})}\Big)\\
= & \frac{\pi_{k}(a_{k}|X,\overline{M}_{k})}{\pi_{k-1}(a_{k}|X,\overline{M}_{k-1})}\prod_{j=1}^{k-1}\Big(\frac{\pi_{j}(a_{j}|X,\overline{M}_{j})}{\pi_{j-1}(a_{j}|X,\overline{M}_{j-1})}\Big)\\
= & \prod_{j=1}^{k}\frac{\pi_{j}(a_{j}|X,\overline{M}_{j})}{\pi_{j-1}(a_{j}|X,\overline{M}_{j-1})}
\end{align*}
and that 
\begin{align*}
 & \frac{\pi_{k}(a_{k+1}|X,\overline{M}_{k})}{\pi_{0}(a_{k+1}|X)}\prod_{j=1}^{k}\frac{f_{j}(M_{j}|X,a_{j},\overline{M}_{j-1})}{f_{j}(M_{j}|X,a_{k+1},\overline{M}_{j-1})}\\
= & \frac{\pi_{k}(a_{k+1}|X,\overline{M}_{k})}{\pi_{0}(a_{k+1}|X)}\prod_{j=1}^{k}\Big(\frac{\pi_{j}(a_{j}|X,\overline{M}_{j})}{\pi_{j}(a_{k+1}|X,\overline{M}_{j})}\cdot\frac{\pi_{j-1}(a_{k+1}|X,\overline{M}_{j-1})}{\pi_{j-1}(a_{j}|X,\overline{M}_{j-1})}\Big)\\
= & \frac{\pi_{k}(a_{k+1}|X,\overline{M}_{k})}{\pi_{0}(a_{k+1}|X)}\prod_{j=1}^{k}\Big(\frac{\pi_{j-1}(a_{k+1}|X,\overline{M}_{j-1})}{\pi_{j}(a_{k+1}|X,\overline{M}_{j})}\Big)\prod_{j=1}^{k}\Big(\frac{\pi_{j}(a_{j}|X,\overline{M}_{j})}{\pi_{j-1}(a_{j}|X,\overline{M}_{j-1})}\Big)\\
= & \prod_{j=1}^{k}\frac{\pi_{j}(a_{j}|X,\overline{M}_{j})}{\pi_{j-1}(a_{j}|X,\overline{M}_{j-1})}.
\end{align*}
Finally, if $\eta_{1}^{*}=(\pi_{0},f_{1},\ldots f_{K},\mu_{K}^{*})$,
we have
\begin{align*}
 & \mathbb{E}[m_{1}(O;\eta_{1}^{*})]\\
= & \mathbb{E}\big[\frac{\mathbb{I}(A=a_{K+1})}{\pi_{0}(a_{K+1}|X)}\Big(\prod_{j=1}^{K}\frac{f_{j}(M_{j}|X,a_{j},\overline{M}_{j-1})}{f_{j}(M_{j}|X,a_{K+1},\overline{M}_{j-1})}\Big)\big(Y-\mu_{K}^{*}(X,\overline{M}_{K})\big)\\
 & +\sum_{k=1}^{K}\frac{\mathbb{I}(A=a_{k})}{\pi_{0}(a_{k}|X)}\Big(\prod_{j=1}^{k}\frac{f_{j}(M_{j}|X,a_{j},\overline{M}_{j-1})}{f_{j}(M_{j}|X,a_{k},\overline{M}_{j-1})}\Big)\big(\mu_{k}^{*}(X,\overline{M}_{k})-\mu_{k-1}^{*}(X,\overline{M}_{k-1})\big)+\mu_{0}^{*}(X)\big]\\
= & \mathbb{E}\Big[\frac{\mathbb{I}(A=a_{K+1})}{\pi_{0}(a_{K+1}|X)}\Big(\prod_{j=1}^{K}\frac{f_{j}(M_{j}|X,a_{j},\overline{M}_{j-1})}{f_{j}(M_{j}|X,a_{K+1},\overline{M}_{j-1})}\Big)Y\\
 & +\sum_{k=1}^{K}\mu_{k}^{*}(X,\overline{M}_{k})\underbrace{\mathbb{E}\big[\Big(\frac{\mathbb{I}(A=a_{k})}{\pi_{0}(a_{k}|X)}\prod_{j=1}^{k-1}\frac{f_{j}(M_{j}|X,a_{j},\overline{M}_{j-1})}{f_{j}(M_{j}|X,a_{k},\overline{M}_{j-1})}-\frac{\mathbb{I}(A=a_{k+1})}{\pi_{0}(a_{k+1}|X)}\prod_{j=1}^{k}\frac{f_{j}(M_{j}|X,a_{j},\overline{M}_{j-1})}{f_{j}(M_{j}|X,a_{k+1},\overline{M}_{j-1})}\Big)|X,\overline{M}_{k}\big]}_{=0\quad(\textup{same as the previous case})}\\
 & +\mu_{0}^{*}(X)\underbrace{\mathbb{E}\big[1-\frac{\mathbb{I}(A=a_{1})}{\pi_{0}(a_{1}|X)}||X\big]}_{=0}\Big]\\
= & \mathbb{E}\Big[\frac{\mathbb{I}(A=a_{K+1})}{\pi_{0}(a_{K+1}|X)}\Big(\prod_{j=1}^{K}\frac{f_{j}(M_{j}|X,a_{j},\overline{M}_{j-1})}{f_{j}(M_{j}|X,a_{K+1},\overline{M}_{j-1})}\Big)Y\Big]\\
= & \theta_{\overline{a}}.
\end{align*}
Now consider $\hat{\theta}_{\overline{a}}^{\textup{eif}_{2}}=\text{\ensuremath{\mathbb{P}_{n}[m_{2}(O;\hat{\eta}_{2})]}}$,
where $m_{2}(O;\hat{\eta}_{2})$ denotes the quantity inside $\mathbb{P}_{n}[\cdot]$
in equation \eqref{eq:EIF2}, and $\hat{\eta}_{2}=(\hat{\pi}_{0},\ldots\hat{\pi}_{K},\hat{\mu}_{0},\ldots\hat{\mu}_{K})$.
In the meantime, let $\eta_{2}=(\pi_{0},\ldots\pi_{K},\mu_{0},\ldots\mu_{K})$
denote the truth and $\eta_{2}^{*}=(\pi_{0}^{*},\ldots\pi_{K}^{*},\mu_{0}^{*},\ldots\mu_{K}^{*})$
denote the probability limit of $\hat{\eta}_{2}$. A first-order Taylor
expansion of $\hat{\theta}_{\overline{a}}^{\textup{eif}_{2}}$ yields
\[
\hat{\theta}_{\overline{a}}^{\textup{eif}_{2}}=\mathbb{P}_{n}\big[m_{2}(O;\eta_{2}^{*})\big]+o_{p}(1).
\]
Hence it suffices to show $\mathbb{E}[m_{2}(O;\eta_{2}^{*})]=\theta_{\overline{a}}$
if 
\[
\eta_{2}^{*}=(\pi_{0},\ldots\pi_{k'-1},\pi_{k'}^{*},\ldots\pi_{K}^{*},\mu_{0}^{*},\ldots\mu_{k'-1}^{*},\mu_{k'},\ldots\mu_{K})
\]
for every $k'\in\{0,\ldots K+1\}$.

First, if $k'=0$, then all the outcome models are correctly specified,
which implies 
\begin{align*}
 & \mathbb{E}[m_{2}(O;\eta_{2}^{*})]\\
= & \mathbb{E}\big[\frac{\mathbb{I}(A=a_{K+1})}{\pi_{0}^{*}(a_{1}|X)}\Big(\prod_{j=1}^{K}\frac{\pi_{j}^{*}(a_{j}|X,\overline{M}_{j})}{\pi_{j}^{*}(a_{j+1}|X,\overline{M}_{j})}\Big)\big(Y-\mu_{K}(X,\overline{M}_{K})\big)\\
 & +\sum_{k=1}^{K}\frac{\mathbb{I}(A=a_{k})}{\pi_{0}^{*}(a_{1}|X)}\Big(\prod_{j=1}^{k-1}\frac{\pi_{j}^{*}(a_{j}|X,\overline{M}_{j})}{\pi_{j}^{*}(a_{j+1}|X,\overline{M}_{j})}\Big)\big(\mu_{k}(X,\overline{M}_{k})-\mu_{k-1}(X,\overline{M}_{k-1})\big)\\
 & +\mu_{0}(X)\big]\\
= & \mathbb{E}\Big[\frac{\mathbb{I}(A=a_{K+1})}{\pi_{0}^{*}(a_{1}|X)}\Big(\prod_{j=1}^{K}\frac{\pi_{j}^{*}(a_{j}|X,\overline{M}_{j})}{\pi_{j}^{*}(a_{j+1}|X,\overline{M}_{j})}\Big)\underbrace{\mathbb{E}\big[Y-\mu_{K}(X,\overline{M}_{K})\big|X,A=a_{K+1},\overline{M}_{K}\big]}_{=0}\\
 & +\sum_{k=1}^{K}\frac{\mathbb{I}(A=a_{k})}{\pi_{0}^{*}(a_{1}|X)}\Big(\prod_{j=1}^{k-1}\frac{\pi_{j}^{*}(a_{j}|X,\overline{M}_{j})}{\pi_{j}^{*}(a_{j+1}|X,\overline{M}_{j})}\Big)\underbrace{\mathbb{E}\big[\mu_{k}(X,\overline{M}_{k})-\mu_{k-1}(X,\overline{M}_{k-1})\big|X,A=a_{k},\overline{M}_{k-1}\big]}_{=0}\\
 & +\mu_{0}(X)\Big]\\
= & \mathbb{E}[\mu_{0}(X)]\\
= & \theta_{\overline{a}}.
\end{align*}
Second, if $k'\in\{1,\ldots K-1\}$, we have
\begin{align*}
 & \mathbb{E}[m_{2}(O;\eta_{2}^{*})]\\
= & \mathbb{E}\big[\frac{\mathbb{I}(A=a_{K+1})}{\pi_{0}^{*}(a_{1}|X)}\Big(\prod_{j=1}^{K}\frac{\pi_{j}^{*}(a_{j}|X,\overline{M}_{j})}{\pi_{j}^{*}(a_{j+1}|X,\overline{M}_{j})}\Big)\big(Y-\mu_{K}(X,\overline{M}_{K})\big)\\
 & +\sum_{k=k'+1}^{K}\frac{\mathbb{I}(A=a_{k})}{\pi_{0}^{*}(a_{1}|X)}\Big(\prod_{j=1}^{k-1}\frac{\pi_{j}^{*}(a_{j}|X,\overline{M}_{j})}{\pi_{j}^{*}(a_{j+1}|X,\overline{M}_{j})}\Big)\big(\mu_{k}(X,\overline{M}_{k})-\mu_{k-1}(X,\overline{M}_{k-1})\big)\\
 & +\frac{\mathbb{I}(A=a_{k'})}{\pi_{0}(a_{1}|X)}\Big(\prod_{j=1}^{k'-1}\frac{\pi_{j}(a_{j}|X,\overline{M}_{j})}{\pi_{j}(a_{j+1}|X,\overline{M}_{j})}\Big)\big(\mu_{k'}(X,\overline{M}_{k'})-\mu_{k'-1}^{*}(X,\overline{M}_{k'-1})\big)\\
 & +\sum_{k=1}^{k'-1}\frac{\mathbb{I}(A=a_{k})}{\pi_{0}(a_{1}|X)}\Big(\prod_{j=1}^{k-1}\frac{\pi_{j}(a_{j}|X,\overline{M}_{j})}{\pi_{j}(a_{j+1}|X,\overline{M}_{j})}\Big)\big(\mu_{k}^{*}(X,\overline{M}_{k})-\mu_{k-1}^{*}(X,\overline{M}_{k-1})\big)\\
 & +\mu_{0}^{*}(X)\big]\\
= & \mathbb{E}\Big[\frac{\mathbb{I}(A=a_{K+1})}{\pi_{0}^{*}(a_{1}|X)}\Big(\prod_{j=1}^{K}\frac{\pi_{j}^{*}(a_{j}|X,\overline{M}_{j})}{\pi_{j}^{*}(a_{j+1}|X,\overline{M}_{j})}\Big)\underbrace{\mathbb{E}\big[Y-\mu_{K}(X,\overline{M}_{K})\big|X,A=a_{K+1},\overline{M}_{K}\big]}_{=0}\\
 & +\sum_{k=k'+1}^{K}\frac{\mathbb{I}(A=a_{k})}{\pi_{0}^{*}(a_{1}|X)}\Big(\prod_{j=1}^{k-1}\frac{\pi_{j}^{*}(a_{j}|X,\overline{M}_{j})}{\pi_{j}^{*}(a_{j+1}|X,\overline{M}_{j})}\Big)\underbrace{\mathbb{E}\big[\mu_{k}(X,\overline{M}_{k})-\mu_{k-1}(X,\overline{M}_{k-1})\big|X,A=a_{k},\overline{M}_{k-1}\big]}_{=0}\\
 & +\frac{\mathbb{I}(A=a_{k'})}{\pi_{0}(a_{1}|X)}\Big(\prod_{j=1}^{k'-1}\frac{\pi_{j}(a_{j}|X,\overline{M}_{j})}{\pi_{j}(a_{j+1}|X,\overline{M}_{j})}\Big)\mu_{k'}(X,\overline{M}_{k'})\\
 & +\sum_{k=1}^{k'-1}\mu_{k}^{*}(X,\overline{M}_{k})\mathbb{E}\big[\frac{\mathbb{I}(A=a_{k})}{\pi_{0}(a_{1}|X)}\Big(\prod_{j=1}^{k-1}\frac{\pi_{j}(a_{j}|X,\overline{M}_{j})}{\pi_{j}(a_{j+1}|X,\overline{M}_{j})}\Big)-\frac{\mathbb{I}(A=a_{k+1})}{\pi_{0}(a_{1}|X)}\Big(\prod_{j=1}^{k}\frac{\pi_{j}(a_{j}|X,\overline{M}_{j})}{\pi_{j}(a_{j+1}|X,\overline{M}_{j})}\Big)|X,\overline{M}_{k}\big]\\
 & +\mu_{0}^{*}(X)\underbrace{\mathbb{E}\big[1-\frac{\mathbb{I}(A=a_{1})}{\pi_{0}(a_{1}|X)}|X\big]}_{=0}\Big]\\
= & \mathbb{E}\Big[\frac{\mathbb{I}(A=a_{k'})}{\pi_{0}(a_{1}|X)}\Big(\prod_{j=1}^{k'-1}\frac{\pi_{j}(a_{j}|X,\overline{M}_{j})}{\pi_{j}(a_{j+1}|X,\overline{M}_{j})}\Big)\mu_{k'}(X,\overline{M}_{k'})\Big]\\
 & +\mathbb{E}\Big[\sum_{k=1}^{k'-1}\mu_{k}^{*}(X,\overline{M}_{k})\Big(\frac{\pi_{k}(a_{k}|X,\overline{M}_{k})}{\pi_{0}(a_{1}|X)}\prod_{j=1}^{k-1}\frac{\pi_{j}(a_{j}|X,\overline{M}_{j})}{\pi_{j}(a_{j+1}|X,\overline{M}_{j})}-\frac{\pi_{k}(a_{k+1}|X,\overline{M}_{k})}{\pi_{0}(a_{1}|X)}\prod_{j=1}^{k}\frac{\pi_{j}(a_{j}|X,\overline{M}_{j})}{\pi_{j}(a_{j+1}|X,\overline{M}_{j})}\Big)\Big]\\
= & \underbrace{\mathbb{E}\Big[\frac{\mathbb{I}(A=a_{k'})}{\pi_{0}(a_{1}|X)}\Big(\prod_{j=1}^{k'-1}\frac{\pi_{j}(a_{j}|X,\overline{M}_{j})}{\pi_{j}(a_{j+1}|X,\overline{M}_{j})}\Big)\mu_{k'}(X,\overline{M}_{k'})\Big]}_{=\theta_{\overline{a}}}\\
 & +\mathbb{E}\Big[\sum_{k=1}^{k'-1}\mu_{k}^{*}(X,\overline{M}_{k})\underbrace{\Big(\prod_{j=1}^{k}\frac{\pi_{j}(a_{j}|X,\overline{M}_{j})}{\pi_{j-1}(a_{j}|X,\overline{M}_{j-1})}-\prod_{j=1}^{k}\frac{\pi_{j}(a_{j}|X,\overline{M}_{j})}{\pi_{j-1}(a_{j}|X,\overline{M}_{j-1})}\Big)}_{=0}\Big]\\
= & \theta_{\overline{a}}.
\end{align*}
Finally, if $k'=K$, we have
\begin{align*}
 & \mathbb{E}[m_{2}(O;\eta_{2}^{*})]\\
= & \mathbb{E}\big[\frac{\mathbb{I}(A=a_{K+1})}{\pi_{0}(a_{1}|X)}\Big(\prod_{j=1}^{K}\frac{\pi_{j}(a_{j}|X,\overline{M}_{j})}{\pi_{j}(a_{j+1}|X,\overline{M}_{j})}\Big)\big(Y-\mu_{K}^{*}(X,\overline{M}_{K})\big)\\
 & +\sum_{k=1}^{K}\frac{\mathbb{I}(A=a_{k})}{\pi_{0}(a_{1}|X)}\Big(\prod_{j=1}^{k-1}\frac{\pi_{j}(a_{j}|X,\overline{M}_{j})}{\pi_{j}(a_{j+1}|X,\overline{M}_{j})}\Big)\big(\mu_{k}^{*}(X,\overline{M}_{k})-\mu_{k-1}^{*}(X,\overline{M}_{k-1})\big)+\mu_{0}^{*}(X)\big]\\
= & \mathbb{E}\Big[\frac{\mathbb{I}(A=a_{K+1})}{\pi_{0}(a_{1}|X)}\Big(\prod_{j=1}^{K}\frac{\pi_{j}(a_{j}|X,\overline{M}_{j})}{\pi_{j}(a_{j+1}|X,\overline{M}_{j})}\Big)Y\\
 & +\sum_{k=1}^{K}\mu_{k}^{*}(X,\overline{M}_{k})\underbrace{\mathbb{E}\big[\Big(\frac{\mathbb{I}(A=a_{k})}{\pi_{0}(a_{1}|X)}\prod_{j=1}^{k-1}\frac{\pi_{j}(a_{j}|X,\overline{M}_{j})}{\pi_{j}(a_{j+1}|X,\overline{M}_{j})}-\frac{\mathbb{I}(A=a_{k+1})}{\pi_{0}(a_{1}|X)}\Big(\prod_{j=1}^{k+1}\frac{\pi_{j}(a_{j}|X,\overline{M}_{j})}{\pi_{j}(a_{j+1}|X,\overline{M}_{j})}\Big)|X,\overline{M}_{k}\big]}_{=0\quad(\textup{same as the previous case})}\\
 & +\mu_{0}^{*}(X)\underbrace{\mathbb{E}\big[1-\frac{\mathbb{I}(A=a_{1})}{\pi_{0}(a_{1}|X)}||X\big]}_{=0}\Big]\\
= & \mathbb{E}\Big[\frac{\mathbb{I}(A=a_{K+1})}{\pi_{0}(a_{1}|X)}\Big(\prod_{j=1}^{K}\frac{\pi_{j}(a_{j}|X,\overline{M}_{j})}{\pi_{j}(a_{j+1}|X,\overline{M}_{j})}\Big)Y\Big]\\
= & \theta_{\overline{a}}.
\end{align*}

\subsection{Data-Adaptive Estimation of Nuisance Parameters\label{subsec:D2_Data-Adaptive-Estimation}}

Let us start with $\hat{\theta}_{\overline{a}}^{\textup{eif}_{2}}=\text{\ensuremath{\mathbb{P}_{n}[m_{2}(O;\hat{\eta}_{2})]}}$.
Let $\tilde{\eta}_{2}=(\hat{\pi}_{0},\ldots\hat{\pi}_{K},\mu_{0},\ldots\mu_{K})$
denote a combination of estimated treatment models $\hat{\pi}_{j}$
and true outcome models $\mu_{j}$ ($0\leq j\leq K+1$), and let $Pg=\int gdP$
denote the expectation of a function $g$ of observed data $O$ at
the true model $P$. As before, denote by $\eta_{2}^{*}$ the probability
limit of $\hat{\eta}_{2}$. $\hat{\theta}_{\overline{a}}^{\textup{eif}_{2}}$
can now be written as
\begin{align}
 & \hat{\theta}_{\overline{a}}^{\textup{eif}_{2}}-\theta_{\overline{a}}\nonumber \\
= & \mathbb{P}_{n}[m_{2}(O;\hat{\eta}_{2})]-P[m_{2}(O;\eta_{2})]\nonumber \\
= & (\mathbb{P}_{n}-P)m_{2}(O;\eta_{2}^{*})+P[m_{2}(O;\hat{\eta}_{2})-m_{2}(O;\eta_{2})]+(\mathbb{P}_{n}-P)[m_{2}(O;\hat{\eta}_{2})-m_{2}(O;\eta_{2}^{*})]\\
= & (\mathbb{P}_{n}-P)\underbrace{[m_{2}(O;\eta_{2}^{*})-\theta_{\overline{a}}]}_{\stackrel{\Delta}{=}\varphi_{\overline{a}}(O;\eta_{2}^{*})}+P[m_{2}(O;\hat{\eta}_{2})-m_{2}(O;\eta_{2})]+(\mathbb{P}_{n}-P)[m_{2}(O;\hat{\eta}_{2})-m_{2}(O;\eta_{2}^{*})]\\
= & \mathbb{P}_{n}\varphi_{\overline{a}}(O;\eta_{2}^{*})-P\varphi_{\overline{a}}(O;\eta_{2}^{*})+\underbrace{P[m_{2}(O;\hat{\eta}_{2})-m_{2}(O;\eta_{2})]}_{\stackrel{\Delta}{=}R_{2}(\hat{\eta}_{2})}+(\mathbb{P}_{n}-P)[m_{2}(O;\hat{\eta}_{2})-m_{2}(O;\eta_{2}^{*})]\label{eq:expansion}
\end{align}
In equation \eqref{eq:expansion}, the last term is an empirical process
term that will be $o_{p}(n^{-1/2})$ either when parametric models
are used to estimate the nuisance functions or when cross-fitting
is used to induce independence between $\hat{\eta}_{2}$ and $O$
(\citealt{chernozhukov2018double}). Thus it remains to analyze the
first three terms: $\mathbb{P}_{n}\varphi_{\overline{a}}(O;\eta_{2}^{*})$,
$P\varphi_{\overline{a}}(O;\eta_{2}^{*})$, and $R_{2}(\hat{\eta}_{2})=P[m_{2}(O;\hat{\eta}_{2})-m_{2}(O;\eta_{2})]$.

First, from our proofs in Section \ref{subsec:D1_Parametric-Estimation},
we know that when $\eta_{2}^{*}=(\pi_{0},\ldots\pi_{k'-1},\pi_{k'}^{*},\ldots\pi_{K}^{*},\mu_{0}^{*},\ldots\mu_{k'-1}^{*},\mu_{k'},\ldots\mu_{K})$
for some $k'$, i.e., when the first $k'$ treatment models and the
last $K-k'+1$ outcome models are consistently estimated, $P\varphi_{\overline{a}}(O;\eta_{2}^{*})=0$.
Because in this case, $\mathbb{P}_{n}\varphi_{\overline{a}}(O;\eta_{2}^{*})\stackrel{p}{\to}P\varphi_{\overline{a}}(O;\eta_{2}^{*})=0$
by the law of large numbers, it suffices to show $R_{2}(\hat{\eta}_{2})=o_{p}(1)$
to establish the consistency of $\hat{\theta}_{\overline{a}}^{\textup{eif}_{2}}$.
Second, in the case where $\eta_{2}^{*}=\eta_{2}$, i.e., when all
of the $2(K+1)$ nuisance functions are consistently estimated, the
first two terms in equation \eqref{eq:expansion} reduces to $\mathbb{P}_{n}\varphi_{\overline{a}}(O;\eta_{2})$,
i.e., the sample average of the efficient influence function, which
has an asymptotic variance of $\mathbb{E}[\big(\varphi_{\overline{a}}(O)\big)^{2}]$.
Thus, in this case, $\hat{\theta}_{\overline{a}}^{\textup{eif}_{2}}$
will be asymptotically normal and semiparametric efficient as long
as $R_{2}(\hat{\eta}_{2})=o_{p}(n^{-1/2}).$

To analyze $R_{2}(\hat{\eta}_{2})$, we first observe that 
\begin{align*}
P[m_{2}(O;\tilde{\eta}_{2})] & =P\big[\frac{\mathbb{I}(A=a_{K+1})}{\hat{\pi}_{0}(a_{1}|X)}\Big(\prod_{j=1}^{K}\frac{\hat{\pi}_{j}(a_{j}|X,\overline{M}_{j})}{\hat{\pi}_{j}(a_{j+1}|X,\overline{M}_{j})}\Big)\big(Y-\mu_{K}(X,\overline{M}_{K})\big)\\
 & +\sum_{k=1}^{K}\frac{\mathbb{I}(A=a_{k})}{\hat{\pi}_{0}(a_{1}|X)}\Big(\prod_{j=1}^{k-1}\frac{\hat{\pi}_{j}(a_{j}|X,\overline{M}_{j})}{\hat{\pi}_{j}(a_{j+1}|X,\overline{M}_{j})}\Big)\big(\mu_{k}(X,\overline{M}_{k})-\mu_{k-1}(X,\overline{M}_{k-1})\big)+\mu_{0}(X)\big]\\
 & =P\Big[\frac{\pi_{0}(a_{K+1}|X,\overline{M}_{K})}{\hat{\pi}_{0}(a_{1}|X)}\Big(\prod_{j=1}^{K}\frac{\hat{\pi}_{j}(a_{j}|X,\overline{M}_{j})}{\hat{\pi}_{j}(a_{j+1}|X,\overline{M}_{j})}\Big)\underbrace{\mathbb{E}\big[Y-\mu_{K}(X,\overline{M}_{K})\big|X,A=a_{K+1},\overline{M}_{K}\big]}_{=0}\\
 & +\sum_{k=1}^{K}\frac{\pi_{0}(a_{k}|X,\overline{M}_{k-1})}{\hat{\pi}_{0}(a_{1}|X)}\Big(\prod_{j=1}^{k-1}\frac{\hat{\pi}_{j}(a_{j}|X,\overline{M}_{j})}{\hat{\pi}_{j}^ {}(a_{j+1}|X,\overline{M}_{j})}\Big)\underbrace{\mathbb{E}\big[\mu_{k}(X,\overline{M}_{k})-\mu_{k-1}(X,\overline{M}_{k-1})\big|X,A=a_{k},\overline{M}_{k-1}\big]}_{=0}\\
 & +\mu_{0}(X)\Big]\\
 & =\text{\ensuremath{P[\mu_{0}(X)]}}\\
 & =P[m_{2}(O;\eta_{2})].
\end{align*}
Then, by substituting $m_{2}(O;\tilde{\eta}_{2})$ for $m_{2}(O;\eta_{2})$
in $R_{2}(\hat{\eta}_{2})$, rearranging terms, and applying the Cauchy-Schwartz
inequality, we obtain
\begin{align}
R_{2}(\hat{\eta}_{2}) & =P[m_{2}(O;\hat{\eta}_{2})-m_{2}(O;\tilde{\eta}_{2})]\nonumber \\
 & =P\Big[\frac{\big(\hat{\pi}_{0}(a_{1}|X)-\pi_{0}(a_{1}|X)\big)\big(\hat{\mu}_{0}(X)-\mu_{0}(X)\big)}{\hat{\pi}_{0}(a_{1}|X)}\Big]\nonumber \\
 & +\sum_{k=1}^{K}P\Big[\Big(\prod_{j=1}^{k}\frac{\hat{\pi}_{j}(a_{j}|X,\overline{M}_{j})}{\hat{\pi}_{j}(a_{j+1}|X,\overline{M}_{j})}\Big)\frac{\big(\hat{\pi}_{k}(a_{k+1}|X,\overline{M}_{k})-\pi_{k}(a_{k+1}|X,\overline{M}_{k})\big)\big(\hat{\mu}_{k}(X,\overline{M}_{k})-\mu_{k}(X,\overline{M}_{k})\big)}{\hat{\pi}_{0}(a_{1}|X)}\Big]\nonumber \\
 & -\sum_{k=1}^{K}P\Big[\Big(\prod_{j=1}^{k-1}\frac{\hat{\pi}_{j}(a_{j}|X,\overline{M}_{j})}{\hat{\pi}_{j}(a_{j+1}|X,\overline{M}_{j})}\Big)\frac{\big(\hat{\pi}_{k}(a_{k}|X,\overline{M}_{k})-\pi_{k}(a_{k}|X,\overline{M}_{k})\big)\big(\hat{\mu}_{k}(X,\overline{M}_{k})-\mu_{k}(X,\overline{M}_{k})\big)}{\hat{\pi}_{0}(a_{1}|X)}\Big]\nonumber \\
 & =\sum_{k=0}^{K}O_{p}\big(\Vert\hat{\pi}_{k}(a_{k+1}|X,\overline{M}_{k})-\pi_{k}(a_{k+1}|X,\overline{M}_{k})\Vert\cdot\Vert\hat{\mu}_{k}(X,\overline{M}_{k})-\mu_{k}(X,\overline{M}_{k})\Vert\big)\nonumber \\
 & +\sum_{k=1}^{K}O_{p}\big(\Vert\hat{\pi}_{k}(a_{k}|X,\overline{M}_{k})-\pi_{k}(a_{k}|X,\overline{M}_{k})\Vert\cdot\Vert\hat{\mu}_{k}(X,\overline{M}_{k})-\mu_{k}(X,\overline{M}_{k})\Vert\big)\label{eq:R2}
\end{align}
where $\Vert g\Vert=(\int g^{T}gdP)^{1/2}$. The last equality uses
the positivity assumption that $\hat{\pi}_{k}(a|X,\overline{M}_{k})$
is bounded away from zero for all $k$ and $a$. Thus, assuming that
the empirical process term is on the order of $o_{p}(n^{-1/2})$ (e.g.,
via cross-fitting), we can write equation \eqref{eq:expansion} as
\[
\hat{\theta}_{\overline{a}}^{\textup{eif}_{2}}-\theta_{\overline{a}}=\mathbb{P}_{n}\varphi_{\overline{a}}(O;\eta_{2}^{*})-P\varphi_{\overline{a}}(O;\eta_{2}^{*})+\sum_{k=0}^{K}O_{p}(\Vert\hat{\pi}_{k}-\pi_{k}\Vert)\cdot O_{p}(\Vert\hat{\mu}_{k}-\mu_{k}\Vert)+o_{p}(n^{-1/2}),
\]
where $\pi_{k}=(\pi_{k}(0|X,\overline{M}_{k}),\,\pi_{k}(1|X,\overline{M}_{k}))^{T}$.
Clearly, when there exists a $k'$ such that the first $k'$ treatment
models and the last $K-k'+1$ outcome models are consistently estimated,
$\sum_{k=0}^{K}O_{p}(\Vert\hat{\pi}_{k}-\pi_{k}\Vert)\cdot O_{p}(\Vert\hat{\mu}_{k}-\mu_{k}\Vert)=o_{p}(1)$.
In this case, since $\mathbb{P}_{n}\varphi_{\overline{a}}(O;\eta_{2}^{*})-P\varphi_{\overline{a}}(O;\eta_{2}^{*})=\mathbb{P}_{n}\varphi_{\overline{a}}(O;\eta_{2}^{*})=o_{p}(1)$,
$\hat{\theta}_{\overline{a}}^{\textup{eif}_{2}}$ is consistent. When
$\eta_{2}^{*}=\eta_{2}$ and $\sum_{k=0}^{K}O_{p}(\Vert\hat{\pi}_{k}-\pi_{k}\Vert)\cdot O_{p}(\Vert\hat{\mu}_{k}-\mu_{k}\Vert)=o_{p}(n^{-1/2})$,
we have $\hat{\theta}_{\overline{a}}^{\textup{eif}_{2}}-\theta_{\overline{a}}=\mathbb{P}_{n}\varphi_{\overline{a}}(O;\eta_{2})+o_{p}(n^{-1/2})$,
implying that $\hat{\theta}_{\overline{a}}^{\textup{eif}_{2}}$ is
CAN and semiparametric efficient. If the nuisance functions are estimated
via parametric models and their parameter estimates are all $\sqrt{n}$-consistent,
$\sum_{k=0}^{K}O_{p}(\Vert\hat{\pi}_{k}-\pi_{k}\Vert)\cdot O_{p}(\Vert\hat{\mu}_{k}-\mu_{k}\Vert)=\sum_{k=0}^{K}O_{p}(n^{-1/2})\cdot O_{p}(n^{-1/2})=o_{p}(n^{-1/2})$,
hence the second part of Theorem 3.

Now let us consider $\hat{\theta}_{\overline{a}}^{\textup{eif}_{1}}$.
In a similar vein, we can write $\hat{\theta}_{\overline{a}}^{\textup{eif}_{1}}-\theta_{\overline{a}}$
as 
\[
\hat{\theta}_{\overline{a}}^{\textup{eif}_{1}}-\theta_{\overline{a}}=\mathbb{P}_{n}\varphi_{\overline{a}}(O;\eta_{1}^{*})-P\varphi_{\overline{a}}(O;\eta_{1}^{*})+\underbrace{\sum_{k=0}^{K}O_{p}(\Vert\check{\pi}_{k}-\pi_{k}\Vert)\cdot O_{p}(\Vert\check{\mu}_{k}-\mu_{k}\Vert)}_{\stackrel{\Delta}{=}R_{2}(\hat{\eta}_{1})}+o_{p}(n^{-1/2}),
\]
where $\check{\pi}_{k}$ and $\check{\mu}_{k}$ are estimates of $\pi_{k}$
and $\mu_{k}$ constructed from $\hat{\eta}_{1}=\{\hat{\pi}_{0},\hat{f}_{1},\ldots\hat{f}_{K},\hat{\mu}_{K}\}$.
First, from our proofs in Section \ref{subsec:D1_Parametric-Estimation},
we know that when $K+1$ of the $K+2$ nuisance functions in $\eta_{1}$
are consistently estimated, $P\varphi_{\overline{a}}(O;\eta_{1}^{*})=0$.
Since in this case $\mathbb{P}_{n}\varphi_{\overline{a}}(O;\eta_{1}^{*})\stackrel{p}{\to}P\varphi_{\overline{a}}(O;\eta_{1}^{*})=0$,
it suffices to show $R_{2}(\hat{\eta}_{1})=o_{p}(1)$ to establish
the consistency of $\hat{\theta}_{\overline{a}}^{\textup{eif}_{1}}$.
Second, in the case where $\eta_{1}^{*}=\eta_{1}$, i.e., when all
of the $K+2$ nuisance functions are consistently estimated, the first
two terms in equation \eqref{eq:expansion} reduces to $\mathbb{P}_{n}\varphi_{\overline{a}}(O;\eta_{1})$,
i.e., the sample average of the efficient influence function, which
has an asymptotic variance of $\mathbb{E}[\big(\varphi_{\overline{a}}(O)\big)^{2}]$.
Thus, in this case, $\hat{\theta}_{\overline{a}}^{\textup{eif}_{1}}$
will be asymptotically normal and semiparametric efficient as long
as $R_{2}(\hat{\eta}_{1})=o_{p}(n^{-1/2}).$

We first note that for any $a$, $\check{\pi}_{k}(a|X,\overline{M}_{k})-\pi_{k}(a|X,\overline{M}_{k})$
can be decomposed as 
\begin{align*}
 & \check{\pi}_{k}(a|X,\overline{M}_{k})-\pi_{k}(a|X,\overline{M}_{k})\\
= & \frac{\hat{p}(\overline{M}_{k}|X,a)\hat{\pi}_{0}(a|X)}{\sum_{a'}\hat{p}(\overline{M}_{k}|X,a')\hat{\pi}_{0}(a'|X)}-\frac{p(\overline{M}_{k}|X,a)\pi_{0}(a|X)}{\sum_{a'}p(\overline{M}_{k}|X,a')\pi_{0}(a'|X)}\\
= & \underbrace{\frac{\hat{p}(\overline{M}_{k}|X,a)\big(\hat{\pi}_{0}(a|X)-\pi_{0}(a|X)\big)}{\sum_{a'}\hat{p}(\overline{M}_{k}|X,a')\hat{\pi}_{0}(a'|X)}}_{\stackrel{\Delta}{=}\Delta_{\pi}^{1}}+\underbrace{\frac{\big(\hat{p}(\overline{M}_{k}|X,a)-p(\overline{M}_{k}|X,a)\big)\pi_{0}(a|X)}{\sum_{a'}\hat{p}(\overline{M}_{k}|X,a')\hat{\pi}_{0}(a'|X)}}_{\stackrel{\Delta}{=}\Delta_{\pi}^{2}}+\\
 & \underbrace{\frac{p(\overline{M}_{k}|X,a)\pi_{0}(a|X)\sum_{a'}\big(p(\overline{M}_{k}|X,a')\pi_{0}(a'|X)-\hat{p}(\overline{M}_{k}|X,a')\hat{\pi}_{0}(a'|X)\big)}{\sum_{a'}\hat{p}(\overline{M}_{k}|X,a')\hat{\pi}_{0}(a'|X)\sum_{a'}p(\overline{M}_{k}|X,a')\pi_{0}(a'|X)}}_{\stackrel{\Delta}{=}\Delta_{\pi}^{3}}.
\end{align*}
By the positivity assumption, we have $\Vert\Delta_{\pi}^{1}\Vert=O_{p}(\Vert\hat{\pi}_{0}-\pi_{0}\Vert)$.
Using the factorization $p(\overline{M}_{k}|X,a)=\prod_{j=1}^{k}p(M_{j}|X,a,\overline{M}_{j-1})$,
$\Vert\Delta_{\pi}^{2}\Vert$ can be expressed as 
\begin{align*}
\Vert\Delta_{\pi}^{2}\Vert & =\Big\Vert\frac{\pi_{0}(a|X)\big(\prod_{j=1}^{k}\hat{f}_{j}(M_{j}|X,a,\overline{M}_{j-1})-\prod_{j=1}^{k}f_{j}(M_{j}|X,a,\overline{M}_{j-1})\big)}{\sum_{a'}\hat{p}(\overline{M}_{k}|X,a')\hat{\pi}_{0}(a'|X)}\Big\Vert\\
 & =\Big\Vert\frac{\pi_{0}(a|X)}{\sum_{a'}\hat{p}(\overline{M}_{k}|X,a')\hat{\pi}_{0}(a'|X)}\cdot\\
 & \sum_{l=1}^{k}\big(\prod_{j=1}^{l-1}\hat{f}_{j}(M_{j}|X,a,\overline{M}_{j-1})\prod_{j=l+1}^{k}f_{j}(M_{j}|X,a,\overline{M}_{j-1})\big)\big(\hat{f}_{l}(M_{l}|X,a,\overline{M}_{l-1})-f_{l}(M_{l}|X,a,\overline{M}_{l-1})\big)\Big\Vert\\
 & =\sum_{l=1}^{k}O_{p}(\Vert\hat{f}_{l}-f_{l}\Vert),
\end{align*}
where $f_{l}=(f_{l}(M_{l}|X,0,\overline{M}_{l-1}),\,f_{l}(M_{l}|X,1,\overline{M}_{l-1}))^{T}$.
By a similar logic, $\Vert\Delta_{\pi}^{3}\Vert$ can be written as
\[
\Vert\Delta_{\pi}^{3}\Vert=O_{p}(\Vert\hat{\pi}_{0}-\pi_{0}\Vert)+\sum_{l=1}^{k}O_{p}(\Vert\hat{f}_{l}-f_{l}\Vert).
\]
 In sum, we have
\begin{align}
\Vert\check{\pi}_{k}-\pi_{k}\Vert & =O_{p}(\Vert\hat{\pi}_{0}-\pi_{0}\Vert)+\sum_{l=1}^{k}O_{p}(\Vert\hat{f}_{l}-f_{l}\Vert).\label{eq:pi_bounds}
\end{align}
Now consider $\Vert\check{\mu}_{k}-\mu_{k}\Vert$. Using the fact
that
\begin{align*}
\mu_{k}(x,\overline{m}_{k}) & =\int\mu_{K}(x,\overline{m}_{k})\big(\prod_{j=k+1}^{K}p(m_{j}|x,a_{j},\overline{m}_{j-1})dm_{j}\big),
\end{align*}
we can decompose $\check{\mu}_{k}(x,\overline{m}_{k})-\mu_{k}(x,\overline{m}_{k})$
into
\begin{align*}
 & \check{\mu}_{k}(x,\overline{m}_{k})-\mu_{k}(x,\overline{m}_{k})\\
= & \int\big(\hat{\mu}_{K}(x,\overline{m}_{K})-\mu_{K}(x,\overline{m}_{K})\big)\big(\prod_{j=k+1}^{K}\hat{f}_{j}(m_{j}|x,a_{j},\overline{m}_{j-1})dm_{j}\big)\\
 & +\sum_{l=k+1}^{K}\int\mu_{K}(x,\overline{m}_{K})\big(\big(\hat{f}_{l}(m_{l}|x,a_{l},\overline{m}_{l-1})-f_{l}(m_{l}|x,a_{l},\overline{m}_{l-1})\big)dm_{l}\big)\cdot\\
 & \big(\prod_{j=k+1}^{l-1}\hat{f}_{j}(m_{j}|x,a_{j},\overline{m}_{j-1})dm_{j}\big)\big(\prod_{j=l+1}^{K}f_{j}(m_{j}|x,a_{j},\overline{m}_{j-1})dm_{j}\big)\\
= & \int\big(\hat{\mu}_{K}(x,\overline{m}_{K})-\mu_{K}(x,\overline{m}_{K})\big)\underbrace{\Big(\frac{\prod_{j=k+1}^{K}\hat{f}_{j}(m_{j}|x,a_{j},\overline{m}_{j-1})}{\prod_{j=k+1}^{K}f_{j}(m_{j}|x,\overline{m}_{j-1})}\Big)}_{\stackrel{\Delta}{=}g(x,\overline{m}_{K})}\big(\prod_{j=k+1}^{K}f_{j}(m_{j}|x,\overline{m}_{j-1})dm_{j}\big)\\
 & +\sum_{l=k+1}^{K}\int\big(\hat{f}_{l}(m_{l}|x,a_{l},\overline{m}_{l-1})-f_{l}(m_{l}|x,a_{l},\overline{m}_{l-1})\big)\\
 & \underbrace{\frac{\mu_{K}(x,\overline{m}_{K})\big(\prod_{j=k+1}^{l-1}\hat{f}_{j}(m_{j}|x,a_{j},\overline{m}_{j-1})\big)\big(\prod_{j=l+1}^{K}f_{j}(m_{j}|x,a_{j},\overline{m}_{j-1})\big)}{\prod_{j=k+1}^{K}f_{j}(m_{j}|x,\overline{m}_{j-1})}}_{\stackrel{\Delta}{=}h_{l}(x,\overline{m}_{K})}\big(\prod_{j=k+1}^{K}f_{j}(m_{j}|x,\overline{m}_{j-1})dm_{j}\big)
\end{align*}
Using the notation $dP_{2}=\prod_{j=k+1}^{K}f_{j}(m_{j}|x,\overline{m}_{j-1})dm_{j}$
and $dP_{1}=dP_{X}(x)\cdot\prod_{j=1}^{k}f_{j}(m_{j}|x,\overline{m}_{j-1})dm_{j}$,
we have
\begin{align}
 & \Vert\check{\mu}_{k}-\mu_{k}\Vert\nonumber \\
= & \big\Vert\int(\hat{\mu}_{K}-\mu_{K})gdP_{2}+\sum_{l=k+1}^{K}\int(\hat{f}_{l}-f_{l})h_{l}dP_{2}\big\Vert_{P_{1}}\nonumber \\
\leq & \big\Vert\int(\hat{\mu}_{K}-\mu_{K})gdP_{2}\big\Vert_{P_{1}}+\sum_{l=k+1}^{K}\big\Vert\int(\hat{f}_{l}-f_{l})h_{l}dP_{2}\big\Vert_{P_{1}}\nonumber \\
= & \big[\int\big(\int(\hat{\mu}_{K}-\mu_{K})gdP_{2}\big)^{2}dP_{1}\big]^{1/2}+\sum_{l=k+1}^{K}\big[\int\big(\int(\hat{f}_{l}-f_{l})h_{l}dP_{2}\big)^{2}dP_{1}\big]^{1/2}\nonumber \\
\leq & \big[\int\big(\int(\hat{\mu}_{K}-\mu_{K})^{2}dP_{2}\big)\big(\int g^{2}dP_{2}\big)dP_{1}\big]^{1/2}+\sum_{l=k+1}^{K}\big[\int\big(\int(\hat{f}_{l}-f_{l})^{2}dP_{2}\big)\big(\int h_{l}^{2}dP_{2}\big)dP_{1}\big]^{1/2}\;(\textup{Cauchy-Schwartz})\nonumber \\
\leq & \big[\int(\hat{\mu}_{K}-\mu_{K})^{2}dP_{2}dP_{1}\cdot\Vert\int g^{2}dP_{2}\Vert_{P_{1},\infty}\big]^{1/2}+\sum_{l=k+1}^{K}\big[\int(\hat{f}_{l}-f_{l})^{2}dP_{2}dP_{1}\cdot\Vert\int h_{l}^{2}dP_{2}\Vert_{P_{1},\infty}\big]^{1/2}\nonumber \\
= & O_{p}(\Vert\hat{\mu}_{K}-\mu_{K}\Vert)+\sum_{l=k+1}^{K}O_{p}(\Vert\hat{f}_{l}-f_{l}\Vert).\label{eq:mu_bounds}
\end{align}
The last equality uses the assumption that $\mu_{K}(X,\overline{M}_{K})$
(and hence $\int h_{l}^{2}dP_{2}$) is bounded.

From equations \eqref{eq:pi_bounds}-\eqref{eq:mu_bounds}, we have
\begin{align}
R_{2}(\hat{\eta}_{1})= & \sum_{k=0}^{K}O_{p}(\Vert\check{\pi}_{k}-\pi_{k}\Vert)\cdot O_{p}(\Vert\check{\mu}_{k}-\mu_{k}\Vert)\nonumber \\
= & \sum_{k=0}^{K}\big(O_{p}(\Vert\hat{\pi}_{0}-\pi_{0}\Vert)+\sum_{l=1}^{k}O_{p}(\Vert\hat{f}_{l}-f_{l}\Vert)\big)\big(O_{p}(\Vert\hat{\mu}_{K}-\mu_{K}\Vert)+\sum_{l=k+1}^{K}O_{p}(\Vert\hat{f}_{l}-f_{l}\Vert)\big)\label{eq:R2_eta1}
\end{align}
Clearly, when $K+1$ of the $K+2$ nuisance functions in $\eta_{1}$
are consistently estimated, $R_{2}(\hat{\eta}_{1})=o_{p}(1)$. In
this case, since $\mathbb{P}_{n}\varphi_{\overline{a}}(O;\eta_{1}^{*})-P\varphi_{\overline{a}}(O;\eta_{1}^{*})=\mathbb{P}_{n}\varphi_{\overline{a}}(O;\eta_{1}^{*})=o_{p}(1)$,
$\hat{\theta}_{\overline{a}}^{\textup{eif}_{1}}$ is consistent. Moreover,
equation \eqref{eq:R2_eta1} suggests that $R_{2}(\hat{\eta}_{1})=o_{p}(n^{-1/2})$
if $\sum\limits _{u,v\in\hat{\eta}_{1};u\neq v}r_{n}(u)r_{n}(v)=o(n^{-1/2})$.
If the nuisance functions are estimated via parametric models and
their parameter estimates are all $\sqrt{n}$-consistent, $R_{2}(\hat{\eta}_{1})=\sum_{k=0}^{K}O_{p}(n^{-1/2})\cdot O_{p}(n^{-1/2})=o_{p}(n^{-1/2})$,
hence the first part of Theorem 3.

\section{Multiply Robust Decomposition of Between-group Disparities\label{sec:Decomposition-of-Between-group}}

The multiply robust semiparametric estimators can also be used to
estimate \textit{noncausal} decompositions of between-group disparities
(\citealt{fortin2011decomposition}). For example, social scientists
in the United States have a long-standing interest in decomposing
the black-white income gap into components that are attributable to
racial differences in various ascriptive and achieved characteristics.
Using linear structural equation models, \citet{duncan1968inheritance}
decomposed the total black-white income gap into components that reflect
black-white differences in family background, academic performance
(net of family background), educational attainment (net of family
background and academic performance), occupational attainment (net
of family background, academic performance, and educational attainment),
and a ``residual'' component that cannot be explained by the above
characteristics. Although proposed prior to \citet{blinder1973wage}
and \citet{oaxaca1973male}, Duncan's decomposition can be viewed
as a generalization of the Blinder-Oaxaca decomposition widely used
in labor economics.

Duncan's decomposition is similar in form to equation \eqref{eq:g-decomp1},
but it is defined in terms of the statistical parameters $\theta_{\overline{a}}$
rather than the causal parameters $\psi_{\overline{a}}$. Moreover,
the left-hand side is now the black-white income gap rather than the
average causal effect of a manipulable intervention, and, therefore,
there are no pretreatment confounders. It should be noted that this
decomposition is different from causal mediation analysis for a randomized
trial, in which case pretreatment covariates may still be needed to
adjust for potential confounding of the mediator-mediator and mediator-outcome
relationships. The components associated with Duncan's decomposition,
by contrast, are purely statistical parameters and should not be interpreted
causally.

Consequently, in the context of decomposing between-group disparities,
the functional $\theta_{\overline{0}_{k},\underline{1}_{k+1}}$can
be estimated as
\begin{equation}
\hat{\theta}_{\overline{0}_{k},\underline{1}_{k+1}}^{\textup{eif}_{2}}=\mathbb{P}_{n}\big[\frac{\mathbb{I}(A=1)}{\hat{\pi}_{0}(0)}\frac{\hat{\pi}_{k}(0|\overline{M}_{k})}{\hat{\pi}_{k}(1|\overline{M}_{k})}\big(Y-\hat{\mu}_{k}(\overline{M}_{k})\big)+\frac{\mathbb{I}(A=0)}{\hat{\pi}_{0}(0)}\big(\hat{\mu}_{k}(\overline{M}_{k})-\hat{\mu}_{0,k}\big)+\hat{\mu}_{0,k}\big],\label{eq:dFormula}
\end{equation}
where $\pi_{0}(0)=\Pr[A=0]$, $\mu_{k}(\overline{M}_{k})=\mathbb{E}[Y|A=1,\overline{M}_{k}]$,
and $\mu_{0,k}=\mathbb{E}[\mu_{k}(\overline{M}_{k})|A=0]$. Since
$\hat{\pi}_{0}(0)$ can be estimated by the sample average of $1-A$
and $\mu_{0,k}$ the sample average of $\hat{\mu}_{k}(\overline{M}_{k})$
among units with $A=0$, equation \eqref{eq:dFormula} involves estimating
only two nuisance functions: $\hat{\pi}_{k}(a|\overline{m}_{k})$
and $\hat{\mu}_{k}(\overline{m}_{k})$. It follows from Theorem 3
that $\hat{\theta}_{\overline{0}_{k},\underline{1}_{k+1}}^{\textup{eif}_{2}}$
is now doubly robust --- it is consistent if either $\hat{\pi}_{k}(a|\overline{m}_{k})$
or $\hat{\mu}_{k}(\overline{m}_{k})$ is consistent.

To implement the full decomposition, we need to estimate $\theta_{\overline{0}_{k},\underline{1}_{k+1}}$
for each $k\in0,1,\ldots K+1$, i.e., estimate the vector-valued parameter
$\bm{\theta}_{\textup{decomp}}=(\theta_{\underline{1}_{1},}\theta_{0,\underline{1}_{2}},\ldots\theta_{\overline{0}_{K},1},\theta_{\overline{0}_{K+1}})$.
Since $\theta_{\underline{1}_{1}}$ and $\theta_{\overline{0}_{K+1}}$
can be estimated by the sample analogs of $\mathbb{E}[Y|A=1]$ and
$\mathbb{E}[Y|A=0]$ and $\hat{\theta}_{\overline{0}_{k},\underline{1}_{k+1}}^{\textup{eif}_{2}}$
is doubly robust with respect to $\hat{\pi}_{k}$ and $\hat{\mu}_{k}$,
the semiparametric estimator $\hat{\bm{\theta}}_{\textup{decomp}}^{\textup{eif}_{2}}=(\hat{\theta}_{\underline{1}_{1}}^{\textup{eif}_{2}},\hat{\theta}_{0,\underline{1}_{2}}^{\textup{eif}_{2}},\ldots\hat{\theta}_{\overline{0}_{K},1}^{\textup{eif}_{2}},\hat{\theta}_{\overline{0}_{K+1}}^{\textup{eif}_{2}})$
is $2^{K}$-robust: it is consistent if for each $k\in[K]$, either
$\hat{\pi}_{k}$ or $\hat{\mu}_{k}$ is consistent. Note that in this
case, the functions $\mu_{k}(\overline{M}_{k})=\mathbb{E}[Y|A=1,\overline{M}_{k}]$
are not estimated iteratively, but separately for each $k$.
\begin{cor}
Define $\hat{\bm{\theta}}_{\textup{decomp}}^{\textup{eif}_{2}}=(\hat{\theta}_{\underline{1}_{1}}^{\textup{eif}_{2}},\hat{\theta}_{0,\underline{1}_{2}}^{\textup{eif}_{2}},\ldots\hat{\theta}_{\overline{0}_{K},1}^{\textup{eif}_{2}},\hat{\theta}_{\overline{0}_{K+1}}^{\textup{eif}_{2}})$.
Suppose $X=\varnothing$, and that all assumptions required for Theorem
4 hold. When the nuisance functions $(\hat{\pi}_{1},\ldots\hat{\pi}_{K},\hat{\mu}_{1},\ldots\hat{\mu}_{K})$
are estimated via parametric models, $\hat{\bm{\theta}}_{\textup{decomp}}^{\textup{eif}_{2}}$
is CAN if for each $k\in[K]$, either $\hat{\pi}_{k}$ or $\hat{\mu}_{k}$
is correctly specified and its estimates are $\sqrt{n}$-consistent.
$\hat{\bm{\theta}}_{\textup{decomp}}^{\textup{eif}_{2}}$ is semiparametric
efficient if all of the nuisance functions are correctly specified
and their parameter estimates $\sqrt{n}$-consistent. When the nuisance
functions are estimated via data-adaptive methods and cross-fitting,
$\hat{\bm{\theta}}_{\textup{decomp}}^{\textup{eif}_{2}}$ is semiparametric
efficient if all of the nuisance functions are consistently estimated
and $\sum_{k=1}^{K}r_{n}(\hat{\pi}_{k})r_{n}(\hat{\mu}_{k})=o(n^{-1/2})$.
\end{cor}

\section{Additional Details of the Simulation Study\label{sec:Simulation-Details}}

The variables $X_{1},X_{2},X_{3},X_{4},A,M_{1},M_{2},Y$ in the simulation
study are generated via the following model:
\begin{align*}
(U_{1},U_{2},U_{3},U_{XY}) & \sim N(0,I_{4}),\\
X_{j} & \sim N((U_{1},U_{2},U_{3},U_{XY})\beta_{X_{j}},1),\quad j=1,2,3,4,\\
A & \sim\textup{Bernoulli}\big(\textup{logit}^{-1}[(1,X_{1},X_{2},X_{3},X_{4})\beta_{A}]\big),\\
M_{1} & \sim N\big((1,X_{1},X_{2},X_{3},X_{4},A)\beta_{M_{1}},1\big),\\
M_{2} & \sim N\big((1,X_{1},X_{2},X_{3},X_{4},A,M_{1})\beta_{M_{2}},1\big),\\
Y & \sim N\big((1,U_{XY},X_{1},X_{2},X_{3},X_{4},A,M_{1},M_{2})\beta_{Y},1\big).
\end{align*}
The coefficients $\beta_{X_{j}}(\text{1\ensuremath{\leq j\leq4}})$
and $\beta_{Y}$ are drawn from $\textup{Uniform}[-1,1]$, the coefficients
$\beta_{A}$ are drawn from $\textup{Uniform}[-0.5,0.5]$, and the
coefficients $\beta_{M_{1}}$ and $\beta_{M_{2}}$ are drawn from
$\textup{Uniform}[0,0.5]$. Specifically, 
\begin{align*}
\beta_{X_{1}} & =(0.77,-0.86,0.35,0.88),\\
\beta_{X_{2}} & =(-0.99,-0.72,-0.1,0.54),\\
\beta_{X_{3}} & =(-0.74,0.1,0.91,0.46),\\
\beta_{X_{4}} & =(-0.21,-0.43,-0.21,-0.7),\\
\beta_{A} & =(-0.36,-0.08,-0.06,0.4,-0.14),\\
\beta_{M_{1}} & =(0,0.3,0.42,0.48,0.28,0.41),\\
\beta_{M_{2}} & =(0.04,0.2,0.09,0.12,0.39,0.34,0.24),\\
\beta_{Y} & =(-0.27,-0.1,0.25,0.2,-0.08,0.78,0.76,-0.4,0.96).
\end{align*}
It can be shown that under the above model, the six nuisance functions
$\pi_{0}(a|x)$, $\pi_{1}(a|x,m_{1})$, $\pi_{2}(a|x,m_{1},m_{2})$,
$\mu_{0}(x)$, $\mu_{1}(x,m_{1})$, and $\mu_{2}(x,m_{1},m_{2})$
for any $\theta_{a_{1},a_{2},a}$ can be consistently estimated via
the following GLMs:
\begin{align*}
\pi_{0}(1|X) & =\textup{logit}^{-1}[(1,X_{1},X_{2},X_{3},X_{4})\gamma_{0}],\\
\pi_{1}(1|X,M_{1}) & =\textup{logit}^{-1}[(1,X_{1},X_{2},X_{3},X_{4},M_{1})\gamma_{1}],\\
\pi_{2}(1|X,M_{1},M_{2}) & =\textup{logit}^{-1}[(1,X_{1},X_{2},X_{3},X_{4},M_{1},M_{2})\gamma_{2}],\\
\mathbb{E}[Y|X,A,M_{1},M_{2}] & =(1,X_{1},X_{2},X_{3},X_{4},A,M_{1},M_{2})\alpha_{2},\quad\mu_{2}(X,M_{1},M_{2})=\mathbb{E}[Y|X,A=a,M_{1},M_{2}],\\
\mathbb{E}[\mu_{2}(X,M_{1},M_{2})|X,A,M_{1}] & =(1,X_{1},X_{2},X_{3},X_{4},A,M_{1})\alpha_{1},\quad\mu_{1}(X,M_{1})=\mathbb{E}[\mu_{2}(X,M_{1},M_{2})|X,A=a_{2},M_{1}],\\
\mathbb{E}[\mu_{1}(X,M_{1})|X,A] & =(1,X_{1},X_{2},X_{3},X_{4},A)\alpha_{0},\quad\mu_{0}(X)=\mathbb{E}[\mu_{1}(X,M_{1})|X,A=a_{1}].
\end{align*}
To demonstrate the multiple robustness of the EIF-based estimators,
we use a set of ``false covariates'' $Z=\big(X_{1},e^{X_{2}/2},(X_{3}/X_{1})^{1\text{/3}},X_{4}/(e^{X_{1}/2}+1)\big)$
to fit a misspecified model for each of the nuisance functions:
\begin{align*}
\pi_{0}(1|Z) & =\textup{logit}^{-1}[(1,Z_{1},Z_{2},Z_{3},Z_{4})\tilde{\gamma}_{0}],\\
\pi_{1}(1|Z,M_{1}) & =\textup{logit}^{-1}[(1,Z_{1},Z_{2},Z_{3},Z_{4},M_{1})\tilde{\gamma}_{1}],\\
\pi_{2}(1|Z,M_{1},M_{2}) & =\textup{logit}^{-1}[(1,Z_{1},Z_{2},Z_{3},Z_{4},M_{1},M_{2})\tilde{\gamma}_{2}],\\
\mathbb{E}[Y|Z,A,M_{1},M_{2}] & =(1,Z_{1},Z_{2},Z_{3},Z_{4},A,M_{1},M_{2})\tilde{\alpha}_{2},\quad\mu_{2}(Z,M_{1},M_{2})=\mathbb{E}[Y|Z,A=a,M_{1},M_{2}],\\
\mathbb{E}[\mu_{2}(Z,M_{1},M_{2})|Z,A,M_{1}] & =(1,Z_{1},Z_{2},Z_{3},Z_{4},A,M_{1})\tilde{\alpha}_{1},\quad\mu_{1}(Z,M_{1})=\mathbb{E}[\mu_{2}(Z,M_{1},M_{2})|Z,A=a_{2},M_{1}],\\
\mathbb{E}[\mu_{1}(Z,M_{1})|Z,A] & =(1,Z_{1},Z_{2},Z_{3},Z_{4},A)\tilde{\alpha}_{0},\quad\mu_{0}(Z)=\mathbb{E}[\mu_{1}(Z,M_{1})|Z,A=a_{1}].
\end{align*}
Each of the five cases described in Section \ref{sec:A-Simulation-Study}
reflects a combination of estimated nuisance functions from these
correctly and incorrectly specified models. For example, in case (a),
all parametric estimators of $\textup{cPSE}_{M_{2}}$ use correctly
specified models for $\pi_{0}(1|x),\pi_{1}(1|x,m_{1}),\pi_{2}(1|x,m_{1},m_{2})$
and incorrectly specified models for $\mu_{0}(x)$, $\mu_{1}(x,m_{1})$,
and $\mu_{2}(x,m_{1},m_{2})$.

\section{Additional Details of the NLSY97 Data\label{sec:NLSY97}}

\renewcommand{\thefigure}{\thesection\arabic{figure}}
\setcounter{figure}{0} 
\renewcommand{\thetable}{\thesection\arabic{table}}
\setcounter{table}{0} 

\noindent The data source for the empirical example comes from the
National Longitudinal Survey of Youth, 1997 cohort (NLSY97). The NLSY97
began with a nationally representative sample of 8,984 men and women
residing in the United States at ages 12-17 in 1997. These individuals
were interviewed annually through 2011 and biennially thereafter.
Table \ref{tab:Group-specific-Means} reports the sample means of
the pretreatment covariates $X$, the mediators $M_{1}$ and $M_{2}$,
and the outcome $Y$ described in the main text, both overall and
separately for treated and untreated units (i.e., college goers and
non-college-goers). Parental education is measured using mother's
years of schooling; when mother's years of schooling is unavailable,
it is measured using father's years of schooling. Parental income
is measured as the average annual parental income from 1997 to 2001.
The mediator $M_{2}$, which gauges civic and political interest,
includes four components: volunteerism, community participation, donation
activity, and political interest. Volunteerism represents the respondent's
self-reported frequency of volunteering work over the past 12 months
(1: None; 2: 1 - 4 times; 3: 5 - 11 times; 4: 12 times or more). Community
participation represents the respondent's self-reported frequency
of attending a meeting or event for a political, environmental, or
community group (1: None; 2: 1 - 4 times; 3: 5 - 11 times; 4: 12 times
or more). Donation activity is a dichotomous variable indicating whether
the respondent donated money to a political, environmental, or community
cause over the past 12 months. Political interest represents the respondent's
self-reported frequency of following government and public affairs
(1: hardly at all; 2: only now and then; 3: some of the time; 4: most
of the time). Volunteerism, community participation, and donation
activity were measured in 2007, and political interest was measured
in both 2008 and 2010. For simplicity, we use the average of the 2008
and 2010 measures of political interest in our analyses (Treating
them as separate variables leads to almost identical results).
\begin{table}
\caption{Overall and group-specific means in pretreatment covariates, mediators,
and outcome.\label{tab:Group-specific-Means}}
\smallskip{}

\noindent \begin{centering}
\begin{tabular}{>{\raggedright}m{2.3cm}lccc}
\hline 
\noalign{\vskip-0.1cm}
 & \multirow{1}{*}{} & Overall & Non-College-Goers & College Goers\tabularnewline[-0.1cm]
\hline 
\noalign{\vskip-0.1cm}
\multirow{21}{2.3cm}{Pretreatment Covariates ($X$)} & Age at 1997 & 15.98 & 16.02 & 15.96\tabularnewline[-0.1cm]
\noalign{\vskip-0.1cm}
 & Female & 0.5 & 0.42 & 0.55\tabularnewline[-0.1cm]
\noalign{\vskip-0.1cm}
 & Black & 0.16 & 0.22 & 0.13\tabularnewline[-0.1cm]
\noalign{\vskip-0.1cm}
 & Hispanic & 0.12 & 0.15 & 0.1\tabularnewline[-0.1cm]
\noalign{\vskip-0.1cm}
 & Parental Education & 13.08 & 12.05 & 13.71\tabularnewline[-0.1cm]
\noalign{\vskip-0.1cm}
 & Parental Income & 86,520 & 60,706 & 102,568\tabularnewline[-0.1cm]
\noalign{\vskip-0.1cm}
 & Parental Assets & 119,242 & 62,573 & 154,550\tabularnewline[-0.1cm]
\noalign{\vskip-0.1cm}
 & Lived with Both Biological Parents & 0.53 & 0.39 & 0.62\tabularnewline[-0.1cm]
\noalign{\vskip-0.1cm}
 & Presence of a Father Figure & 0.76 & 0.68 & 0.8\tabularnewline[-0.1cm]
\noalign{\vskip-0.1cm}
 & Lived in Rural Area & 0.27 & 0.29 & 0.26\tabularnewline[-0.1cm]
\noalign{\vskip-0.1cm}
 & Lived in the South & 0.37 & 0.39 & 0.35\tabularnewline[-0.1cm]
\noalign{\vskip-0.1cm}
 & ASVAB Percentile Score & 53.4 & 37.26 & 62.72\tabularnewline[-0.1cm]
\noalign{\vskip-0.1cm}
 & High School GPA & 2.9 & 2.5 & 3.16\tabularnewline[-0.1cm]
\noalign{\vskip-0.1cm}
 & Substance Use Index & 1.36 & 1.56 & 1.23\tabularnewline[-0.1cm]
\noalign{\vskip-0.1cm}
 & Delinquency Index & 1.54 & 2.06 & 1.22\tabularnewline[-0.1cm]
\noalign{\vskip-0.1cm}
 & Had Children by Age 18 & 0.06 & 0.11 & 0.02\tabularnewline[-0.1cm]
\noalign{\vskip-0.1cm}
 & 75\%+ of Peers Expected College & 0.56 & 0.41 & 0.66\tabularnewline[-0.1cm]
\noalign{\vskip-0.1cm}
 & 90\%+ of Peers Expected College & 0.19 & 0.12 & 0.24\tabularnewline[-0.1cm]
\noalign{\vskip-0.1cm}
 & Property Ever Stolen at School & 0.24 & 0.27 & 0.22\tabularnewline[-0.1cm]
\noalign{\vskip-0.1cm}
 & Ever Threatened at School & 0.19 & 0.27 & 0.14\tabularnewline[-0.1cm]
\noalign{\vskip-0.1cm}
 & Ever in a Fight at School & 0.12 & 0.18 & 0.08\tabularnewline[-0.1cm]
\hline 
\noalign{\vskip-0.1cm}
\multirow{1}{2.3cm}{Mediator $M_{1}$} & Average Earnings in 2006-2009 & 33,600 & 25,082 & 38,899\tabularnewline[-0.1cm]
\hline 
\noalign{\vskip-0.1cm}
\multirow{4}{2.3cm}{Mediator $M_{2}$} & Volunteerism & 1.57 & 1.46 & 1.64\tabularnewline[-0.1cm]
\noalign{\vskip-0.1cm}
 & Community Participation & 1.26 & 1.17 & 1.32\tabularnewline[-0.1cm]
\noalign{\vskip-0.1cm}
 & Donation Activity & 0.3 & 0.22 & 0.35\tabularnewline[-0.1cm]
\noalign{\vskip-0.1cm}
 & Political Interest & 2.63 & 2.34 & 2.81\tabularnewline[-0.1cm]
\hline 
\noalign{\vskip-0.1cm}
Outcome ($Y$) & Voted in the 2010 General Election & 0.45 & 0.3 & 0.54\tabularnewline[-0.1cm]
\hline 
\noalign{\vskip-0.1cm}
\multicolumn{2}{l}{Sample Size} & 2,976 & 1,240 & 1,736\tabularnewline[-0.1cm]
\hline 
\end{tabular}\smallskip{}
\par\end{centering}
\raggedright{}\quad Note: All statistics are calculated using NLSY97
sampling weights.
\end{table}

To gain a basic understanding of the treatment-mediator and mediator-outcome
relationships in this dataset, we fit a linear regression model for
each component of the mediators and for the outcome given their antecedent
variables (including the pretreatment covariates). These models, if
correctly specified, will identify the causal effects of $A$ on $M_{1}$,
$(A,M_{1})$ on $M_{2}$, and $(A,M_{1},M_{2})$ on $Y$ under the
conditional independence assumptions described in Section \ref{subsec:Two Mediators}.
The coefficients of these regression models are shown in Table \ref{tab:Regression-Models}.
The first column indicates a substantively strong and statistically
significant effect of college attendance on log earnings: adjusting
for pretreatment covariates, attending college by age 20 is associated
with a 44.3 percent increase ($e^{0.367}-1=0.443$) in estimated earnings
from 2006 to 2009. The next four columns suggest that the direct effects
of college attendance on volunteerism, community participation, and
donation activity (i.e., $A\to M_{2}$) are relatively small and not
statistically significant. The estimated direct effect of college
attendance on political interest, by contrast, is much larger and
statistically significant. The last column shows statistically significant
effects of volunteerism, community participation, and political interest
on voting (at the $p<0.05$ level). The estimated effect of political
interest is particularly strong: a one unit increase in the four-point
scale of political interest is associated with a 14.8 percentage point
increase in the estimated probability of voting. The coefficient of
college attendance in the last model can be interpreted as the direct
effect of college on voting (i.e., $A\to Y$), i.e., the effect that
operates neither through economic status nor through civic and political
interest. The estimate, 11.7 percentage points, is comparable to our
semiparametric estimates reported in the main text.
\begin{table}
\caption{Regression models for the mediators and the outcome.\label{tab:Regression-Models}}

\noindent \begin{centering}
\begin{tabular}{>{\raggedright}m{2.1cm}|>{\centering}m{1.8cm}|>{\centering}m{2.1cm}>{\centering}m{2.2cm}>{\centering}m{1.8cm}>{\centering}m{1.8cm}|>{\centering}m{1.5cm}}
\hline 
 & $M_{1}$ & \multicolumn{4}{c|}{$M_{2}$} & $Y$\tabularnewline
 & Log  Earnings & Volunteerism & Community Participation & Donation Activity & Political Interest & Voting\tabularnewline
\hline 
College Attendance & 0.367 (0.055) & 0.038

(0.046) & 0.039

(0.028) & 0.036 (0.023) & 0.259 (0.046) & 0.117 (0.023)\tabularnewline
Log Earnings &   & 0.001

(0.017) & 0.008

(0.011) & 0.036 (0.009) & 0.050 (0.016) & 0.016 (0.009)\tabularnewline
Volunteerism &   &   &   &   &   & 0.028 (0.013)\tabularnewline
Community Participation &   &   &   &   &   & 0.041 (0.019)\tabularnewline
Donation Activity &   &   &   &   &   & -0.007 (0.024)\tabularnewline
Political Interest &   &   &   &   &   & 0.148 (0.010)\tabularnewline
\hline 
\end{tabular}\smallskip{}
\par\end{centering}
\raggedright{}\quad Note: Regression coefficients for the pretreatment covariates are omitted. Numbers in\\
\quad parentheses are heteroskedasticity-robust standard errors, which are adjusted for multiple\\
\quad imputation via Rubin's (1987)  method.
\end{table}

\end{document}